\documentclass[twocolumn]{aastex63}
\usepackage{epsfig}
\usepackage{natbib}
\usepackage{amsmath}
\usepackage{amsfonts}
\usepackage{hyperref}
\usepackage{amsfonts,graphicx,subfigure,color,gensymb,longtable}
\usepackage{afterpage}

\usepackage{xcolor}
\definecolor{forestgreen}{rgb}{0.13, 0.75, 0.13}

\newcommand{\red}{\color{red}}

\newcommand{\p}{^\prime}

\def\LCDM{$\Lambda$CDM}

\def\fLCDMOMPan{$0.334\pm0.018$}
\def\fLCDMOLPan{$0.666\pm0.018$}
\def\fLCDMHPanSH{$73.6\pm1.1$}

\def\oLCDMOMPan{$0.306\pm0.057$}
\def\oLCDMOLPan{$0.625\pm0.084$}
\def\oLCDMHPanSH{$73.4\pm1.1$}

\def\fomimprov{$2$}

\def\wPanSH{$-0.90\pm 0.14$}
\def\OMPanSH{$0.309^{+0.063}_{-0.069}$}
\def\OLPanSH{$0.691^{+0.069}_{-0.063}$}
\def\HPanSH{$73.5\pm1.1$}

\def\wPanPlanck{$-0.982^{+0.022}_{-0.038}$}
\def\OMPanPlanck{$0.325^{+0.010}_{-0.008}$}
\def\OLPanPlanck{$0.675^{+0.008}_{-0.010}$}
\def\HPanPlanck{$66.49^{+0.50}_{-0.83}$}

\def\wPanPlanckBAO{$-0.978^{+0.024}_{-0.031}$}
\def\OMPanPlanckBAO{$0.316^{+0.005}_{-0.008}$}
\def\OLPanPlanckBAO{$0.684^{+0.008}_{-0.005}$}
\def\HPanPlanckBAO{$66.87^{+1.00}_{-0.32}$}

\def\wPanPlanckgBAO{$-0.974^{+0.025}_{-0.031}$}
\def\OMPanPlanckgBAO{$0.319^{+0.006}_{-0.007}$}
\def\OLPanPlanckgBAO{$0.681^{+0.007}_{-0.006}$}
\def\HPanPlanckgBAO{$66.78^{+0.76}_{-0.50}$}

\def\wwawnPan{$-0.93\pm0.15$}
\def\wwawaPan{$-0.1^{+0.9}_{-2.0}$}
\def\wwaOMPan{$0.403^{+0.054}_{-0.098}$}
\def\wwaOLPan{$0.597^{+0.098}_{-0.054}$}
\def\wwaHPanSH{$73.3\pm1.1$}

\def\wwawPanPlanck{$-0.851^{+0.092}_{-0.099}$}
\def\wwawaPanPlanck{$-0.70^{+0.49}_{-0.51}$}
\def\wwaOMPanPlanck{$0.318^{+0.012}_{-0.014}$}
\def\wwaOLPanPlanck{$0.682^{+0.014}_{-0.012}$}
\def\wwaHPanPlanck{$67.4^{+1.1}_{-1.2}$}

\def\wwaHPanPlanckBAO{$67.41^{+0.52}_{-0.82}$}
\def\wwaOLPanPlanckBAO{$0.684^{+0.005}_{-0.009}$}
\def\wwawPanPlanckBAO{$-0.841^{+0.066}_{-0.061}$}

\def\wwawaPanPlanckBAO{$-0.65^{+0.28}_{-0.32}$}

\def\wwaOMPanPlanckBAO{$0.316^{+0.009}_{-0.005}$}

\def\wwaHPanPlanckgBAO{$67.12^{+0.71}_{-0.69}$}
\def\wwaOLPanPlanckgBAO{$0.682^{+0.006}_{-0.008}$}
\def\wwawPanPlanckgBAO{$-0.878^{+0.063}_{-0.069}$}
\def\wwawaPanPlanckgBAO{$-0.45^{+0.29}_{-0.32}$}
\def\wwaOMPanPlanckgBAO{$0.318^{+0.009}_{-0.006}$}

\def\nunique{1550}

\usepackage{color}
\usepackage{soul} 
\usepackage{comment}

\def\sc{14}

\def\red{\color{red}}

\begin{document}

\title{The Pantheon+ Analysis: Cosmological Constraints}

\author{Dillon Brout}
\affil{Center for Astrophysics, Harvard \& Smithsonian, 60 Garden Street, Cambridge, MA 02138, USA}
\affil{NASA Einstein Fellow}
\email{dillon.brout@cfa.harvard.edu}

\author{Dan Scolnic}
\affil{Department of Physics, Duke University, Durham, NC, 27708, USA}

\author{Brodie Popovic}
\affil{Department of Physics, Duke University, Durham, NC, 27708, USA}

\author{Adam G.\ Riess}
\affil{Space Telescope Science Institute, 3700 San Martin Drive, Baltimore, MD 21218, USA}
\affil{Department of Physics and Astronomy, Johns Hopkins University, Baltimore, MD 21218 USA}

\author{Anthony Carr}
\affil{School of Mathematics and Physics, University of Queensland, Brisbane, QLD 4072, Australia}

\author{Joe Zuntz}
\affil{Institute for Astronomy, University of Edinburgh, Edinburgh EH9 3HJ, United Kingdom}

\author{Rick Kessler}
\affil{Kavli Institute for Cosmological Physics, University of Chicago, Chicago, IL 60637, USA}
\affil{Department of Astronomy and Astrophysics, University of Chicago, Chicago, IL 60637, USA}


\author{Tamara M. Davis}
\affil{School of Mathematics and Physics, University of Queensland, Brisbane, QLD 4072, Australia}

\author{Samuel Hinton}
\affil{School of Mathematics and Physics, University of Queensland, Brisbane, QLD 4072, Australia}

\author{David Jones}
\affil{Department of Astronomy and Astrophysics, University of California, Santa Cruz, CA 92064, USA}
\affil{NASA Einstein Fellow}

\author{W. D'Arcy Kenworthy}
\affil{Department of Physics and Astronomy, Johns Hopkins University, Baltimore, MD 21218 USA}

\author{Erik R. Peterson}
\affil{Department of Physics, Duke University, Durham, NC, 27708, USA}

\author{Khaled Said}
\affil{School of Mathematics and Physics, University of Queensland, Brisbane, QLD 4072, Australia}

\author{Georgie Taylor}
\affil{Research School of Astronomy and Astrophysics, Australian National University, Canberra, Australia}

\author{Noor Ali}
\affil{Umeå University, 901 87, Umeå, Sweden}

\author{Patrick Armstrong}
\affil{Mt. Stromlo Observatory, The Research School of Astronomy and Astrophysics, Australian National University, ACT 2601, Australia}

\author{Pranav Charvu}
\affil{Department of Physics, Duke University, Durham, NC, 27708, USA}

\author{Arianna Dwomoh}
\affil{Department of Physics, Duke University, Durham, NC, 27708, USA}

\author{Cole Meldorf}
\affil{Department of Astronomy and Astrophysics, University of Chicago, Chicago, IL 60637, USA}

\author{Antonella Palmese}
\affil{Department of Physics, University of California, Berkeley, CA 94720-7300, USA}

\author{Helen Qu}
\affil{Department of Physics and Astronomy, University of Pennsylvania, Philadelphia, PA 19104, USA}

\author{Benjamin M.\ Rose}
\affil{Department of Physics, Duke University, Durham, NC, 27708, USA}

\author{Bruno Sanchez}
\affil{Department of Physics, Duke University, Durham, NC, 27708, USA}

\author{Christopher W. Stubbs}
\affil{Department of Physics, Harvard University 17 Oxford Street, Cambridge, MA 02138, USA}
\affil{Center for Astrophysics, Harvard \& Smithsonian, 60 Garden Street, Cambridge, MA 02138, USA}

\author{Maria Vincenzi}
\affil{Department of Physics, Duke University, Durham, NC, 27708, USA}

\author{Charlotte M. Wood}
\affil{Department of Physics and Astronomy, University of Notre Dame, Notre Dame, IN 46556, USA}


\author{Peter J.\ Brown}
\affil{Department of Physics and Astronomy, Texas A\&M University, 4242 TAMU, College Station, TX 77843, USA}
\affil{George P.\ and Cynthia Woods Mitchell Institute for Fundamental Physics \& Astronomy, College Station, TX 77843, USA}

\author{Rebecca Chen}
\affil{Department of Physics, Duke University, Durham, NC, 27708, USA}

\author{Ken Chambers}
\affil{Institute of Astronomy, University of Hawaii, 2680 Woodlawn Drive, Honolulu, HI 96822, USA}

\author{David A.\ Coulter}
\affil{Department of Astronomy and Astrophysics, University of California, Santa Cruz, CA 92064, USA}

\author{Mi Dai}
\affil{Department of Physics and Astronomy, Johns Hopkins University, Baltimore, MD 21218 USA}

\author{Georgios Dimitriadis}
\affil{School of Physics, Trinity College Dublin, The University of Dublin, Dublin 2, Ireland}

\author{Alexei V.\ Filippenko}
\affil{Department of Astronomy, University of California, Berkeley, CA 94720-3411, USA}

\author{Ryan J.\ Foley}
\affil{Department of Astronomy and Astrophysics, University of California, Santa Cruz, CA 92064, USA}

\author{Saurabh W.\ Jha}
\affiliation{Department of Physics and Astronomy, Rutgers, the State University of New Jersey, Piscataway, NJ 08854, USA }

\author{Lisa Kelsey}
\affil{Institute of Cosmology and Gravitation, University of Portsmouth, Portsmouth, PO1 3FX, UK}

\author{Robert P.\ Kirshner}
\affil{Gordon and Betty Moore Foundation, Palo Alto, CA 94304, USA}
\affil{Center for Astrophysics, Harvard \& Smithsonian, 60 Garden Street, Cambridge, MA 02138, USA}

\author{Anais M{\"o}ller}
\affil{Centre for Astrophysics \& Supercomputing, Swinburne University of Technology, Victoria 3122, Australia}
\affil{LPC, Université Clermont Auvergne, CNRS/IN2P3, F-63000 Clermont-Ferrand, France}

\author{Jessie Muir}
\affil{Perimeter Institute for Theoretical Physics, 31 Caroline St. North, Waterloo, ON N2L 2Y5, Canada}

\author{Seshadri Nadathur}
\affil{Department of Physics \& Astronomy, University College London, Gower Street, London, WC1E 6BT, UK}

\author{Yen-Chen Pan}
\affil{Graduate Institute of Astronomy, National Central University, 32001 Jhongli, Taiwan}

\author{Armin Rest}
\affil{Space Telescope Science Institute, 3700 San Martin Drive, Baltimore, MD 21218, USA}

\author{Cesar Rojas-Bravo}
\affil{Department of Astronomy and Astrophysics, University of California, Santa Cruz, CA 92064, USA}

\author{Masao Sako}
\affil{Department of Physics and Astronomy, University of Pennsylvania, Philadelphia, PA 19104, USA}

\author{Matthew R.\ Siebert}
\affil{Department of Astronomy and Astrophysics, University of California, Santa Cruz, CA 92064, USA}

\author{Mat Smith}
\affil{Université de Lyon, Université Claude Bernard Lyon 1, CNRS/IN2P3, IP2I Lyon, F-69622, Villeurbanne, France}

\author{Benjamin E.\ Stahl}
\affil{Department of Astronomy, University of California, Berkeley, CA 94720-3411, USA}

\author{Phil Wiseman}
\affil{School of Physics and Astronomy, University of Southampton, Southampton SO17 1BJ, UK}

\acceptjournal{the Astrophysical Journal}


\vspace{8mm}
\begin{abstract}

We present constraints on cosmological parameters from the Pantheon+ analysis of 1701 light curves of \nunique\ distinct Type Ia supernovae (SNe~Ia) ranging in redshift from $z=0.001$ to 2.26. This work features an increased sample size from the addition of multiple cross-calibrated photometric systems of SNe covering an increased redshift span, and improved treatments of systematic uncertainties in comparison to the original Pantheon analysis which together result in a factor of $2\times$ improvement in cosmological constraining power. For a Flat$\Lambda$CDM model, we find $\Omega_M=$~\fLCDMOMPan\ from SNe Ia alone. For a Flat$w_0$CDM model, we measure $w_0=$~\wPanSH\ from SNe~Ia alone, H$_0=$~\HPanSH~{km\,s$^{-1}$\,Mpc$^{-1}$} when including the Cepheid host distances and covariance (SH0ES), and $w_0=$~\wPanPlanckBAO\ when combining the SN likelihood with Planck constraints from the cosmic microwave background (CMB) and baryon acoustic oscillations (BAO); both $w_0$ values are consistent with a cosmological constant. We also present the most precise measurements to date on the evolution of dark energy in a Flat$w_0w_a$CDM universe, and measure $w_a=$~\wwawaPan\ from Pantheon+ SNe~Ia alone, H$_0=$~\wwaHPanSH~{km\,s$^{-1}$\,Mpc$^{-1}$} when including SH0ES Cepheid distances, and $w_a=$~\wwawaPanPlanckBAO\ when combining Pantheon+ SNe~Ia with CMB and BAO data. Finally, we find that systematic uncertainties in the use of SNe~Ia along the distance ladder comprise {less than} one third of the total uncertainty in the measurement of H$_0$ and cannot explain the present ``Hubble tension" between local measurements and early-Universe predictions from the cosmological model.

\end{abstract}



\bigskip
\section{Introduction}

Type Ia supernovae (SNe~Ia) anchor the standard model of cosmology with their unmatched ability to map the past 10 billion years of expansion history.  
SNe~Ia provided the first evidence of the accelerating expansion of the Universe \citep{riess98,perlmutter99}, and they remain invaluable because they are (1) bright enough to be seen at large cosmic distances, (2) common enough to be found in large numbers, and (3) can be standardized to $\sim 0.1$\,mag precision in brightness {or $\sim 5$\% in distance per object.} 

Statistical leverage from large samples of SNe~Ia has grown rapidly over the last 3 decades, and well-calibrated and standardized compilations of these samples have facilitated measurements of the {\it relative} expansion history across the redshift range $0<z<1$ characterized by the equation-of-state parameter of dark energy ($w = P/(\rho c^2)$), and the measurement of the Hubble constant H$_0$, the current expansion rate {determined from {\it absolute} distances}. Measurements of $w$ are constrained from the comparison of standardized SN~Ia magnitudes over a wide range of redshifts obtained from different surveys with different observing-depth strategies. Measurements of H$_0$ require very nearby ($<50$\,Mpc, $\sim1$ discovered per year) SNe~Ia found by multiple surveys in galaxies that host calibrated primary distance indicators [e.g., Cepheids, tip of the red giant branch (TRGB)] which are then compared to SNe in the Hubble flow, often from the same surveys. 

However, simply combining several subsamples into a large sample of SNe~Ia does not provide meaningful gains without rigorous cross-calibration, self-consistent analysis of their light curves and redshifts, and characterization of their numerous sources of related uncertainties or covariance. As samples and compilations grow, ever greater attention must be paid to the control of systematic uncertainties which would otherwise dominate sample uncertainties. 

This analysis, Pantheon+, is the successor to the original Pantheon analysis \citep{Scolnic18} and builds on the analysis framework of the original Pantheon to combine an even larger number of SN~Ia samples and include those that are in galaxies with measured Cepheid distances {in order to be able to simultaneously constrain parameters describing the full expansion history (e.g., $\Omega_M$, $w_0$, $w_a$) with the local expansion rate (H$_0$)}. The original Pantheon compilation of 1048 SNe~Ia was used to measure a value (from SNe~Ia alone) of $w=-1.090\pm0.220$. \cite{Riess16}, in their measurement of the local expansion rate H$_0$, used a pre-release version of Pantheon based on \cite{Supercal} and further augmented the sample as Pantheon did not extend to reach the low redshifts of the primary distance indicators at $z<0.01$. 

Although there was significant overlap in data and analysis between the Pantheon measurement of $w$ and the H$_0$ measurement of \cite{Riess16}, \cite{Riess16} included several Cepheid calibrator SNe Ia that were not included in Pantheon and the fitting for H$_0$ and parameters describing the expansion history were done independently rather than simultaneously.
\cite{dhawan20H0} later established a framework for considering the covariance between SNe in primary distance indicator hosts and SNe in the Hubble flow. We build on that framework, which was developed originally for a redshift-binned Hubble diagram, and in this paper we create the first unbinned sample with covariance extending down to $z=0.001$ that can be used {to propagate correlated systematics for simultaneous measurements of H$_0$, $\Omega_M$, $w_0$, and $w_a$.} We analyze the largest set of cosmologically viable SN~Ia light curves to date, include low-redshift samples to extend the lower bound in redshift to $0.001$ which contains the primary distance indicators (SNe in SH0ES Cepheid host galaxies), propagate systematic uncertainties for both primary distance indicators and higher-redshift SNe simultaneously, and leverage the large strides made in the field of SN~Ia cosmology since the original Pantheon. 

\begin{figure*}
        \centering 
	    \includegraphics[width=.85\textwidth]{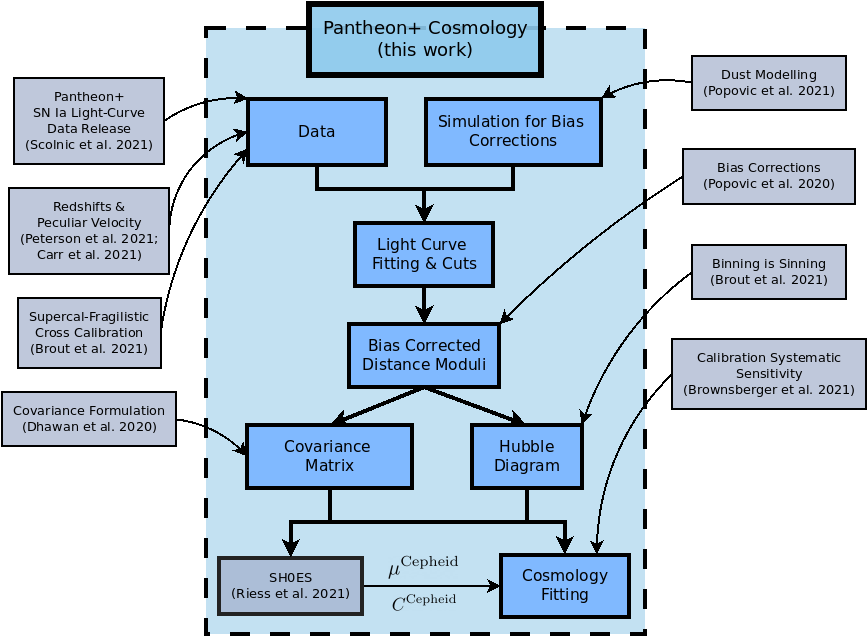}  
        \caption{ Analysis roadmap of this work and supporting/complementary Pantheon+ and SH0ES papers. Components of the analysis here are shown in blue. The companion paper R22, which provides a constraint on H$_0$, requires the Hubble diagram and covariance computed in this work. Likewise, measurements of H$_0$ in this work require the R22 Cepheid distance and covariance. Supporting papers are shown in gray boxes.}
        \label{fig:roadmap} \vspace{2mm}
        
    \end{figure*}

This paper is the culmination of a series of papers that comprise the Pantheon+ analysis.  
A graphic of an overview of the numerous Pantheon+ supporting analyses, on which this paper heavily relies, is shown in Fig.~\ref{fig:roadmap}.
Details of each paper pertinent to this analysis are described in Section~\ref{sec:Inputs}. In short, these papers include \cite[][hereafter S22]{plussample}, which describes the sample of 1701 cosmologically viable SN~Ia light curves {of \nunique\ distinct SNe}, which we will refer to as ``the Pantheon+ sample." The redshifts and peculiar velocities of the SNe used here are given by \cite{carr21} and a comprehensive analysis of peculiar velocities is presented by \cite{peterson21}. The cross-calibration of the different photometric systems used in this analysis can be found in \cite[][hereafter Fragilistic]{fragilistic}, and calibration-related systematic uncertainty limits are determined by \cite{brownsberger21}. The underlying SN~Ia populations describing the dataset are given by \cite{popovic21a}. The model for intrinsic {brightness} variations was developed by \cite{bs20} and then improved and evaluated by \cite{popovic21b}. The novel systematic framework for simultaneous measurement of H$_0$ and cosmology is developed by \cite{dhawan20H0}, and {improved methodology for systematic uncertainties is described by} \cite{binning21}. 

In this work we discuss briefly the aforementioned papers in the context of their use in this analysis, evaluate several additional systematic uncertainties not addressed in these works, measure cosmological parameters, {examine additional signals in the Hubble diagram}, and compile systematic uncertainty budgets on cosmological parameters. A companion paper by the SH0ES Team \citep[][hereafter R22]{SH0ES21} combines {from this work 277 Hubble flow (0.023$<z<$0.15) SNe~Ia and 42 SNe~Ia in Cepheid-calibrator hosts, their relative distances, and their covariance, with the absolute distances of primary distance anchors (Cepheids and TRGB) from R22 in order to measure H$_0$ under the assumption of Flat$\Lambda$CDM. Similarly, in this work we utilize the full Pantheon+ sample of \nunique\ SNe~Ia in combination with the R22 Cepheid host distances to show the impact of cosmological models with more freedom than those used in R22 as well as the impact of SN-related systematic uncertainties on inferences of H$_0$.}

An important aspect of this work is the public release of the data and simulations used here that allow for the reproduction of multiple different stages of this analysis. In Appendix~\ref{appendix:dr}, we present the numerous products that will be made available, including SN distances, redshifts, uncertainties, covariance, and extensive \texttt{SNANA} simulations \citep{Kessler2009} of the data that model astrophysical effects, cosmological effects, and the observation/telescope effects of each survey down to the level of cadence, weather history, etc. We encourage the community to validate alternate analyses {of the publicly released Pantheon+ sample} on these simulations.

The structure of the paper is as follows.  In Section~\ref{sec:Method}, we describe the methodology from fitting SN light curves to constraining cosmological parameters. Section~\ref{sec:Inputs} summarizes all of the inputs to the analysis including the data sample, calibration, and redshifts. In Section~\ref{sec:Results}, we describe the cosmological results. Sections~\ref{sec:Discussion} and~\ref{sec:Conclusions} are our discussions and conclusions, respectively.

\section{Methodology of Constraining Cosmological Parameters with SNe~Ia}
\label{sec:Method}

\subsection{Measuring Distances to SNe~Ia}
To standardize the SN~Ia brightnesses we fit light curves using \texttt{SNANA} with the SALT2 model as originally developed by \cite{Guy10} { and updated in \citealp[hereafter SALT2-B22]{fragilistic}}. For each SN, the SALT2 light-curve fit returns four parameters: the light-curve amplitude $x_0$ where $m_B \equiv -2.5{\rm log}_{10}(x_0)$; $x_1$, the stretch parameter corresponding to light-curve width; $c$, the light-curve color {that includes contributions from both intrinsic color and dust}; and $t_0$, the time of peak brightness. {Extinction due to Milky Way dust is accounted for in the SALT2 light-curve fitting.}
From the parameters $m_B$ and $x_1,c$, we standardize the SN brightnesses and infer distance modului ($\mu$), used in the Hubble diagram, with a modified version of the \citet{Tripp98} distance estimator. 
Following \cite[][hereafter BBC]{BBC}, the distance modulus is defined as
\begin{equation}
    \mu=m_B + \alpha x_1 - \beta c - M  - \delta_{\rm bias} +\delta_{\rm host},~\label{eq:Tripp}
\end{equation}
where $\alpha$ and $\beta$ are global nuisance parameters relating stretch and color (respectively) to luminosity. $M$ is the fiducial magnitude of an SN~Ia, which can be calibrated by setting an absolute distance scale with primary distance anchors such as Cepheids. $\delta_{\rm bias}$ is a correction term\footnote{Past analyses have the opposite sign $ +\delta_{\rm bias}$; however, since the values of $\delta_{\rm bias}$ in the public release are meant to be subtracted, we change the sign compared to previous works.} to account for selection biases that is determined from simulations following \cite{popovic21a}, described in detail in Appendix~\ref{appendix:biascor}.
$\delta_{\rm host}$ is the luminosity correction (step) for residual correlations between the standardized brightness of an SN~Ia and the host-galaxy mass,
\begin{equation}
\delta_{\rm host} = \gamma \times (1 + e^{(M_\star - S)/ \tau_{M_\star}})^{-1} - \frac{\gamma}{2}~, \label{eq:BCmu2}
\end{equation}
where $\gamma$ is the magnitude of the SN~Ia luminosity differences between SNe in high ($M_\star>10^{10}\,M_\odot$) and low ($M_\star<10^{10}\,M_\odot$) stellar mass galaxies and where `hostless' SNe have been assumed to reside in galaxies with low stellar mass. $M_\star$ is the inferred stellar mass measured in units of solar mass ($M_\odot$) from spectral energy distribution (SED) fitting to the photometry of each host galaxy, $S$ is the step location (nominal analysis assumes $S=10^{10}\,M_\odot$), and $\tau_{M\star}$ describes the width of the step.

The total distance modulus error, $\sigma_\mu$, for SN $i$ is described as 
\begin{multline}
        \sigma_{\mu,i}^2 = f(z_i,c_i,M_{\star,i})\sigma^2_{{\rm meas},i} + \sigma^2_{\rm floor}(z_i,c_i,M_{\star,i}) \\ + \sigma^2_{{\rm lens},i} + \sigma^2_{{z},i} + \sigma^2_{{\rm vpec},i \, ,} 
    \label{eq:TrippErr}
\end{multline}
where $\sigma_{\rm meas}$ is the measurement uncertainty of SALT2 light-curve fit parameters and their associated covariances (see Eq.~3 of \citealt{BBC}) resulting from photometric uncertainties. The measurement uncertainty is scaled by $f(z_i,c_i,M_{\star,i})$ specific to each survey in order to account for selection effects that can reduce the observed scatter at the limits of each sample. The uncertainty contribution from gravitational lensing as given by \cite{jonsson10} is $\sigma_{\textrm{lens}} = 0.055z$. {We note that, as discussed by \cite{Kessler18}, the correct lensing distribution is utilized in simulations.} The nominal distance modulus uncertainty contribution due to the combination of redshift measurement uncertainty ($\sigma_{z}$) and peculiar velocity uncertainty ($\sigma_{\rm vpec}$) have both been converted to distance modulus uncertainty under the assumption of a cosmological model. {\cite{chen21} note that the optimal way to characterize redshift measurement uncertainty at high redshifts (e.g., the DES sample, $z>0.3$) is to float the redshift and use the uncertainty in redshift as a prior in the light-curve fit. However, following previous analyses we fix the redshift and include the associated distance {modulus} uncertainty $\sigma_{z}$ in Eq.~\ref{eq:TrippErr}, which is a correct estimate at low redshifts ($z<0.1$).}
Lastly, $\sigma_{\rm floor}$ represents the floor in standardizability owing to intrinsic unmodeled variations in SNe~Ia such that
\begin{equation}
\label{eq:errmodel}
   \sigma^2_{\rm floor}(z_i,c_i,M_{\star,i}) = \sigma^2_{\rm scat}(z_i,c_i,M_{\star,i}) + \sigma^2_{\rm gray},
\end{equation}
where $\sigma^2_{\rm scat}(z_i,c_i,M_{\star,i})$ is determined from a model that describes intrinsic brightness fluctuations { and $\sigma^2_{\rm gray}$ is a single number representing a gray (color independent) floor in standardizability for all SNe~Ia; $\sigma^2_{\rm gray}$ determined after the BBC fitting process in order to bring the Hubble diagram reduced $\chi^2$ to unity. The details of $\sigma^2_{\rm scat}(z_i,c_i,M_{\star,i})$, its model dependence, and its contribution to systematic uncertainties are discussed in further detail in Section \ref{subsubsection:intrinsic} and Appendix~\ref{appendix:biascor}.}

To determine the distance modulus values of all the SNe, we follow the BBC fitting process with updates to increase the dimensionality of bias corrections in \cite{popovic21a}. The likelihood (as given in Eq.~6 of \citealp{BBC}) results in a cosmology-independent minimization of the free parameters ($\alpha$, $\beta$, $\gamma$, $\sigma_{\textrm{gray}}$) that minimize the scatter in the Hubble diagram. 
While the BBC process was designed for utility for photometric cosmology analyses and uses SN~Ia classification probabilities, the data analyzed here are a spectroscopically confirmed SN~Ia sample, and therefore we set the non-Ia SN probabilities to 0 for the whole sample.

\subsection{The Covariance Matrix}

Following \cite{Conley11}, we compute covariance matrices $C_{\rm stat}$ \& $C_{\rm syst}$ to account for statistical and systematic uncertainties and expected correlations between the SN~Ia light curves in the sample when analyzing cosmological models. {BBC produce both a redshift-binned and an unbinned Hubble diagram, enabling both binned and unbinned covariance matrices.
For the original Pantheon \citep{Scolnic18}, JLA \citep{Betoule14}, and DES3YR \citep{Brout18b}, $C_{\rm stat}$ and $C_{\rm syst}$ were redshift-binned matrices (or smoothed as a function of redshift) citing computational limitations. Following \cite{binning21}, in this work we utilize the unbinned Hubble diagrams to create unbinned covariance matrices.} The Pantheon+ sample \citep{plussample} also includes ``duplicate SNe~Ia," SNe~Ia that have been observed simultaneously by numerous different surveys, so that statistical covariance $C_{\rm stat}$ is computed as
\begin{equation}
        C_{\rm stat}(i,j) =
        \begin{cases}
            \vspace{3mm}
            \sigma_\mu^2 & i=j \\
            \sigma^2_{\rm floor}+ \sigma^2_{\rm lens}+\\ \sigma^2_z+\sigma^2_{\rm vpec} & i\neq j~\&~{\rm SN}_i={\rm SN}_j
        \end{cases} ,
\end{equation}
where each row of the matrix corresponds to an SN \textit{light curve}, the diagonal of $C_{\rm stat}$ is the full distance error ($\sigma_\mu^2$) of the $i^{\rm th}$ light curve, and {where measurement noise from components other than the light curve itself are included as} off-diagonal covariance between entries corresponding to light curves of the same SN (${\rm SN}_i={\rm SN}_j$) observed by two different surveys.

Systematic uncertainties can manifest in three key places in the analysis: (1) from changing aspects affecting the light-curve fitting (e.g.,~survey photometry, calibration, SALT2 model), (2) from changing redshifts that propagate to changes in distance modului relative to a cosmological model, and (3) from changes in the astrophysical or survey-dependent assumptions in the simulations used for bias corrections. For each of these categories we examine all of the known significant sources of systematic uncertainty ($\psi$) with sizes S$_\psi$ which result in residuals in the Hubble diagram relative to our baseline analysis ($\mu_{\rm BASE}$).  In order to compute the effect of systematics, we first define
\begin{equation}
\label{eq:deltamupsi}
\Delta\mu^i_\psi \equiv \mu^i_\psi - \mu^i_{\rm BASE} - ( \mu_{\rm ref}(z_\psi)-\mu_{\rm ref}(z_{\rm BASE}) ) ,
\end{equation} 
where $\mu^i_\psi$ is the set of distances for systematic $\psi$. For systematics that affect redshift, we have included new methodology in Eq.~\ref{eq:deltamupsi} that utilizes a reference cosmological model distance $\mu_{\rm ref}(z)$ corresponding to Flat$\Lambda$CDM ($\Omega_M=0.3, \Omega_\Lambda=0.7$). The $\mu_{\rm mod}(z_\psi)$ and $\mu_{\rm mod}(z_{\rm BASE})$ are the cosmological model distances corresponding to redshifts $z_\psi$ and $z_{\rm BASE}$. In order to propagate redshift effects into a distance$\times$distance covariance matrix, the additional component $\mu_{\rm mod}(z_\psi)-\mu_{\rm mod}(z_{\rm BASE})$ accounts for the difference in inferred model distance. 

Assuming linearity between $\Delta\mu_\psi$ and $\psi$, we compute the derivative for each $\psi$ in order to build a 1701$\times$1701 systematic covariance matrix as,
\begin{equation}
\label{eq:csys}
C^{ij}_{\rm syst} = \sum_{\psi} \frac{\partial \Delta\mu^i_\psi}{\partial {\rm S}_\psi} ~ \frac{\partial \Delta\mu^j_\psi}{\partial {\rm S}_\psi} ~ \sigma_\psi^2,
\end{equation}
which denotes the covariance between the $i^{\rm th}$ and $j^{\rm th}$ {light-curve fit} summed over the different sources of systematic uncertainty ($\psi$) with uncertainty $\sigma_\psi$ (see Section~\ref{sec:Inputs} for details). As shown by \cite{binning21}, the $\sigma_\psi$ serve as priors on the known size of systematic uncertainties, but the data itself can constrain the impact of each systematic under the condition that information has not been collapsed by binning/smoothing (as was done for  the original Pantheon, JLA, and DES3YR). 

Fluctuations of the sample of light curves that pass { the sample quality cuts (Table 2 of S22)} for different systematics result in an ill-defined covariance matrix. {To have a well-defined unbinned covariance matrix requires a subtle treatment in order to ensure that the sample is consistent in both the light-curve fitting and BBC stages across all systematics in the analysis. Quality cuts at the light-curve stage are only applied to the set of SNe based on their values found in the baseline analysis, and this SN sample is used
for all systematic tests. We perform the BBC process twice --- the first iteration to identify the subset of $<1\%$ of SNe for which bias corrections are unable to be computed, and a second iteration using only the common set of SNe that have valid bias corrections in all systematic variants. The final cosmology sample of 1701 light curves that satisfy all criteria is described in detail in S22 (see the ``Systematics" row in Table 2 of S22). }

Finally, the statistical and systematic covariance matrices are combined and used to constrain cosmological models:
\begin{equation}
\label{eq:cstatplussyst}
C_{{\rm stat+syst}} = C_{\rm stat} + C_{{\rm syst}}.
\end{equation}

\subsection{Cosmology}
Constraining cosmological models with SN data using $\chi^2$ {has been used in previous SN~Ia cosmology analyses (e.g., \citealt{riess98,Astier06}) and first included systematic covariance in \cite{Conley11}.} Here we follow closely the formalism of \cite{Conley11} where cosmological parameters are constrained by minimizing a $\chi^2$ likelihood:
\begin{equation}
\label{eq:likelihood}
-2{\rm ln}(\mathcal{L}) = \chi^2 = \Delta \vec{D}^T~C_{\rm stat+syst}^{-1}~\Delta \vec{D} ,
\end{equation}
where $\vec{D}$ is the vector of 1701 SN distance modulus residuals computed as
\begin{equation}
\label{eq:dmu}\Delta D_i = \mu_i - \mu_{{\rm model}}(z_i) ,
\end{equation}
and each SN distance ($\mu_i$) is compared to the predicted model distance given the measured SN/host redshift ($\mu_{{\rm model}}(z_i)$). The model distances are defined as 
\begin{equation}
\mu_{{\rm model}}(z_i) = 5\log(d_L(z_i)/10\,{\rm pc}) ,
\end{equation} 
where $d_L$ is the model-based luminosity distance that includes the parameters describing the expansion history $H(z)$. For a flat cosmology ($\Omega_k=0$) the luminosity distance is described by
\begin{equation}
\label{eq:dl}
d_L(z) = (1+z)c\int_0^{z}\frac{dz^\prime}{H(z^\prime)},
\end{equation}
where $d_L(z)$ is calculated at each step of the cosmological fitting process, and the parameterization of the expansion history (used in Eq.~\ref{eq:dl} and therefore in the likelihood Eq.~\ref{eq:likelihood}) in this work is defined as
\begin{equation}
H(z) = {{\rm H}_0}\ \sqrt[]{\Omega_M(1+z)^3+\Omega_{\Lambda}(1+z)^{3(1+w)}}.
\label{eq:hz}
\end{equation}
See \cite{hogg} for the forms of the expansion history $H(z)$ used in the case that the assumption of flatness is relaxed.

The parameters $M$ (Eq.~\ref{eq:Tripp}) and H$_0$ (Eq.~\ref{eq:hz}) are degenerate when analyzing SNe alone. 
However, we also present constraints that include the recently released SH0ES Cepheid host distance anchors (R22) in the likelihood which facilitates constraints on both $M$ and H$_0$.

When utilizing SH0ES Cepheid host distances, the SN distance residuals are modified to the following:

\begin{equation}
\label{eq:dmuprime}
\Delta D^\prime_i=
        \begin{cases}
            \mu_i - \mu_i^{{\rm Cepheid}} & i \in \text{Cepheid hosts} \\
            \mu_i - \mu_{{\rm model}}(z_i) &\text{otherwise} ,
        \end{cases}
    \end{equation}
where $\mu_i^{{\rm Cepheid}}$ is the Cepheid calibrated host-galaxy distance provided by SH0ES and where $\mu_i - \mu_i^{{\rm Cepheid}}$ is sensitive to the parameters $M$ and H$_0$ and is largely insensitive to $\Omega_M$ or $w$.  We also include the SH0ES Cepheid host-distance covariance matrix ($C^{\rm Cepheid}_{\rm stat+syst}$) presented by R22 such that the likelihood becomes
\begin{equation}
\label{eq:sh0eslikelihood}
-2{\rm ln}(\mathcal{L}^\prime) = \Delta \vec{D^\prime}^T~(C^{\rm SN}_{\rm stat+syst}+C_{\rm stat+syst}^{\rm Cepheid})^{-1}~\Delta \vec{D^\prime} ,
\end{equation}
where $C^{\rm SN}_{\rm stat+syst}$ denotes the SN covariance. 

We evaluate the likelihoods with the PolyChord \citep{polychord} sampler in the CosmoSIS package \citep{cosmosis} using 250 live points, 30 repeats, and an evidence tolerance requirement of 0.1. This resulted in converged chains containing 1000--3000 independent samples.
 We verified the SN-only results with CosmoMC (\citealt{cosmomc}) and with the fast cosmology grid-search program in \texttt{SNANA}. The likelihood for Pantheon+ and R22 Cepheid host distance samples will be made available in the public version of CosmoSIS. In this work we also utilize the additional public likelihoods in CosmoSIS in order to combine with and assess agreement with external cosmological probes: Planck (\citealt{planck}) and baryon acoustic oscillations (BAO, likelihoods discussed in Section~\ref{sec:Results}).

In this work we investigate four cosmological parameterizations:
\begin{itemize}
\item Flat$\Lambda$CDM: $\Omega_M$ is floated and we fix $w=-1$ and $\Omega_M+\Omega_\Lambda= 1$.
\item $\Lambda$CDM: $\Omega_M$ and $\Omega_\Lambda$ are floated and we fix $w=-1$.
\item Flat$w$CDM: $w$ and $\Omega_M$ are floated and we fix $\Omega_M+\Omega_\Lambda =1$.
\item Flat$w_0w_a$CDM: $w=w_0+w_a\,(1+z)$, $\Omega_M$, $w_0$, $w_a$ are floated and we fix $\Omega_M + \Omega_\Lambda =1$.
\end{itemize}

We blind our analysis in two ways simultaneously. First, we blind the binned distance residuals output by the BBC fit as cosmological parameters could be inferred visually from simply looking at the Hubble diagram. Secondly, in order to prevent accidental viewing of the cosmological parameters themselves, the CosmoSIS chains were shifted by unknown values {following the formalism of \cite{Hinton16}.}

\section{Data and Analysis Inputs}
\label{sec:Inputs}
Here we review each component of the dataset and analysis. We discuss the fundamental \uline{\textit{purpose}}, the \uline{\textit{baseline}} treatment {in this analysis}, and the \uline{\textit{systematic}} uncertainties associated with each aspect (if applicable). The impact of systematics in both distance and cosmological inference is shown in Section~\ref{sec:Results}.  We provide a brief overview of this section here.\\

\noindent\textbf{Data}\\
Sec.~\ref{subsubsection:lightcurves}: {SN~Ia Light Curves}\\
Sec.~\ref{subsubsection:redshifts}: {{Redshifts}}\\
Sec.~\ref{subsubsection:pvs}: {{Peculiar Velocities}}\\
Sec.~\ref{subsubsection:hosts}: {{Host-Galaxy Properties}}\\

\noindent\textbf{Calibration and Light-Curve Fitting} \\
Sec.~\ref{subsubsection:calibration}: {{Calibration}}\\
Sec.~\ref{subsubsection:salt2}: {{SALT2 Model}}\\
Sec.~\ref{subsubsection:mw}: {{Milky Way Extinction}}\\

\noindent\textbf{Simulations}\\
Sec.~\ref{subsubsection:model}: {{Survey Modeling}}\\
Sec.~\ref{subsubsection:intrinsic}: {{Intrinsic Scatter Models}}\\
Sec.~\ref{subsubsection:uncertainty}: {{Uncertainty Modeling}}\\
Sec.~\ref{subsubsection:validation}: {{Validation}}

\subsection{Data}

\subsubsection{\textbf{SN~Ia Light Curves}}
\label{subsubsection:lightcurves}

\noindent\uline{\textit{Purpose:}} The flux-calibrated light-curve photometry is fit to determine the SALT2 parameters used in standardization (Eq.~\ref{eq:Tripp}). \\
\uline{\textit{Baseline:}} The light-curve data is described in detail by S22 and references therein. The full set of spectroscopically classified photometric light curves is compiled from 18 different publicly available and privately released samples. In total, 2077 SN light-curve fits converged using SALT2; after quality cuts are applied (Table 2 of S22), this results in 1701 SN light curves {of \nunique\ unique SNe~Ia} usable for cosmological constraints. The {sample} includes a 3.5$\sigma$ {Hubble residual} outlier cut to remove { 5} potential contaminants { that are likely} non-normal Type Ia or misidentified redshifts. {The sample of cosmologically viable light curves includes} 81 light curves of 42 SNe used to calibrate Cepheid brightnesses as utilized by R22. The survey SN photometry compiled in \cite{plussample} and analyzed here is from DES\footnote[1]{Not included in Pantheon 2018} \citep{Brout18a,Smith_2020}, Foundation$^1$ \citep{Foley2018}, PS1 \citep{Scolnic18}, SNLS \citep{Betoule14}, SDSS \citep{Sako11}, {\it HST} \citep{Gilliland1998,Riess2001,Suzuki2012,Riess2018,Riess2004,Riess2007}, Low-$z$ (grouped together as 
LOSS\_1$^1$, \citealp{Ganeshalingam2010}; LOSS\_2$^1$, 
\citealp{Stahl2019}; SOUSA$^1$\footnote{
https://pbrown801.github.io/SOUSA/}, \citealp{sousa14}; CNIa0.02$^1$, \citealp{CNIa0.02}; CSP, \citealp
{krisciunis17}; CfA1, \citealp{Riess99}; CfA2, \citealp{jha06}; CfA3, \citealp{hicken09a}; CfA4, \citealp{Hicken12}, and numerous smaller low-redshift samples$^1$ of 1--2 SNe given by \citealt{Burns18}, \citealt{Burns20}, \citealt{Milne10}, \citealt{Kris2017_df}, \citealt{Stritzinger10}, \citealt{Gall2018}, \citealt{Zhang10}, \citealt{Tsvetkov10}, and \citealt{Kawabata20}.) \\
\uline{\textit{Systematics:}} See Calibration Section~\ref{subsubsection:calibration}.

\subsubsection{\textbf{Redshifts}}
\label{subsubsection:redshifts}

\noindent\uline{\textit{Purpose:}} The {peculiar-velocity corrected CMB-frame} redshift of each SN/host is required to compare the inferred distance to a distance predicted by a cosmological model, as given in Eq.~\ref{eq:dmu}. Additionally, {heliocentric} redshifts are required in the SALT2 light-curve fits in order to shift the model spectrum to match the data. \\
\uline{\textit{Baseline:}} The redshifts for all of the SNe (and their host galaxies, depending on what is available) are provided by \cite{carr21}, who performed a comprehensive review of redshifts for the Pantheon+ samples and {made numerous corrections.} \cite{carr21} report the heliocentric redshifts for each SN and convert the redshift into the CMB frame. The redshifts of the Pantheon+ sample cover a range of $0.001 < z < 2.3$. While redshifts of the 42 Cepheid host calibrator SNe are included, they are not used in the {comparison of} SN~Ia magnitudes to the Cepheid distance scale and are only provided for reference and for SALT2 fitting. \\
\uline{\textit{Systematics:}} Following \cite{carr21}, we apply a coherent shift to each redshift of $+4\times 10^{-5}$. This was conservatively stated by \cite{Calcino17} for the potential size of a local void bias and by \cite{davis19} as a potential measurement bias.

\subsubsection{\textbf{Peculiar Velocities}}
\label{subsubsection:pvs}

\noindent\uline{\textit{Purpose:}} Peculiar motions of galaxies arise from coherent flows, motion of halos, inflow into clusters or superclusters, and intragroup motion. Corrections are applied to the observed redshifts (after light-curve fitting) based on peculiar-velocity maps derived from independent large spectroscopic galaxy surveys.  

\noindent\uline{\textit{Baseline:}} The nominal peculiar velocities used for this analysis were determined by \cite{peterson21} from a comparison of multiple treatments of peculiar-velocity maps and group catalogs. Corrections were applied by \cite{carr21} for the Pantheon+ sample. The baseline corrections are based on 2M++ \citep{Carrick15} with global parameters found in \cite{said20} and combined with group velocities estimated from \cite{tully15} group assignments. The  $\sigma_{\rm vpec}$ in Eq.~\ref{eq:TrippErr} is found { using} 240\,km\,s$^{-1}$ after accounting for uncertainties propagated into the covariance matrix described below. { This $\sigma_{\rm vpec}$ floor is in agreement with what was used in \cite{peterson21} and for the SNe between $0.001<z<0.02$ it is likely a conservative estimate as \cite{kenworthy22} found for the most nearby SN calibrators a floor of $155\pm25$\,km\,s$^{-1}$. This apparent reduction at the lowest redshifts may be due to the peculiar velocity maps having higher fidelity at these redshifts and because Pantheon+ has relatively better virial-group information at these redshifts.}\\

\noindent\uline{\textit{Systematics:}} 
\cite{peterson21} discuss multiple viable alternatives for the treatment of peculiar velocities.
The first approach is to use the 2M++ corrections \citep{Carrick15} integrating over the line-of-sight relation (iLOS) between distance and the measured redshift. We take this variation as the first systematic with $\sigma_\psi^2=0.5$.
The second approach is to use the 2MRS \citep{lilownusser} peculiar-velocity map; however, differences between 2MRS and 2M++ at very low redshift ($z<0.01$) cause numerical stability issues for off-diagonal $C_{\rm syst}$ elements.
We incorporate only the diagonal differences between 2MRS and 2M++ into $C_{\rm syst}$ with $\sigma_\psi^2=0.5$. As a numerically stable estimate of the off-diagonal terms, we use the 2M++ velocities transformed by the slope and offset difference between the 2M++ and 2MRS maps found in \cite{peterson21}.
The two approaches added in quadrature result in an effective $\sigma_\psi^2=1.0$.

\subsubsection{\textbf{Host-Galaxy Properties}}
\label{subsubsection:hosts}

\noindent\uline{\textit{Purpose:}} The observed host-galaxy mass versus SN luminosity relation is used to standardize the SN~Ia brightnesses in two ways. First, simulations of the dataset include correlations between SN color {and SN stretch} and host properties such as dust as a function of host mass following \cite{popovic21b}. Second, a further residual correction is applied in the Tripp Eq.~\ref{eq:Tripp} where the ``mass step" $\gamma$ is fit in the BBC stage. \\
\uline{\textit{Baseline:}} The host-galaxy stellar masses are presented by S22 and references therein. Masses are determined for all host galaxies, and star-formation rates and morphologies are also included the low-$z$ sample. In the baseline analysis we apply the mass step at $10^{10}M_\odot$ following Pantheon and DES3YR. \\
\uline{\textit{Systematics:}} Several independent analyses \citep{Sullivan2010,Childress_2013,Kelsey2020} have suggested that the optimal location of the mass step could range between $10^{9.8}M_\odot$ and $10^{10.2}M_\odot$. We therefore include a systematic uncertainty where the mass step occurs at $10^{10.2}M_\odot$. \\

\subsection{Calibration and Light-Curve Fitting}
\subsubsection{\textbf{Calibration}}
\label{subsubsection:calibration}

\noindent\uline{\textit{Purpose:}} Photometric calibration of each passband in each survey is needed to fit light curves and facilitate comparison of the brightnesses of SNe across different telescopes/instruments/filters. Photometric calibration is also important to homogenize spectrophotometric datasets used in the SALT2 model training. \\
\uline{\textit{Baseline:}} The calibration of all 25 photometric systems used in this work is discussed in Fragilistic \citep{fragilistic}.  The outputs of Fragilistic are a best-fit calibration solution for each of the 105 passbands and a joint $105\times105$ covariance matrix that describes the covariance between the zeropoint calibrations of all passbands that arise from using a single common stellar catalog to tie all surveys together (PS1).   \\
\uline{\textit{Systematics:}} The systematics due to calibration and their impact are discussed in detail in Fragilistic. We estimate the impact of the correlated { filter zero-point and central wavelength} uncertainties by refitting SALT2 light curves  {(with retrained SALT2 models; see next \textbf{SALT2 Model})} using 9 realizations of the 105 zero-points. For each of the 9 realizations a value of $\sigma_\psi^2=1/9$ is adopted such that they add in quadrature to $\sim1$. The uncertainty in modeling the spectrum of the {\it HST} primary standard star C26202 has been tripled to account for the recent update in \cite{Bohlin21}; it is now set to 15\,mmag over 7000\,\AA\ ($\sigma_\psi=3$ for a systematic of 5\,mmag over 7000\,\AA). Lastly, an additional conservative systematic is included { only for the CSP SNe} to account for the 2\% recalibration in CSP tertiary stellar magnitudes from \cite{Stritzinger2010} to \cite{krisciunis17} ($\sigma_\psi=1$).  \\

\subsubsection{\textbf{SALT2 Model}}
\label{subsubsection:salt2}

\noindent\uline{\textit{Purpose:}} The trained SALT2 model is required to fit light curves and determine the light-curve parameters ($m_b$, $c$, $x_1$) for each SN used in Eq.~\ref{eq:Tripp}. \\
\uline{\textit{Baseline:}} We use the Fragilistic calibration solution and newly trained { SALT2-B22 model\footnote{released publicly at pantheonplussh0es.github.io}} which was developed following the formalism of \cite{Guy10} and \cite{taylor21}. { The SALT2 model includes a component of training statistical uncertainty which is incorporated in the fitted light-curve parameters} \\
\uline{\textit{Systematics:}} For each of the 9 correlated realizations of Fragilistic { filter zero-points and central wavelengths} discussed above (for \textbf{Calibration}) we simultaneously retrain the SALT2 model. Additionally, to
conservatively account for a possible systematic from the redevelopment of the SALT2 model training process itself,
we adopt an additional systematic by fitting the dataset with the SALT2 model trained by \cite{Betoule14} and applying a scaling of $\sigma_\psi=1/3$ { (See Section~\ref{sec:Discussion} and Fig.~\ref{fig:muchange} for impact)}.\\

\subsubsection{\textbf{Milky Way Extinction}}
\label{subsubsection:mw}

\noindent\uline{\textit{Purpose:}} Values of the Milky Way (MW) Galactic dust extinction, $E(B-V)_{\rm MW}$, are applied to the SALT2 model spectra during both the model training process and during the data light-curve fitting process. The ``extinction curve" describes the relation between the amount of reddening and extinction as a function of wavelength. \\
\uline{\textit{Baseline:}} We account for MW extinction using maps from \cite{schlegel98}, with a scale of 0.86 following \cite{sfd1}. We assume the MW extinction curve from \cite{Fitzpatrick99} with $R_V=3.1$. \\
\uline{\textit{Systematics:}} Similarly to Pantheon, we adopt a global 5\% uncertainty scaling of $E(B-V)_{\rm MW}$ based on the fact that \cite{sfd2}, in a reanalysis of \cite{sfd1},  derive smaller values of reddening by $ 4\%$, despite using a very similar SDSS footprint ($\sigma_\psi=1$). While \cite{Schlafly11} found that their results prefer the \cite{Fitzpatrick99} extinction curve, we conservatively include an additional systematic uncertainty in the MW extinction curve and {analyze the data (training and light-curve fit)}
 using the \cite{ccm} and apply a systematic scaling of $\sigma_\psi=1/3$ {as this reflects the preference of \cite{Fitzpatrick99} over \cite{ccm}}.

\begin{table}
\centering
\makebox[.2\textwidth]{%
\resizebox{.55\textwidth}{!}{
\begin{tabular}{c|c|c|c}
    Survey & Cadence & DETEFF & SPECEFF  \\
    \hline
        LOW-$z$ & \cite{Scolnic18} & \cite{Kessler19} & \cite{Scolnic18} \\
    FOUND & \cite{Jones19}  & N/A & \cite{Jones19} \\
    SDSS & \cite{Kessler13} & \cite{Kessler2009} & \cite{popovic21a} \\
    PS1 & \cite{Jones18}  & \cite{Jones17} & \cite{Scolnic18} \\
    DES & \cite{Smith20}  & \cite{Kessler2015} & \cite{DES3YR} \\
    SNLS & \cite{Kessler13}  & N/A & \cite{popovic21a} \\
    HST & \cite{Scolnic18} & N/A & N/A  \\
\end{tabular}%
}}
\caption{References for inputs to \texttt{SNANA} simulations used for this analysis.  We give references for the ``Cadence," which describes the observing history; the ``DETEFF," which describes the detection efficiency based on the signal-to-noise ratio (SNR); and the ``SPECEFF," which describes the spectroscopic selection efficiency as a function of SN magnitude.}
\label{tab:SimInputTable}
\end{table}

\vspace{-5mm}

\begin{figure}[h]
    \centering
	\includegraphics[trim=0 70 0 10,clip,width=.45\textwidth]{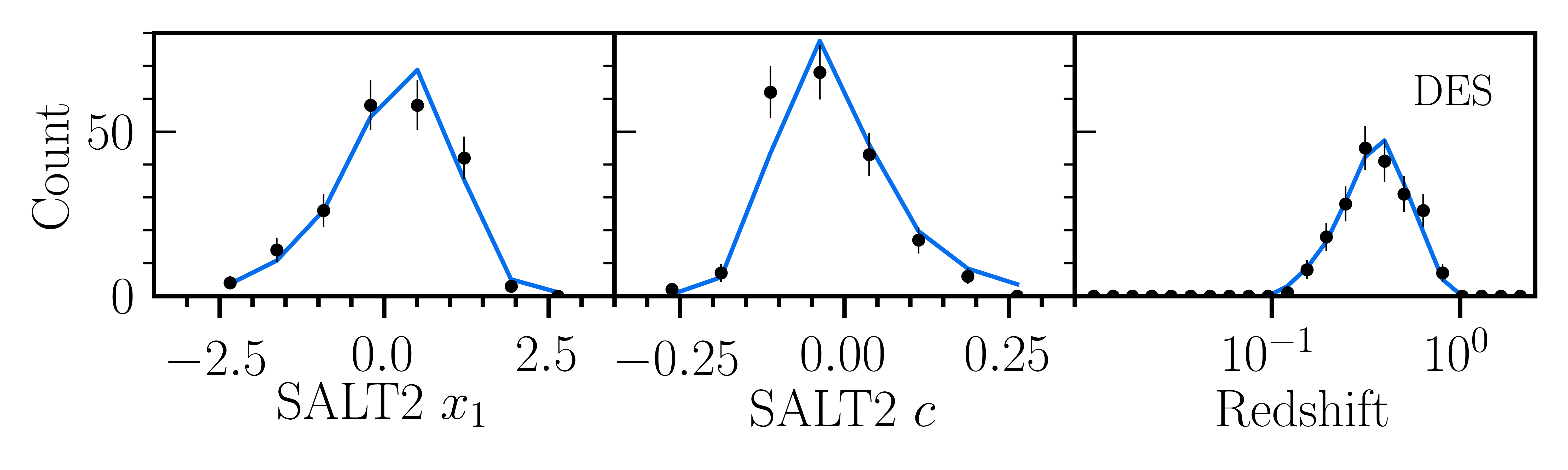} \vspace{0mm}
	\includegraphics[trim=0 70 0 10,clip,width=.45\textwidth]{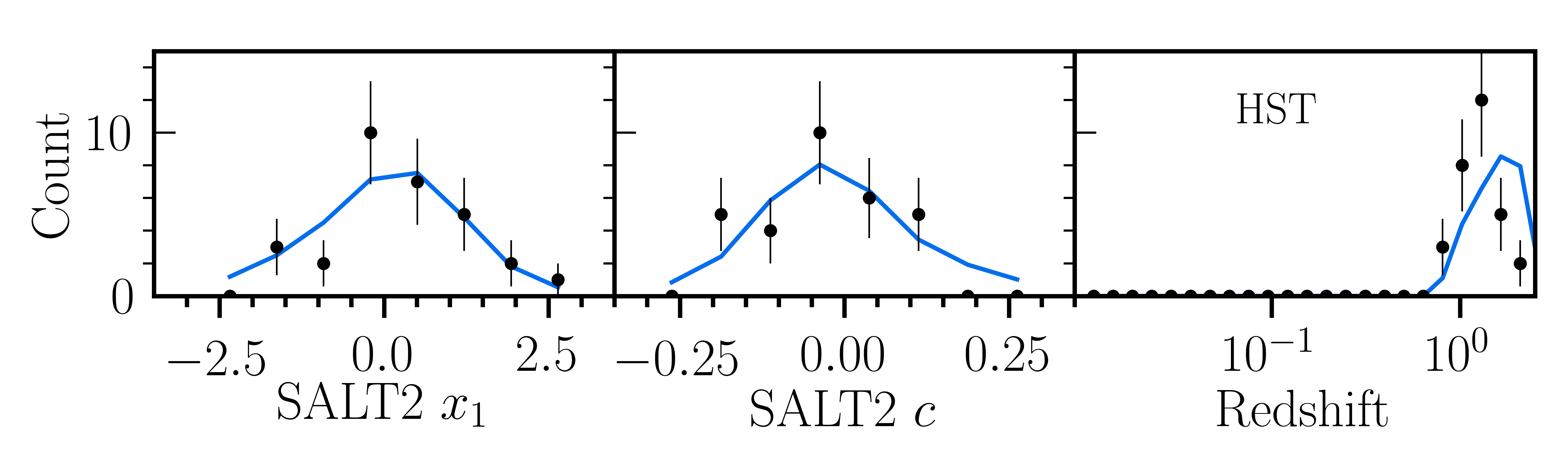} \vspace{0mm}
	\includegraphics[trim=0 70 0 10,clip,width=.45\textwidth]{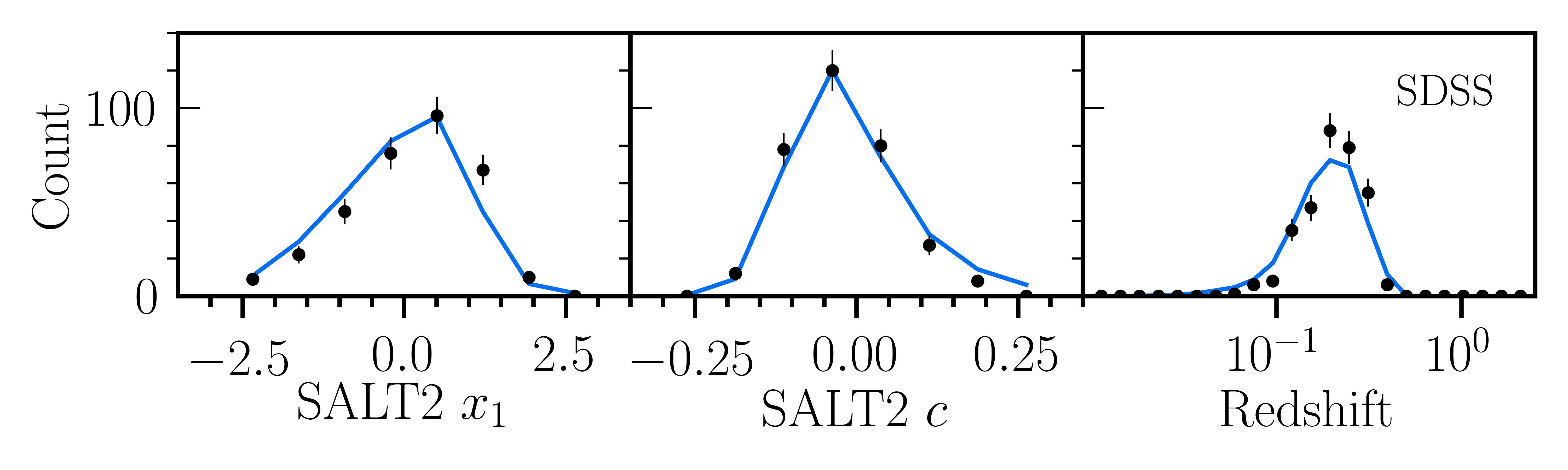} \vspace{0mm}
	\includegraphics[trim=0 70 0 10,clip,width=.45\textwidth]{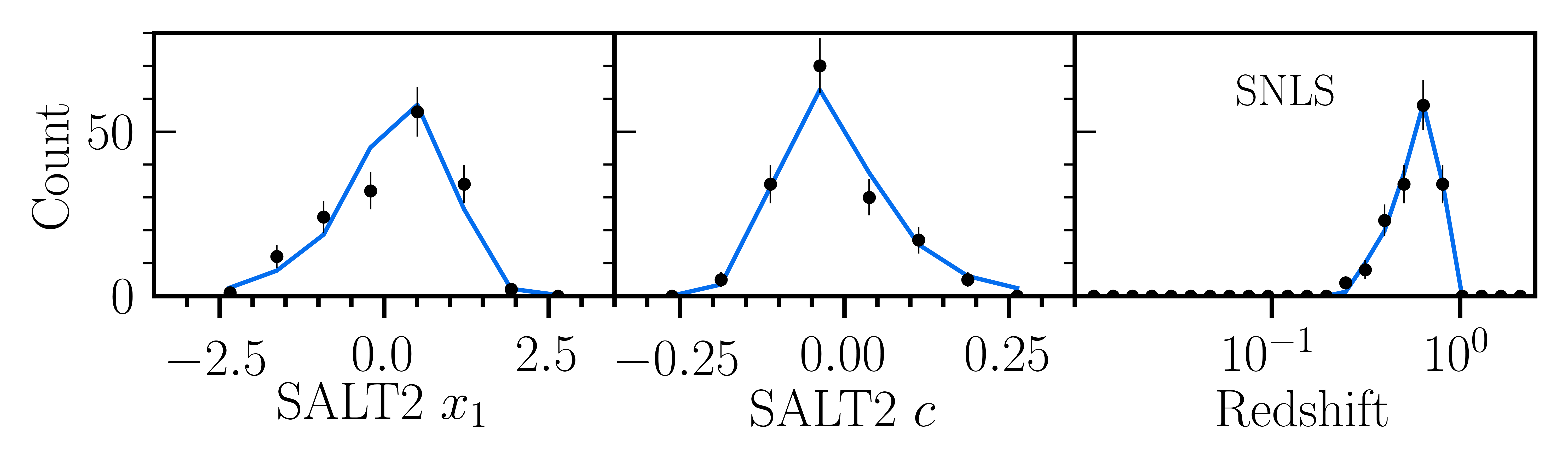} \vspace{0mm}
	\includegraphics[trim=0 70 0 10,clip,width=.45\textwidth]{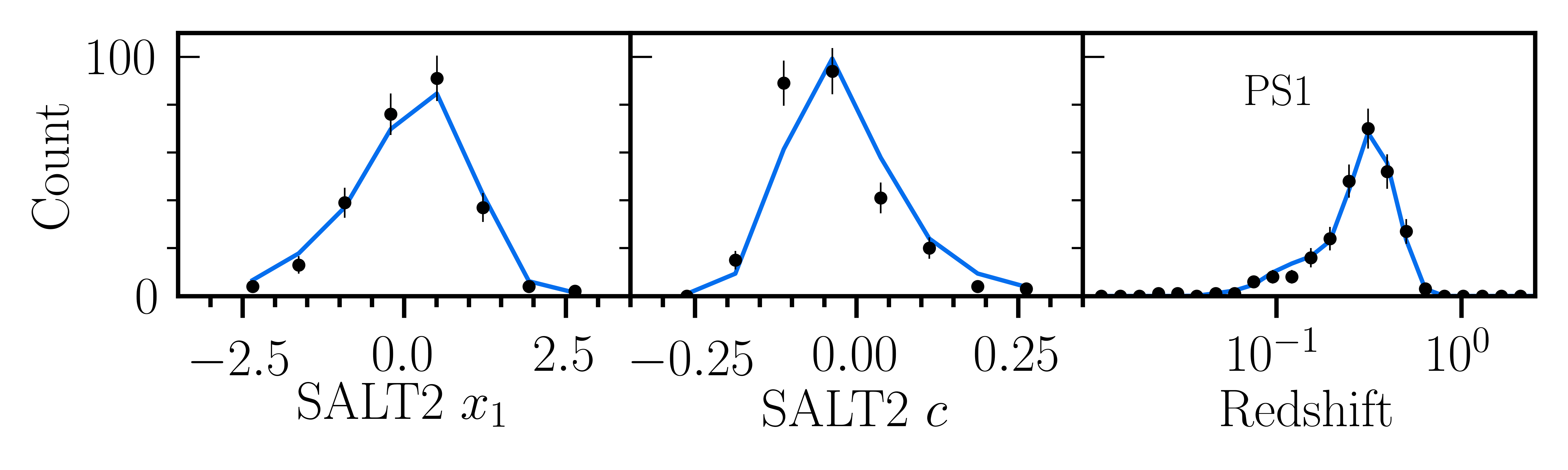} \vspace{0mm}
	\includegraphics[trim=0 70 0 10,clip,width=.45\textwidth]{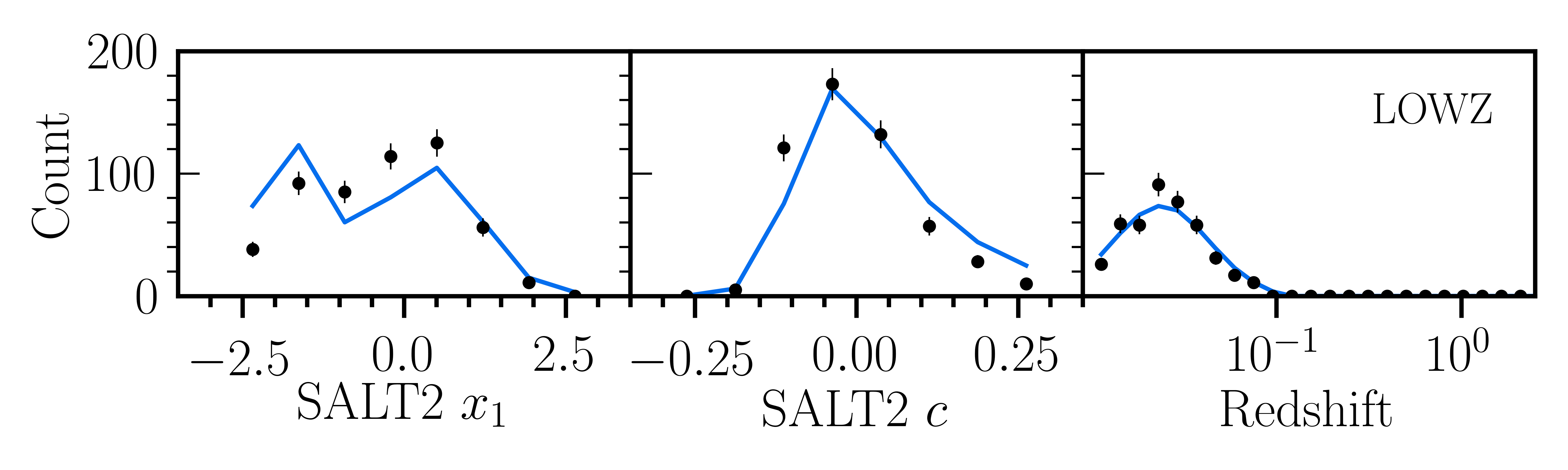} \vspace{0mm}
	\includegraphics[trim=0 0 0 10,clip,width=.45\textwidth]{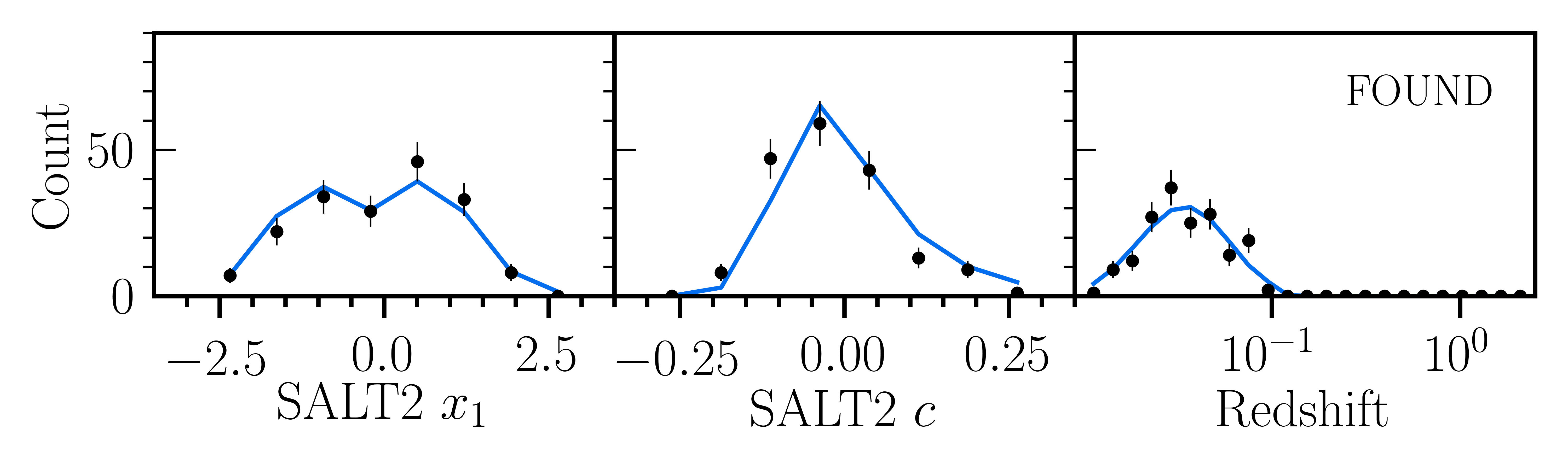} 
      \caption{Comparison between observed data { (black points)} and simulations { (blue lines)} for the largest subsamples in this analysis: DES, {\it HST}, SDSS, SNLS, PS1, LOW-$z$, Foundation (FOUND). We compare three key distributions: the SALT2 light-curve-fit parameters $x_1$ and $c$ are shown as well as the measured redshift.}
    \label{fig:datamc} \vspace{-1mm}
\end{figure}

\bigskip
\subsection{Simulations}
\label{sec:Sims}

\subsubsection{\textbf{Survey Modeling}}
\label{subsubsection:model}

\noindent\uline{\textit{Purpose:}} We utilize catalog-level simulations of large samples of SN~Ia ($>1,000,000$ per survey) light curves. \texttt{SNANA} simulations specific to each survey in our analysis are prescribed by each aspect of acquiring an SN~Ia sample. As detailed in Figure~1 of \cite{Kessler18}, the simulations require three main sets of inputs: \\
\noindent A \textbf{Source Model} for generating SNe with realistic astrophysical properties and applying cosmological effects such as redshifting, dimming, lensing, peculiar velocities, and MW extinction. \\
\noindent A \textbf{Noise Model}, unique to each survey, for applying instrumental and atmospheric noise to determine a detection efficiency (``DETEFF").\\
\noindent A \textbf{Trigger Model}, unique to each survey, {that includes the observing cadence and describes an efficiency as a function of $B$-band peak magnitude for detecting SNe and obtaining a spectroscopic confirmation (``SPECEFF").}. \\
\noindent These simulations for each survey are combined and used to forward model the underlying populations of the SN properties \citep{popovic21b,popovic21a} and to determine the expected biases in measured SN distances that follow from the known selection effects. These biases are corrected in the $\delta_{\rm bias}$ term of Eq.~\ref{eq:Tripp}. \\
\uline{\textit{Baseline:}}  
Depicted in Fig.~\ref{fig:datamc} are the distributions of the key observables ($z$, $x_1$, $c$) for both data and simulations of each survey used in this analysis. We find good agreement between the data and simulations, as described in detail by \cite{popovic21b} and \cite{popovic21a}. We note that the agreement in the redshift dimension is achieved despite not explicitly tuning the redshift distribution of surveys.

We simulate SNe in LOW-$z$ and Foundation down to $z = 0.001$. Novel for this work specifically are the simulations of primary distance indicator hosts of SNe in the range $0.001<z<0.01$ which are assumed to have the same color and stretch populations as those of their respective surveys (LOW-$z$ and FOUND), and specifically over this redshift range they are assumed to be complete with flat spectroscopic selection efficiency. These simulations facilitate bias corrections to the Cepheid calibrator SNe and thus the propagation of modeling systematics to the SNe used in the companion SH0ES analysis (R22). 

{The simulation inputs for survey cadence, DETEFF, and SPECEFF functions 
have been evaluated in many analyses over the past decade. Table~\ref{tab:SimInputTable} shows
a summary of where we obtain these inputs for each survey. 
Survey metadata is used to model the cadence and instrumental properties, if available, such as for 
FOUND, SDSS, PS1, DES, and SNLS. LOW-$z$ data do not provide such
metadata, and thus the cadence and noise properties are extracted
from the data as described in} Section 6 of \cite{Kessler18}  following the procedure developed by \cite{Scolnic18}, which assumes that the LOW-$z$ subset of SNe is magnitude-limited. 
{These are simulations of the CfA and CSP samples, but not of the newer samples included in this work (LOSS, SOUSA, CNIa0.02), thereby implicitly assuming that the CfA and CSP} samples have similar selection effects and therefore distance biases as the newer additions. To simulate SN-host correlations, a catalog of host-galaxy properties and specifically their stellar-mass distributions are taken from \cite{popovic21a}. 
The simulations used for bias corrections for all surveys are performed in \LCDM~($w=-1.0$, $\Omega_M=0.3$, $\Omega_\Lambda=0.7$) with the SALT2-B22 model. \\
\uline{\textit{Systematics:}} We increase the SNR of each simulation by 20\%, resulting in all survey simulated distributions changing by more than $1\sigma$, as a { single} conservative systematic in the determination of the selection biases. \cite{BBC} showed that the sensitivity of the bias corrections to the input cosmology is relatively weak; this was confirmed by \cite{Brout18b} and found to be a negligible contribution to SN~Ia uncertainty budgets. We therefore do not include this as an additional systematic.\\

\subsubsection{\textbf{Intrinsic Scatter Models}}
\label{subsubsection:intrinsic}

\noindent\uline{\textit{Purpose:}} A model of the intrinsic SN brightness variations, called ``intrinsic scatter," is needed to account for the observed residual variation in SN~Ia standardized luminosities that exceeds expectations from measurement uncertainties alone. In addition, models of the true (``parent") populations of SN~Ia SALT2 parameters $c$ and $x_1$ are required for the Source Model in \texttt{SNANA}. The intrinsic scatter model is utilized in the bias-correction simulations. \\
\uline{\textit{Baseline:}} We utilize the { \citealp[hereafter BS21]{bs20}} model that prescribes SN~Ia scatter into two color-dependent components: a standard cosmological color law specific to SNe~Ia and additional dust-based color laws and dust extinctions that vary with each galaxy/SN. This approach is preferred because of its novel replication of the observed relationships between SN color and residual Hubble diagram scatter as well as its ability to replicate the ``mass step" as a function of SN~Ia color. We use the scatter model parameters from BS21 with improvements from \cite{popovic21b} in our baseline bias-correction simulations; because the BS21 model includes within it the parent $c$ population, we also utilize the separate parent population for $x_1$ derived by \cite{popovic21a}. Improving upon \cite{Scolnic16}, \cite{popovic21a} fit for parent populations in bins of mass to account for host--SN~Ia relationships. \cite{popovic21a} split their populations into high- and low-redshift groups, and notably for low-redshift surveys the $x_1$ populations are fitted with a two-Gaussian model to recreate the observed double peak in the $x_1$ distribution.\\
\uline{\textit{Systematics:}} 
{We include two categories of systematics for the intrinsic scatter model 
and parent populations:  (1) different models of intrinsic scatter, and 
(2) determination of parameters for the BS21 model.}
For the former, we use two additional scatter models {from \cite{Kessler13}} that have been used in previous cosmology analyses (JLA, Pantheon, DES3YR). These are (1) the ``G10'' model based on \cite{Guy10} which describes
$\sim 70$\% of the excess Hubble scatter from ``gray" variations and the remaining scatter from wavelength-dependent variations, and (2) the ``C11'' model based on \cite{Chotard11} which describes $\sim 30$\% of the excess Hubble scatter from coherent variations, and the remaining scatter from wavelength-dependent variations. 
{For the G10 and C11 scatter models, bias corrections are performed in 7-D 
as given by \cite{popovic21a}}.
For the systematic uncertainty in the determination of the BS21 model parameters we adopt three different viable sets of dust and intrinsic SN populations from \cite{popovic21b}. These populations are the best-fit (maximum likelihood) parameters (hereafter P21), the mean posterior set of parameters, and a set that represent a 1$\sigma$ fluctuation in the uncertainty. 
{Lastly, while the BS21 and P21 models impact the simulated bias corrections, the SALT2 training and light-curve fitting has not been altered.} 
{ The choice of scatter model is propagated through the simulations used for the bias corrections applied in Eq.~\ref{eq:Tripp} and for the uncertainty modeling in $\sigma_{\rm scat}$ of Eq.~\ref{eq:errmodel}.}

  \begin{figure}[h]
        \centering 
	    \includegraphics[width=.48\textwidth]{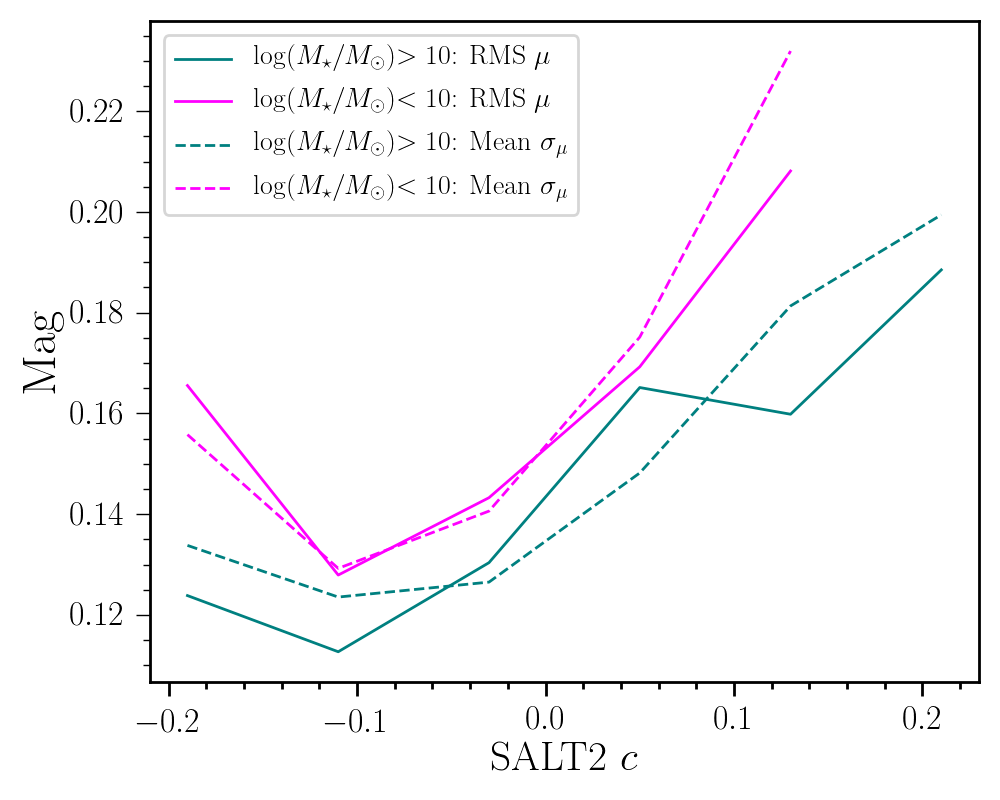}  
        \caption{Pantheon+ distance-modulus uncertainties (shown as dashed lines with mean $\sigma_\mu$ and split on host mass) in comparison to the observed root-mean square (RMS) of the distance-modulus residuals (shown as solid lines as RMS $\mu$ split on host mass), as a function of color. This shows that the distance errors are adequately modeled (Eq.~\ref{eq:errmodel}) {as a function of SN color and host stellar mass}. In previous analyses, the uncertainties were roughly flat as a function of color.}
        \label{fig:bs20errmodeling} 
    \end{figure} 
    
\vspace{2mm}
\subsubsection{\textbf{Distance-Modulus Uncertainty Modeling}}
\label{subsubsection:uncertainty}

\noindent\uline{\textit{Purpose:}} To match the reported SN distance-modulus uncertainties {(Eq.~\ref{eq:TrippErr}) to the scatter in distance that is observed in the data.}   \\
\uline{\textit{Baseline:}} The BS21 model parameters have been fit to the observed scatter in the dataset. We can utilize large BS21 simulations to determine $\sigma_{\rm scat}(z,c,M_\star)$ after accounting for selection effects. The efficacy of this method is shown in Fig.~\ref{fig:bs20errmodeling}, which demonstrates good agreement between the observed RMS of the Hubble residuals and the uncertainties of the distance-modulus values. \\
\uline{\textit{Systematics:}} To conservatively account for how SN cosmology was done in the past (JLA, Pantheon), in Eq.~\ref{eq:TrippErr} we set $\sigma_{\rm scat}(z,c,M_\star)=0$ and allow only a single $\sigma_{\textrm{gray}}$ parameter to { replicate the methodology used with historic intrinsic scatter models (G10 and C11). However, we note that for G10 and C11, the trends in RMS seen for the data in Fig.~\ref{fig:bs20errmodeling} do not match the reported uncertainties.}\\

\subsubsection{\textbf{Validation}}
\label{subsubsection:validation}

\noindent\uline{\textit{Purpose:}} To verify that our analysis can recover input values in data-sized simulated samples and does not produce biases. {Such tests are sensitive to the light-curve fitting and BBC technique (as well as implementation and coding errors); however, they are not sensitive to certain aspects of the analysis such as the assumption of the SALT2 model or photometric calibration.}\\
\uline{\textit{Baseline:}} We perform an end-to-end test of our baseline analysis pipeline from survey photometry catalog-level simulations. We create 20 realizations of each survey in an arbitrary cosmological model ($w=-1$): 10 with the BS21 scatter model and 10 with the G10 scatter model. We perform light-curve fitting, apply bias corrections, compile into 10 Hubble diagrams, and maximize the cosmological likelihoods (Eq.~\ref{eq:likelihood}) using a fast cosmology grid-search program in \texttt{SNANA} \citep{Kessler2009}, with approximate priors from
CMB measurements (Planck \citealt{planck})
to obtain best-fit cosmological parameters and uncertainties. For the BS21 model simulations we recover a mean best-fit $w=-1.012 \pm 0.011$ and for the G10 model simulations we recover a mean best-fit $w=-0.983\pm0.015$; both are within $\sim1\sigma$ of the input cosmology. The 20 realizations are made available publicly\footnote{Will be made available after publication at\\ pantheonplussh0es.github.io} along with bias-correction simulations.\\

\vspace{5mm}

\section{Results}
~\label{sec:Results}

  \begin{figure*}
    \centering \vspace{20mm}
	\includegraphics[width=.98\textwidth]{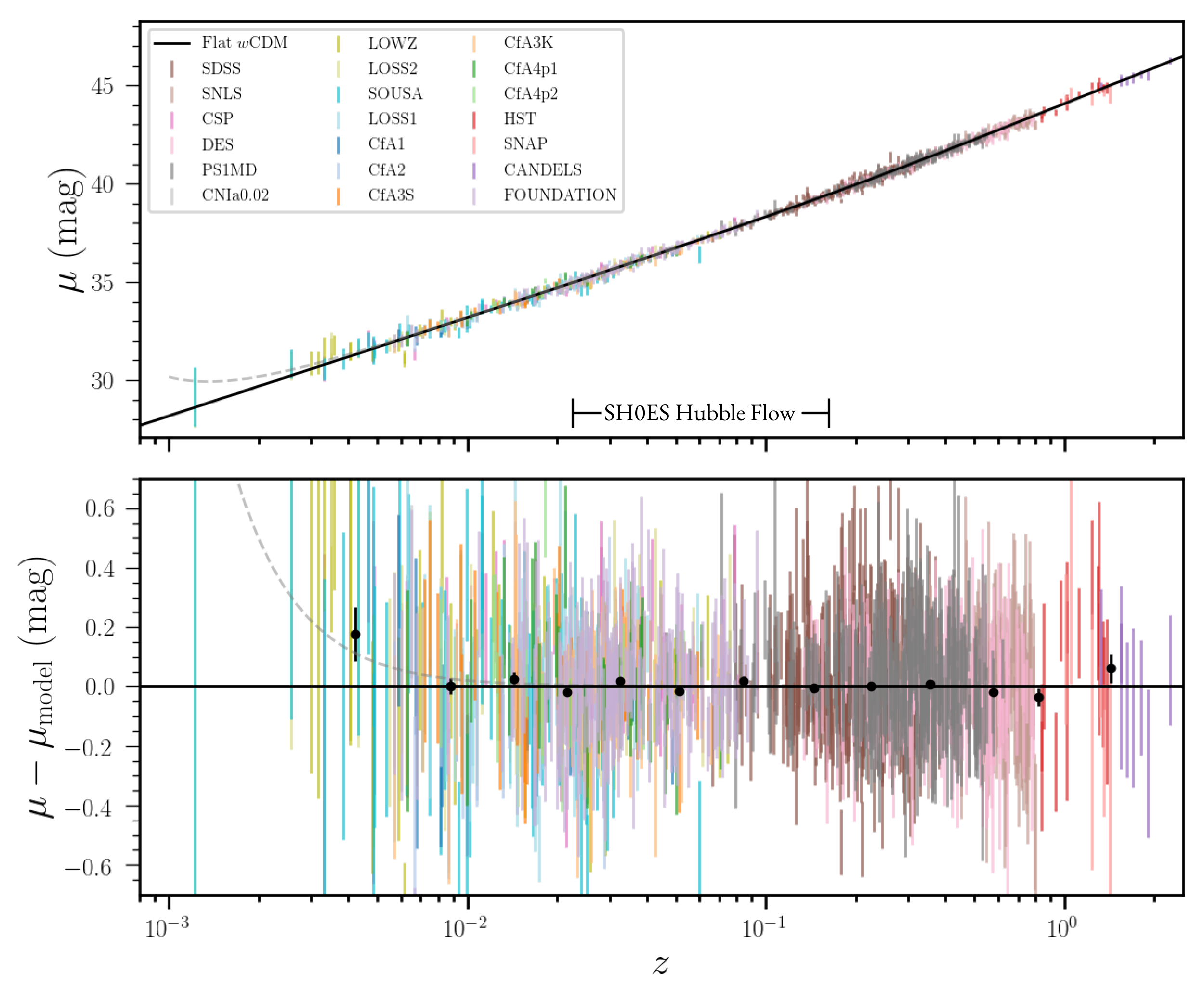} 
      \caption{\textbf{Top panel}: The Pantheon+ ``Hubble diagram" showing the distance modulus $\mu$ versus redshift $z$. The 18 different surveys are each given different colors.  \textbf{Bottom panel}: The distance-modulus residuals relative to a best-fit cosmological model {with binned data for reference (black points). Both the data errors and the binned data errors include only statistical uncertainties.} At $z<0.01$, the sensitivity of peculiar velocities is very large, and the uncertainties shown reflect this uncertainty. { Dashed line is the predicted Hubble residual bias stemming from biased redshifts due to volumetric effects in the very nearby universe (assuming 250~km~s$^{-1}$ uncorrected velocity scatter).}}
    \label{fig:HD} \vspace{30mm}
\end{figure*}

\begin{table}
\vspace{3mm}
\caption{Standardization Parameters and Results}
\centering

\begin{tabular}{lcccccc}
\toprule
 \multicolumn{1}{|c}{}& \multicolumn{4}{|c|}{BBC Fit} & \multicolumn{2}{c|}{CosmoSIS Fit}\\ 
\hline
 Model & $\alpha$ & $\beta$ & $\sigma_{\rm gray}$ & $\gamma$ & RMS & $ln(\mathcal{L})$\\ 
\hline
\vspace{-.2cm}
\\

BS21 & 0.148(4)& 3.09(4) & 0.00 & $-0.003$(7) & 0.171 & $-1635$\\
P21 & 0.145(5) & 3.00(5) & 0.00 & 0.019(10) & 0.171 & $-1674$\\
G10 & 0.153(4) & 2.98(5) & 0.10 & 0.054(7) & 0.173 & $-1676$ \\
C11 & 0.153(4) & 3.44(6) & 0.12 & 0.053(8) & 0.173 & $-1681$ \\

\hline
\label{tab:nuisance}

\end{tabular}
{\raggedright
\ \\
\textbf{Notes}:~The nuisance parameters, as defined in Eq.~\ref{eq:Tripp} and~\ref{eq:TrippErr} are given here for different assumptions about the intrinsic scatter model, as described in Sec.~\ref{sec:Inputs} (Intrinsic Scatter Model).  That $\sigma_{\rm int}\sim0$ and $\gamma\sim0$ for the BS21 and P21 models are due to modeling the scatter and mass-step as part of the BBC process, which is discussed in further detail in Appendix~\ref{appendix:biascor}.  The BS21 is the baseline choice for intrinsic scatter.  The RMS is given in units of mag. The Hubble diagram likelihood values for each model ($\mathcal{L}$) include an uncertainty normalization term.\\}
\end{table}

\subsection{Standardization Parameters}    

The standardization nuisance parameters $\alpha$, $\beta$, $\gamma$, and $\sigma_{\rm gray}$ defined in Eq.~\ref{eq:Tripp} and~\ref{eq:TrippErr} are shown in Table~\ref{tab:nuisance} for each of the scatter models used in this work. The best-fit $\alpha$ are similar across scatter models to within $\sim1\sigma$. The best-fit $\beta$ values differ across models owing to different treatments of SN~Ia color; however, the values for the baseline dust model (BS21) and the P21 dust model are self-consistent. 

As shown in Table~\ref{tab:nuisance}, the additional $\sigma_{\rm gray}$ term for the BS21 and P21 models is found to be zero. {As discussed in Sec.~\ref{sec:Method}, this is consistent with the expectation that if the simulations correctly model the intrinsic scatter and noise of the data, the $\sigma_{\rm scat}(z,c,M_\star)$ term of Eq.~\ref{eq:TrippErr} is sufficient to describe the distance-modulus uncertainties with $\sigma_{\rm gray}=0$.} As discussed in Appendix~\ref{appendix:biascor}, for our G10 and C11 systematic treatment, $\sigma_{\rm scat}(z,c,M_\star)$ is set to 0, and therefore $\sigma_{\rm gray}\approx0.10$ approximates the scatter, {though it does not account for the observed color dependence}.

\begin{figure}
    \centering 
	\includegraphics[width=.49\textwidth]{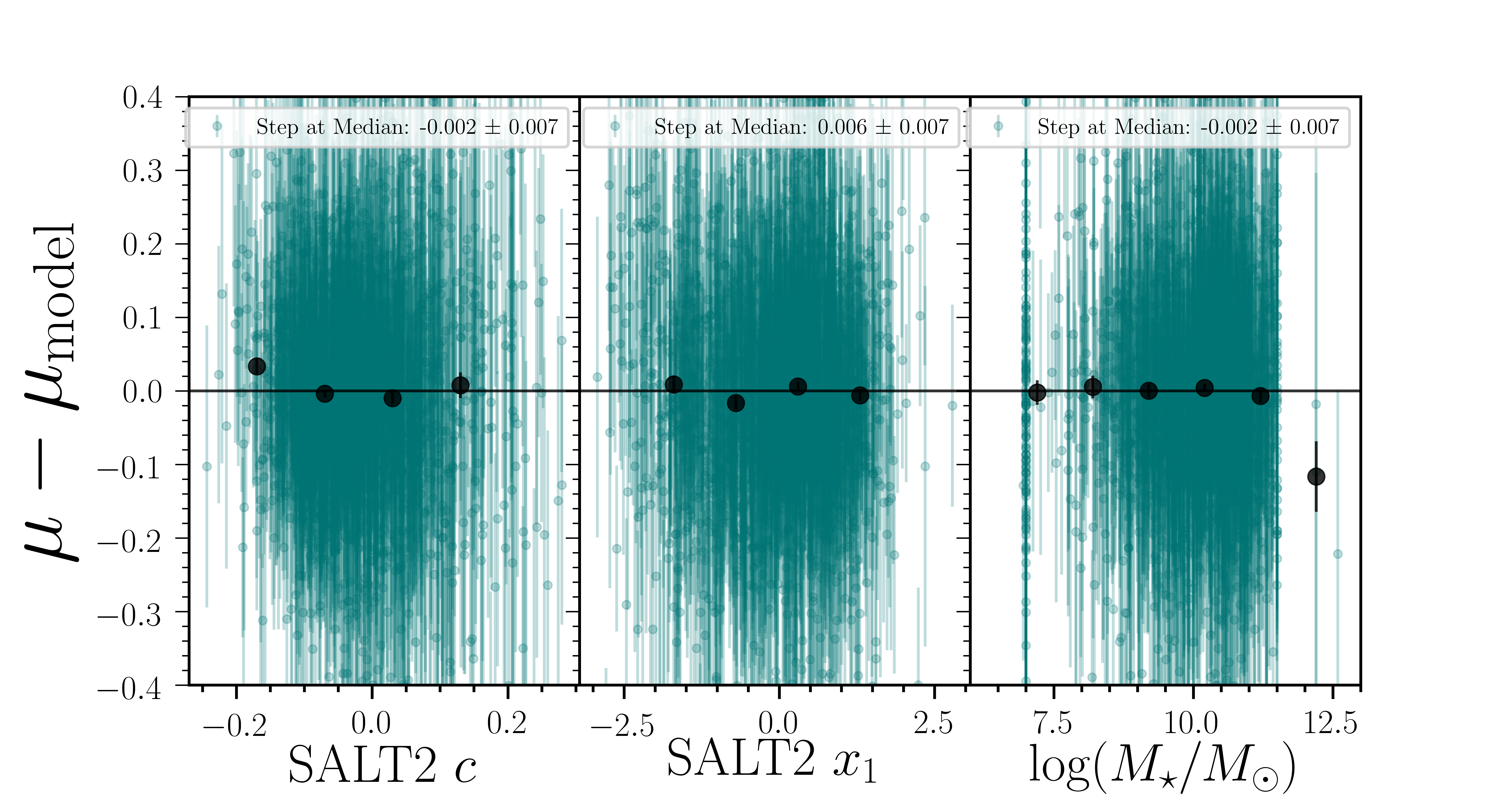} 
      \caption{Pantheon+ sample Hubble diagram residuals (teal) to the best-fit cosmology ($\mu-\mu_{\rm model}$) for the baseline analysis as a function of SALT2 $c$, SALT2 $x_1$, and host-galaxy stellar mass $M_\star$. { Distances ($\mu$) follow Eq.~\ref{eq:Tripp} and include $\alpha$, $\beta$, $\delta_{\rm bias}$, and $\delta_{\rm host}$ corrections.}  Binned data are shown for reference (black). No significant residual correlations are seen.}
    \label{fig:HDres} 
\end{figure}

Table~\ref{tab:nuisance} also shows that the best-fit host stellar mass corrections ($\gamma$) are consistent with zero for BS21 and P21. This is in agreement with the findings of \cite{popovic21b}, that modeling the intrinsic scatter in bias-correction simulations with correlations that match those in the observed data removes the need for {\it ad hoc} corrections in intrinsic brightness  (i.e., $\gamma=0$). { This can also be seen in Fig.~\ref{fig:HDres}}. For the bias correction based on the G10 and C11 models that do not include { any} mass dependence, the resulting $\gamma$ is $\sim0.05$ found at $7\sigma$ confidence.

\begin{figure}
\includegraphics[width=0.48\textwidth]{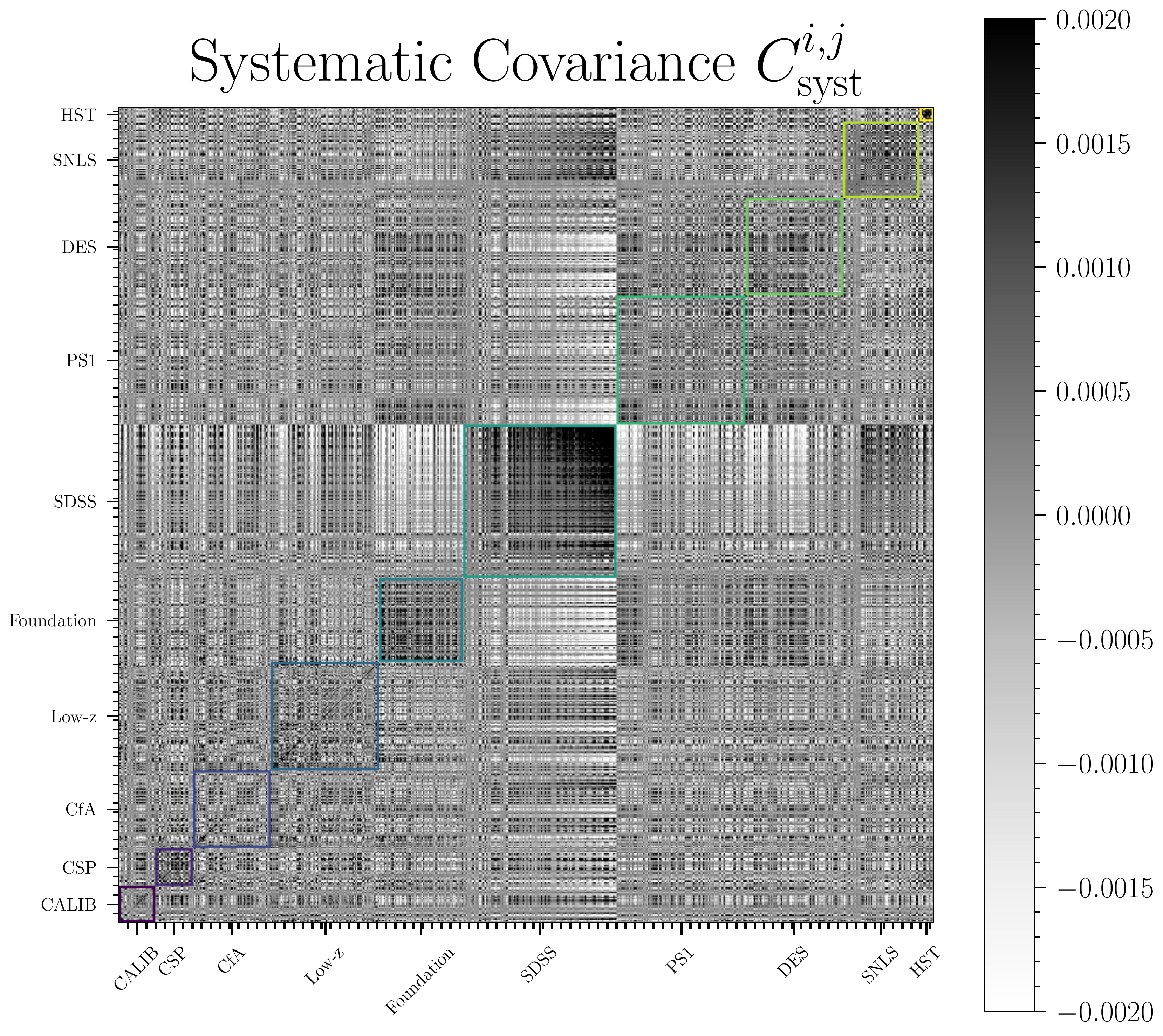}
\caption{The systematic covariance matrix as defined in Eq.~\ref{eq:csys}. To show the inherent structure, the dataset is sorted by survey and within each survey (colored boxes), by redshift. {``CALIB" are the set of 81 SN light curves in the SH0ES Cepheid-calibrator galaxies.} The shading corresponds to the size of the covariance in magnitudes.} 
\label{fig:covmat}
\vspace{.1in}
\end{figure}

\subsection{The Hubble Diagram and Distance Covariance Matrix}

\subsubsection{The Hubble Diagram}
The Pantheon+ Hubble diagram of 1701 SN~Ia light curves compiled from 18 different surveys and ranging in redshift from 0.001 to 2.26 is shown in the top panel of Fig.~\ref{fig:HD}. In the bottom panel of Fig.~\ref{fig:HD} are the residuals to the best-fit cosmology (Eq.~\ref{eq:dmu}). Best-fit cosmological parameters will be presented in the following subsections.

Shown in Table~\ref{tab:nuisance} is the total observed scatter (RMS) in the Hubble diagram residuals to the best-fit model (bottom of Fig~\ref{fig:HD}) for different scatter models.
The BS21 model results in the lowest Hubble diagram RMS and $\chi^2$, a $>5\sigma$ improvement {determined from the difference in likelihoods relative to the} G10 and C11 scatter models.
The observed scatter of $\sim0.17$\,mag is larger than seen in the original Pantheon because Pantheon+ extends to lower redshifts and thus is more impacted by scatter induced by peculiar velocities. If we set the minimum redshift to 0.01, the total scatter is reduced to 0.15\,mag, matching that of Pantheon. {Finally, compared to the original BS21 analysis,
P21 uses a more rigorous fitting process that is optimized to better characterize SN~Ia colors and intrinsic scatter in addition to Hubble residuals. For this reason, the improvements of P21 are not solely described by the cosmological model likelihood $\mathcal{L}$ of Table~\ref{tab:nuisance}. We therefore have included the use of P21 population parameters as a systematic uncertainty. }

\subsubsection{The Very Nearby Hubble Diagram}
\label{sec:verynearby}
We note from Fig.~\ref{fig:HD} that in the very nearby Universe, $z<0.008$ ($v < 2400$\,km\,s$^{-1}$), the mean of the Hubble diagram residuals
is positive by $\sim$ 5\% at $\sim 2 \sigma$ significance.  This is seen {\it after} the use of peculiar-velocity maps from either 2M++ or 2mrs. A similar signal is also seen in the Hubble residuals of the Cepheid distances \citep{kenworthy22}. A bias of roughly this size and direction is expected in the presence of measurement errors and unmodeled peculiar velocities which scatter more objects down from higher redshifts and greater volume than from the reverse.  This effect is significant only for the most nearby galaxies ($z<0.008$). In Fig.~\ref{fig:HD}, we include the prediction (dashed line) for this bias assuming 250~km~s$^{-1}$ uncorrected velocity scatter (not a fit).

In the the 3-rung distance ladder utilized to measure $H_0$ by the SH0ES Team (R22) and in Eq.~\ref{eq:dmuprime} in this work, the nearby ($z<\sim0.01$) Hubble diagram is not used. Rather, only the distance moduli from such nearby SNe are used in the SN-Cepheid absolute distance calibration in the 2$^{\rm nd}$ rung. Furthermore, in the R22 measurement of the Hubble flow, only SNe with \textit{redshifts} $z>0.023$ are used in the 3$^{\rm rd}$ rung to limit sensitivity to peculiar velocities. This approach is insensitive to the volumetric redshift scatter effects and there is no resulting impact on the R22 H$_0$.  However, more local measurements of H$_0$ from, for example, a 2-rung distance ladder using primary distance indicators like Cepheids and TRGB and their host redshifts (mostly at $z \leq 0.01$) are more sensitive to peculiar velocities and the volumetric bias they induce, and are likely to be biased low at the few percent level if not appropriately accounting for this expected bias \citep{kenworthy22}. For measurements of other cosmological parameters (e.g., $w$ or $\Omega_M$) with Pantheon+ described in the following subsections, the mean Hubble residual bias of the Low-z and Foundation sample is $\sim2$~mmag and $\sim1$~mmag (respectively), and is considered to be negligible.

\begin{figure}
\includegraphics[width=0.48\textwidth]{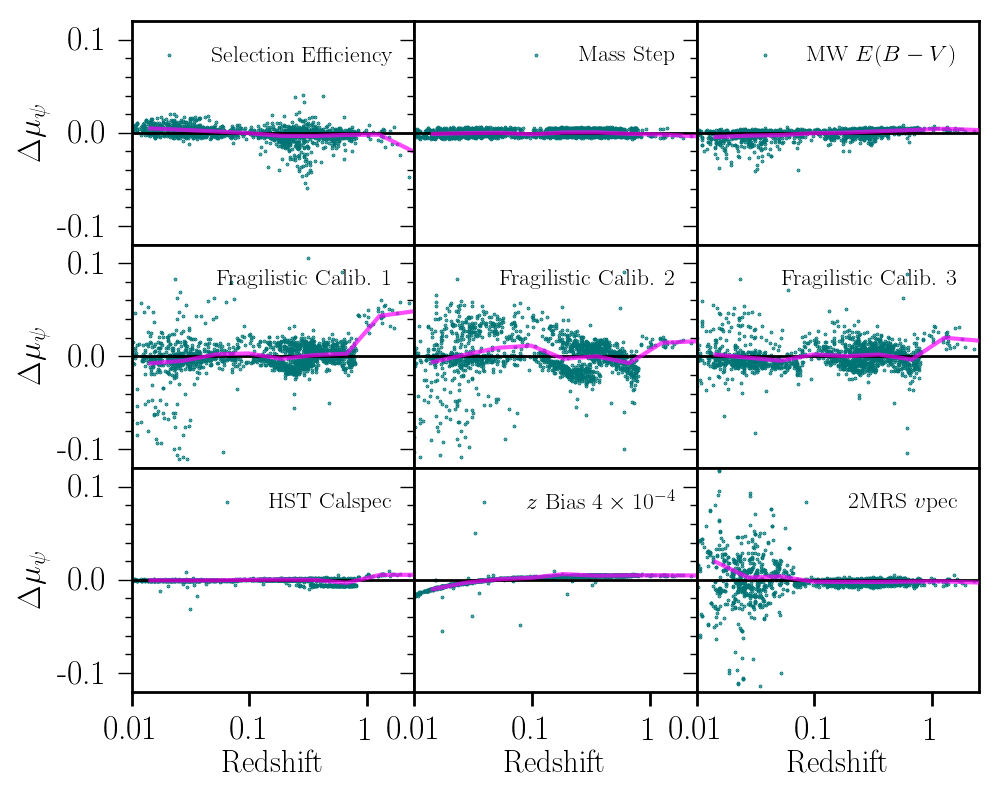}
\caption{Visualizing the impact of a number of the top systematic uncertainties in this analysis.  The $\mu$ residuals are described by Eq.~\ref{eq:deltamupsi}. Each of these systematics is explained in Sec.~\ref{sec:Inputs} and are combined to form the covariance matrix shown in Fig.~\ref{fig:covmat}. { Fragilistic provides 9 systematic sets of trained SALT2 models, zero-point solutions, and filter central wavelengths. Here we show the impact on distance of just the first 3.}}
\label{fig:sysmudif}
\vspace{.1in}
\end{figure}

\subsubsection{The Distance Covariance Matrix}
Built following Eq.~\ref{eq:csys}, the 1701$\times$1701 systematic distance covariance matrix is shown in Fig.~\ref{fig:covmat}.  The sample is sorted by survey and redshift to help visualize the covariances. The Hubble diagram residuals (Eq.~\ref{eq:dmu}) that are used to build the covariance matrix are shown in Fig.~\ref{fig:sysmudif} for several example sources of systematic uncertainty.  As discussed in Appendix~\ref{appendix:dr}, the information used to create the Hubble diagram as well as the covariance matrix is publicly available\footnote{pantheonplussh0es.github.io} and tools to read in this information are in CosmoSIS. { The SDSS subsample contributions to the covariance matrix (Fig.~\ref{fig:covmat}) stand out visually due to their strong spectroscopic selection function.}

\subsection{Constraints on Cosmological Parameters From Pantheon+ and SH0ES}

Parameter constraints from the Pantheon+ {SNe~Ia} and SH0ES {Cepheid host absolute distances} are shown in Table~\ref{tab:ppcosmology} for Flat$\Lambda$CDM, $\Lambda$CDM, Flat$w$CDM, and Flat$w_0w_a$CDM. Unless otherwise stated, constraints on cosmological parameters include both statistical and systematic uncertainties. 
From the Pantheon+ SNe~Ia, for a Flat$\Lambda$CDM model we find $\Omega_M=~$\fLCDMOMPan. We note that SH0ES (R22) utilizes Pantheon+ SNe at $z<0.8$ to constrain the deceleration parameter and find $q_0=-0.51\pm0.024$. In a flat universe  $q_0=\frac{3\Omega_M}{2}-1$, which gives $\Omega_M = 0.326\pm0.016$, consistent with the result for $\Omega_M$ reported in this work. { Results for H$_0$ from the inclusion of the SH0ES Cepheid host distances are discussed below.}

\begin{figure}
\includegraphics[width=0.49\textwidth]{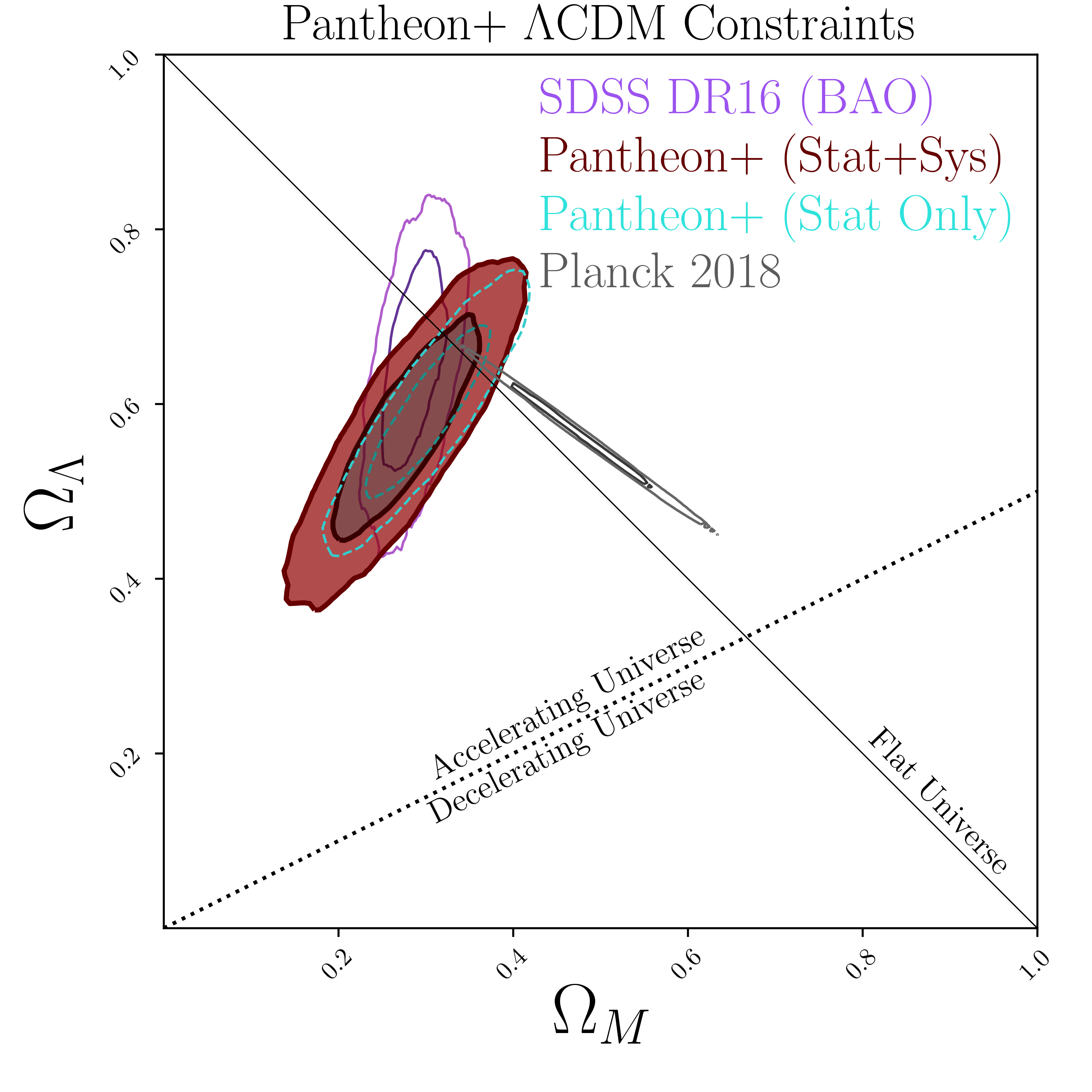}
\caption{Confidence contours at the 68\% and 95\% level for the $\Omega_M$ and $\Omega_\Lambda$ cosmological parameters for the $\Lambda$CDM from the Pantheon+ dataset, as well as from the Planck and combined BAO datasets. The constraints from including both the statistical and systematic uncertainties (shaded red) are shown as well as when only statistical uncertainties are propagated (unfilled dashed).  We include two lines for reference: one for a flat universe, where $\Omega_M+\Omega_\Lambda=1$ and the other that indicates an accelerating universe. }
\label{fig:omol}
\vspace{.1in}
\end{figure}

The constraints on $\Omega_M$ and $\Omega_\Lambda$ for a $\Lambda$CDM model are shown in Fig.~\ref{fig:omol}. We find  $\Omega_M$=~\oLCDMOMPan\ and $\Omega_\Lambda$=~\oLCDMOLPan; a flat universe is within the 68\% confidence region and $\Omega_M=0$ and $\Omega_\Lambda=0$ are together rejected at 4.4$\sigma$ using only the SNe.

\begin{figure*}
\includegraphics[width=0.99\textwidth]{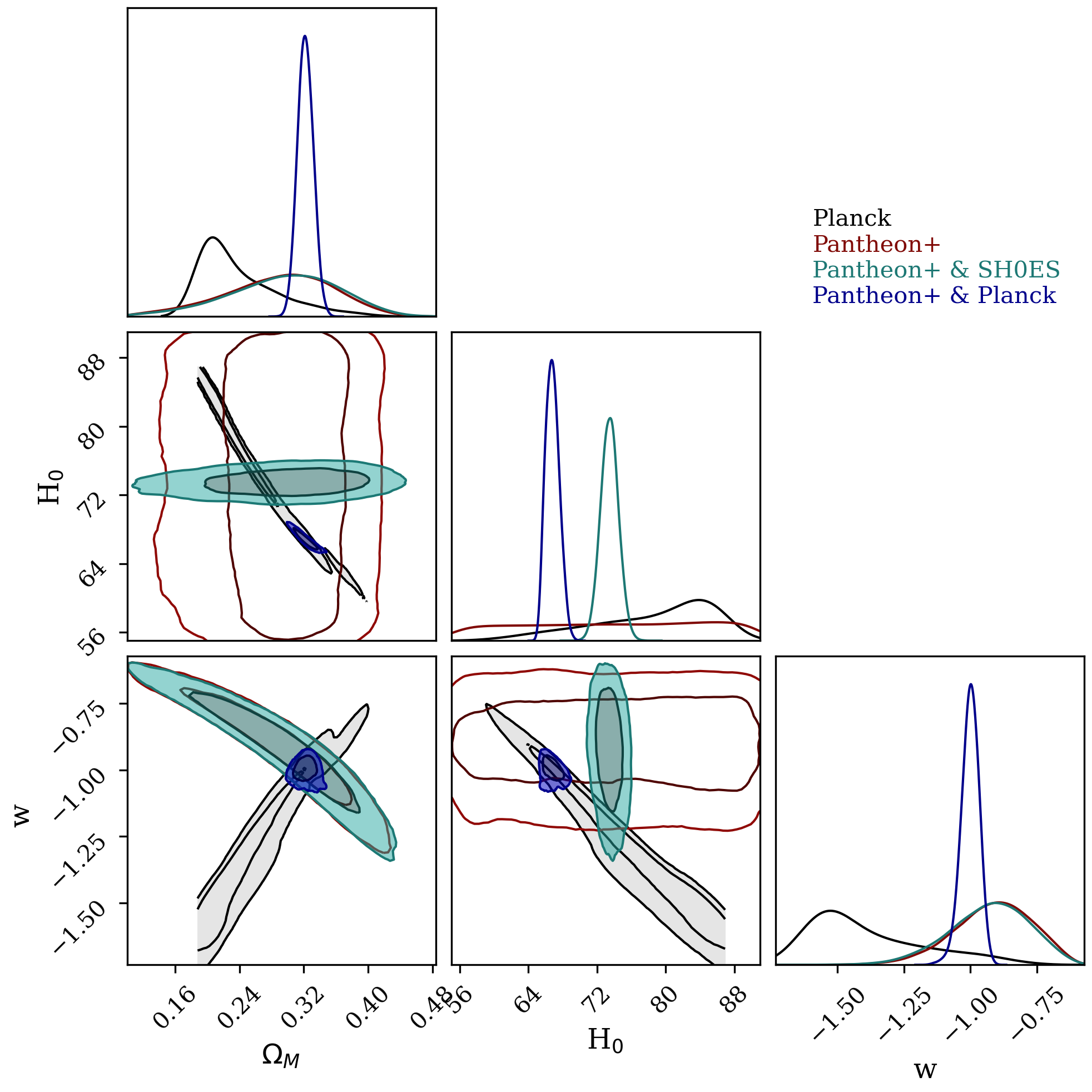}
\caption{$68\%$ and $95\%$ confidence contours for Flat$w$CDM for cosmological parameters $\Omega_M$, H$_0$, and $w$.  The contours from the Pantheon+ (red), Pantheon+ \& SH0ES combined dataset (teal), Planck \cite{planck} TTTEEE-lowE constraints (gray). The combination of Planck and Pantheon+ (blue) is also shown, which is consistent with a cosmological constant.  Planck constraints are bounded by $0.2<\Omega_M<0.4$ for computational speed. {The histograms depict marginalized relative probabilities between probes.}}
\label{fig:omw}
\vspace{.4in}
\end{figure*}

For a Flat$w$CDM model, from the SNe~Ia alone { (not including SH0ES Cepheid calibration)} we find $\Omega_M=~$\OMPanSH\ and $w=$~\wPanSH\ as shown in the third row of Table~\ref{tab:ppcosmology} and in the blue contour of Fig.~\ref{fig:omw}.  This result is consistent within $1\sigma$ of the cosmological constant ($w=-1$).

\begin{figure*}
\includegraphics[width=0.99\textwidth]{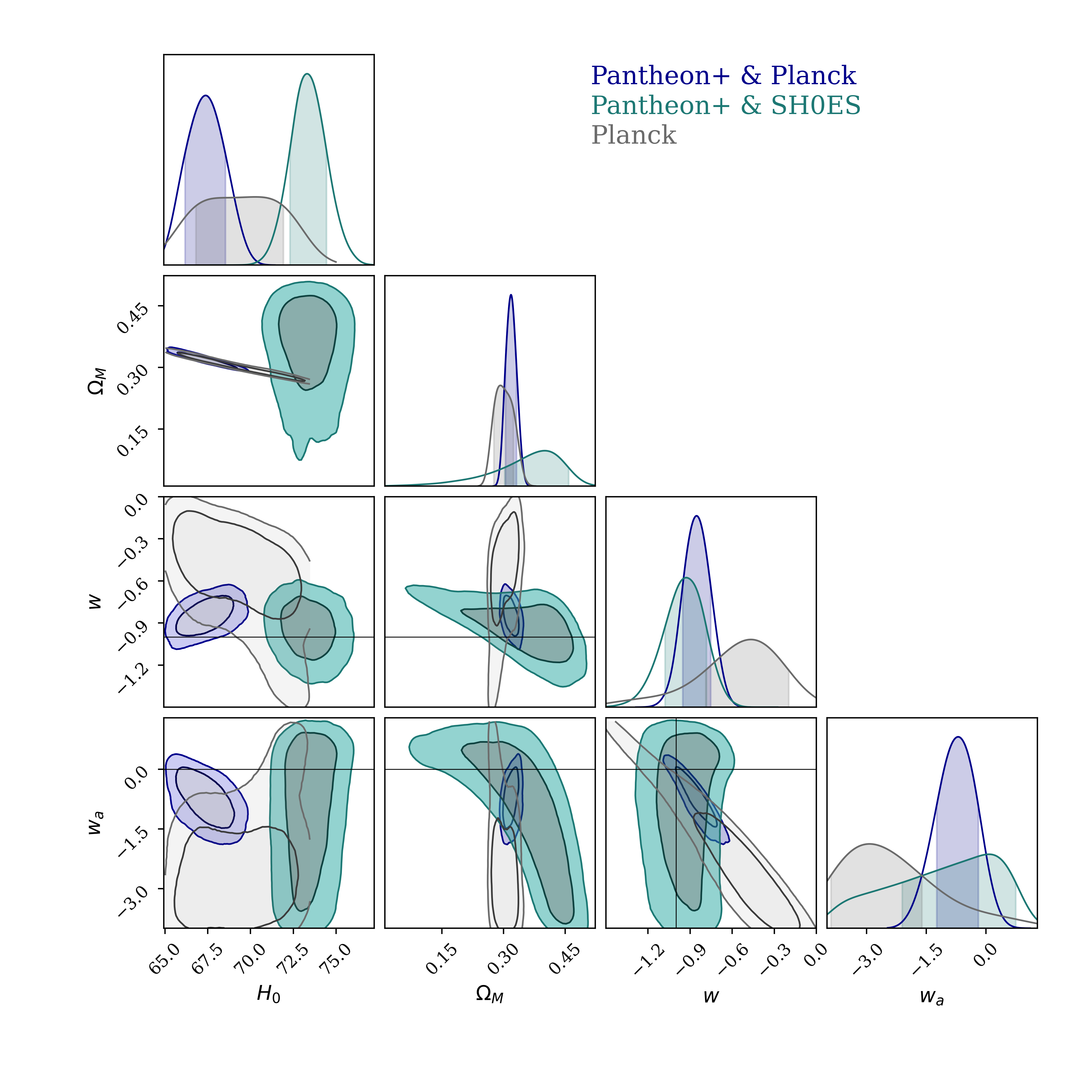}
 \caption{Constraints for Flat$w_0w_a$CDM from the Pantheon+ dataset in combination with SH0ES, Planck TTTEEE-lowE.}
\label{fig:wwa}
\vspace{.4in}
\end{figure*}

For a Flat$w_0w_a$CDM model, from the SNe~Ia alone { (not including SH0ES Cepheid calibration)}  we find $w_0=~$\wwawnPan\ and $w_a=~$\wwawaPan\ as shown in the fourth row of Table~\ref{tab:ppcosmology} and in Fig.~\ref{fig:wwa}. These results are again consistent with a cosmological constant.

Using distances and a stat+syst covariance matrix that extends to the Cepheid calibrators (Eq.~\ref{eq:sh0eslikelihood}) {  and combining the Pantheon+ SNe with the SH0ES Cepheid host distance calibration}, we are able to robustly and simultaneously constrain H$_0$ and other cosmological parameters describing the expansion history.
 {While we use SH0ES Cepheid data and covariance in this work, likewise Pantheon+ distances and covariance are used in Section~5.2 of R22 in order} to fit H$_0$ and $q_0$ in Flat$\Lambda$CDM. As shown in the top Pantheon+ \& SH0ES section of Table~\ref{tab:ppcosmology}, for $\Lambda$CDM, Flat$w$CDM, and Flat$w_0w_a$CDM we find H$_0=$ \oLCDMHPanSH, \HPanSH, and \wwaHPanSH\ km\,s$^{-1}$\,Mpc$^{-1}$, respectively. We note that more complex models do not result in decreased H$_0$ constraining power from the SNe~Ia + Cepheids, while this is not necessarily true for other cosmological probes (Sec.~\ref{sec:externalprobesresults}).

\begin{table*}
    \raggedleft
    \caption{Results for Cosmological Models}
    \label{tab:ppcosmology}
    \begin{tabular}{lccccc}
        \toprule
		 & $\Omega_M$ & $\Omega_\Lambda$ & H$_0$ & $w_0$ & $w_a$ \\ 
		\hline\\
		\textbf{Pantheon+ \& SH0ES} - \textit{All Models}\phantom{AAA}\\
		\hline
		Flat$\Lambda$CDM & \fLCDMOMPan & \fLCDMOLPan & \fLCDMHPanSH &-&-\\ 
		$\Lambda$CDM & \oLCDMOMPan & \oLCDMOLPan & \oLCDMHPanSH&-&-\\ 
		Flat$w$CDM & \OMPanSH & \OLPanSH & \HPanSH & \wPanSH &-\\ 
		Flat$w_0w_a$CDM & \wwaOMPan & \wwaOLPan & \wwaHPanSH & \wwawnPan &\wwawaPan \\ 
		\hline\\
		\textbf{External Probes (No SH0ES)} - \textit{Flat$w$CDM} \\
		\hline		
		Planck \& Pantheon+ & \OMPanPlanck & \OLPanPlanck& \HPanPlanck & \wPanPlanck& -\\ 
        Planck \& galaxyBAO  \& Pantheon+ & \OMPanPlanckgBAO & \OLPanPlanckgBAO& \HPanPlanckgBAO & \wPanPlanckgBAO & -\\ 
        Planck \& allBAO  \& Pantheon+ & \OMPanPlanckBAO & \OLPanPlanckBAO& \HPanPlanckBAO & \wPanPlanckBAO & -\\ 
		\hline\\
		\textbf{External Probes (No SH0ES)} - \textit{Flat$w_0w_a$CDM} \\
		\hline	
		Planck \& Pantheon+ & \wwaOMPanPlanck & \wwaOLPanPlanck & \wwaHPanPlanck & \wwawPanPlanck & \wwawaPanPlanck\\ 
        Planck \& galaxyBAO  \& Pantheon+ & \wwaOMPanPlanckgBAO &\wwaOLPanPlanckgBAO & \wwaHPanPlanckgBAO & \wwawPanPlanckgBAO& \wwawaPanPlanckgBAO \\ 
        Planck \& allBAO  \& Pantheon+ & \wwaOMPanPlanckBAO &\wwaOLPanPlanckBAO & \wwaHPanPlanckBAO & \wwawPanPlanckBAO& \wwawaPanPlanckBAO \\ 
\toprule
    \end{tabular}
    \vspace{3mm}
    \raggedright
    \\ Notes:  Summary of marginalized parameter constraints for Pantheon+ and other external probes. The mean and 68\% confidence limit are provided for each cosmological parameter. A blank value indicates a parameter not used in the cosmological fit.\\
    \vspace{15mm}
\end{table*}

\begin{figure}
\includegraphics[width=0.49\textwidth]{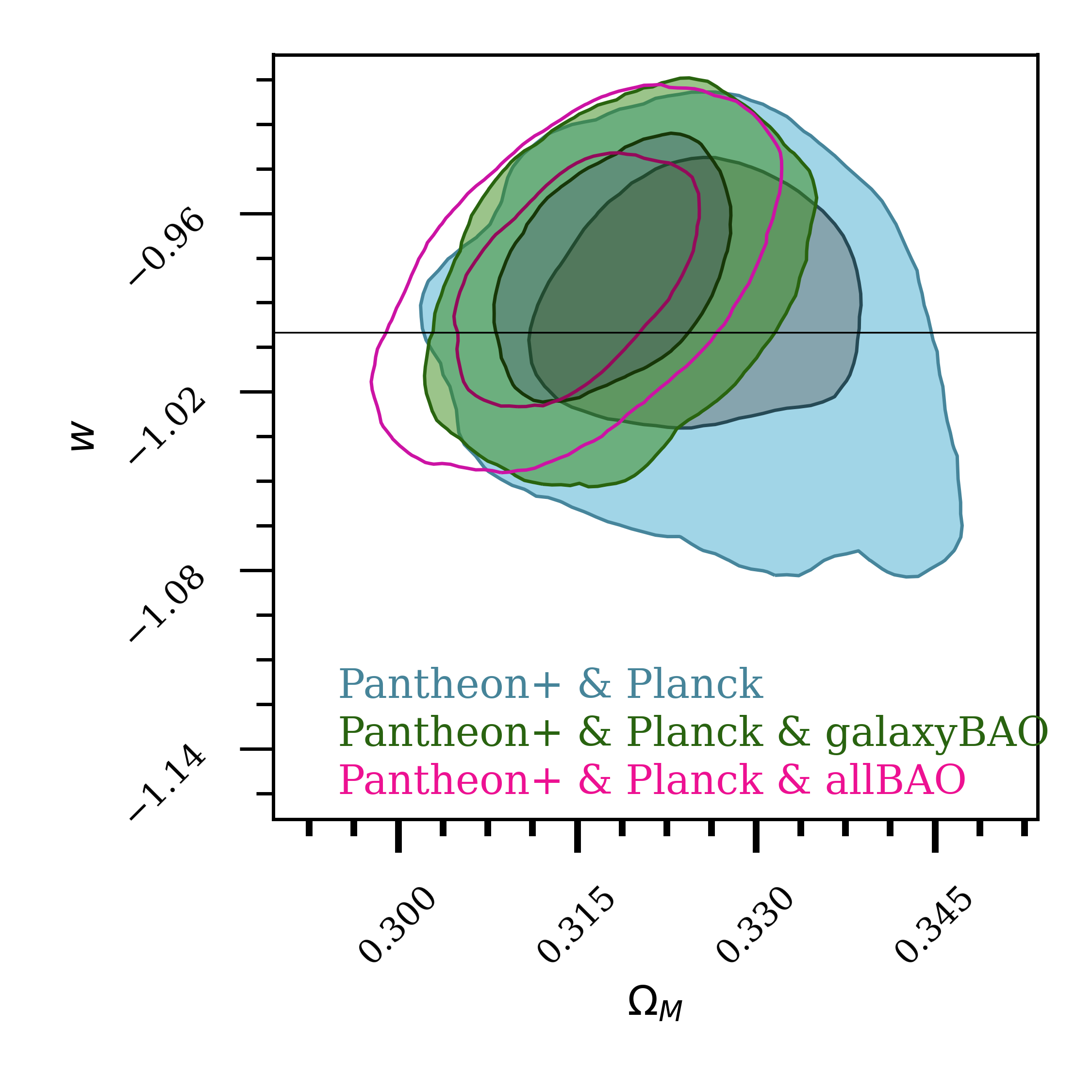}
 \caption{Constraints for Flat$w_0$CDM from the Pantheon+ dataset in combination with Planck \& galaxyBAO or Planck \& allBAO.}
\label{fig:omwzoom}
\end{figure}

\begin{figure}
\includegraphics[width=0.49\textwidth]{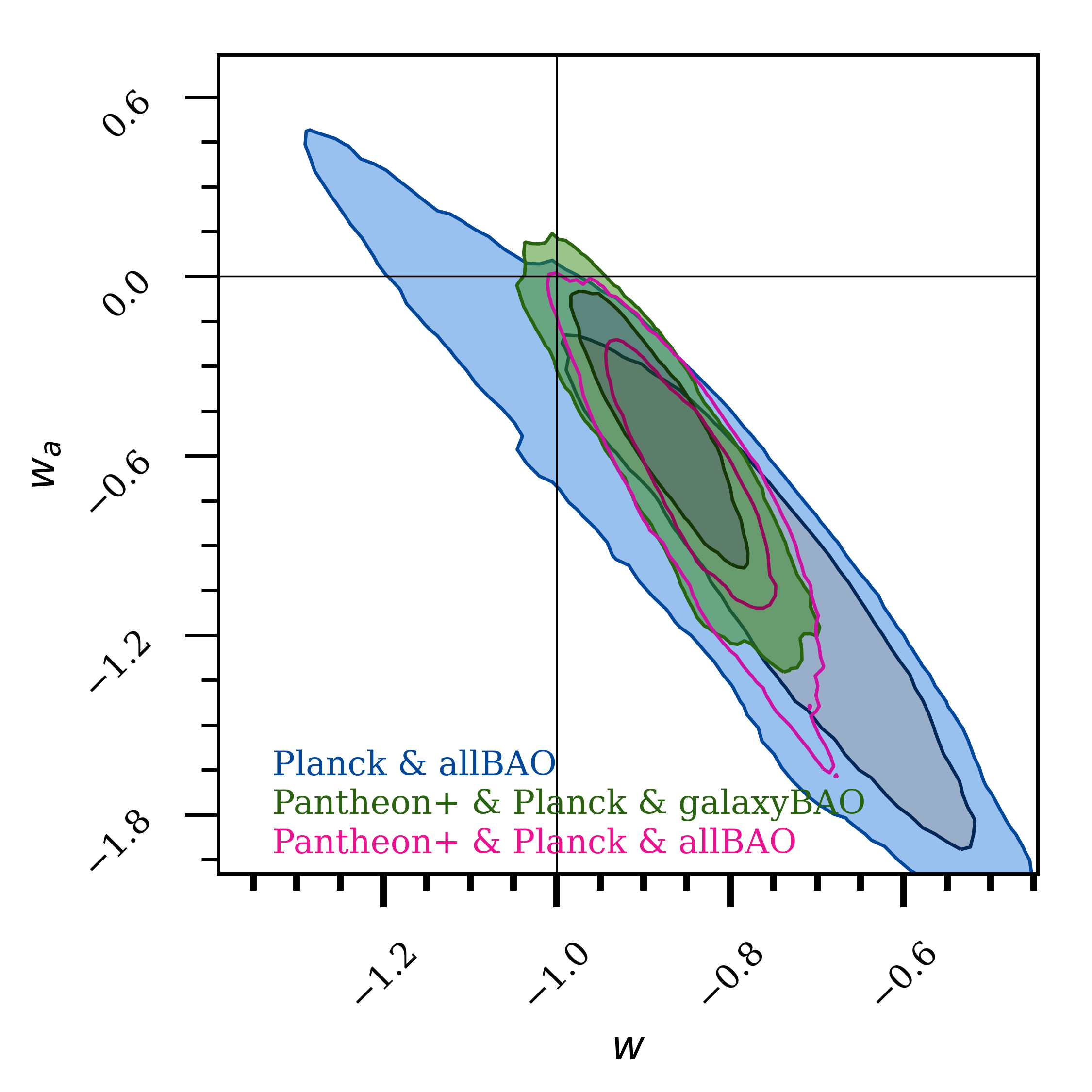}
 \caption{Constraints for Flat$w_0w_a$CDM from the Pantheon+ dataset in combination with Planck \& galaxyBAO or Planck \& allBAO.}
\label{fig:wwazoom}
\end{figure}

\subsection{Constraints on Cosmological Parameters From Multiple Probes}
\label{sec:externalprobesresults}

In this work we combine the Pantheon+ SNe with external cosmological probes: CMB from Planck \citep{planck} TTTEEE-lowE and baryon acoustic oscillations (BAO) from SDSS MGS \citep{bao_mgs}, SDSS BOSS \citep{bossdr12}, SDSS eBOSS LRG \citep{ebosslrg}, SDSS eBOSS ELG \citep{ebosslrg}, SDSS eBOSS QSO \citep{ebossqso}, SDSS eBOSS Lya \citep{ebosslyaa}, all of which have been implemented in CosmoSIS. The aforementioned BAO constraints are denoted ``allBAO"; we also provide constraints from the combination of spectroscopic redshift galaxy-only subset of BAO probes denoted ``galaxyBAO." We report constraints in Table~\ref{tab:ppcosmology} for combinations of datasets that are deemed compatible and discussed below. 

For a Flat$w$CDM model when combining Pantheon+ and Planck we find $w=~$\wPanPlanck\ and $\Omega_M=~$\OMPanPlanck, and when further including allBAO we find $w=~$\wPanPlanckBAO\ and $\Omega_M=~$\OMPanPlanckBAO, both of which are  consistent with the cosmological constant at $\sim3$\% (Fig.~\ref{fig:omwzoom}).  As can be seen in Fig.~\ref{fig:omw}, we do not include SH0ES in combinations with Planck because these measurements are incompatible (R22).

For a Flat$w_0w_a$CDM model when combining Pantheon+ and Planck we find $w_0=~$\wwawPanPlanck\ and $w_a=~$\wwawaPanPlanck,  and when combining Pantheon+, Planck, and BAO we find $w_0=~$\wwawPanPlanckBAO\ and $w_a=~$\wwawaPanPlanckBAO, which is moderately consistent (2$\sigma$) with a cosmological constant (Fig.~\ref{fig:wwazoom}). We note that this result is not driven by any single probe. In Fig.~\ref{fig:wwa} we show constraints for Planck alone and for the combination of Planck \& Pantheon+. While {the broader model freedom of} the Flat$w_0w_a$CDM allows the Planck alone H$_0$ to be consistent with 73\,km\,s$^{-1}$\,Mpc$^{-1}$ {owing to degeneracy between H$_0$ and $w_a$} (see Fig.~\ref{fig:wwa}), after combining Planck with Pantheon+, the H$_0/w_a$ degeneracy is broken (H$_0=~$\wwaHPanPlanck~km\,s$^{-1}$\,Mpc$^{-1}$). Therefore, the inclusion of SH0ES with Planck \& Pantheon+ results in a Bayesian evidence ratio of $-9$, and we deem this set of probes incompatible and do not include them in Fig.~\ref{fig:wwa} nor Table~\ref{tab:ppcosmology}.

\subsection{Impact of Systematics on Cosmological Parameter Fits}

To understand the impact of systematic uncertainties, in Table~\ref{Tab:inputsys} we group the systematics investigated in this work into four main categories: Calibration/SALT2, Redshifts, Astrophysics, and Modeling. The baseline, systematic treatments ($S_\psi$) and scaling priors ($\sigma_\psi$) (as described in detail in Section~\ref{sec:Inputs}) are summarized for each source. The final three columns of Table~\ref{Tab:inputsys} relate to fits of the sample when combined with Planck \cite{planck} in a Flat$w$CDM model when isolating that systematic. We define both the change in best fit ($\Delta w_{\rm sys}$) and the systematic uncertainty contribution to $w$ ($\sigma_w^{\rm sys}$) as follows:
\begin{equation}
    \Delta w_{\rm sys} =  w_{\rm sys} - w_{\rm stat}
\end{equation}
\vspace{-3mm}
\begin{equation}
    \sigma_w^{\rm sys} = \sqrt{\sigma_{w\rm tot}^2-\sigma_{w\rm stat}^2},
\end{equation}
where $w_{\rm sys}$ and $\sigma_{w\rm tot}$ are the cosmological constraints when utilizing $C_{\rm stat+sys}$ and where $w_{\rm stat}$ and $\sigma_{w\rm stat}$ are the statistical-only constraints when utilizing $C_{\rm stat}$. 

We find that the final systematic uncertainty in $w$ ($\sigma_{\rm sys}=0.019$) is comparable yet smaller ($\sim80\%$) than the statistical uncertainty, suggesting that the measurement is not systematics dominated. The largest contribution to the systematic error budget (0.011) is due to the potential for redshift-measurement bias. This is followed by the uncertainties in the Fragilistic calibration offsets and the resulting propagation to SALT2 model-training uncertainties and light-curve fitting uncertainties (0.009). Additionally important is the conservative uncertainty that was applied owing to the usage of the new SALT2 training methodology (0.008) as well as the uncertainty in the MW extinction maps (0.008). 

 Interestingly, numerous systematic uncertainties are found to be negligible (e.g., BS21 Parameters, G10 versus C11) in the cosmological parameter budget. While certain systematics cause redshift-dependent trends as shown in Fig.~\ref{fig:sysmudif}, they also change the relative scatter of the Hubble residuals.  This can most easily be seen for the cosmological likelihood values ($\mathcal{L}$) for the distances with different intrinsic scatter models shown in Table~\ref{tab:nuisance}.  If the baseline analysis is {significantly preferred (larger $\mathcal{L}$) by the data over one of the analysis variants, the impact of that systematic on cosmological constraints} will be reduced,  as is the case for intrinsic scatter.

 As we have built a covariance matrix that includes the Cepheid calibrators, we can measure H$_0$ with and without systematic uncertainties.  For Flat$\Lambda$CDM, we find H$_0=$~\fLCDMHPanSH\,km\,s$^{-1}$\,Mpc$^{-1}$, and when considering only statistical uncertainties { from the SNe alone (excluding Cepheid and physical distance calibration uncertainties) $\sigma^{\rm stat+syst}_{{\rm H}_0}=0.7$\,km\,s$^{-1}$\,Mpc$^{-1}$, and $\sigma^{\rm syst}_{{\rm H}_0}=0.29$\,km\,s$^{-1}$\,Mpc$^{-1}$.} This suggests that SN systematic uncertainties are not dominating the constraint on H$_0$ and cannot explain the $\sim7$\,km\,s$^{-1}$\,Mpc$^{-1}$ difference between Planck and SH0ES.

\begin{figure}
\includegraphics[width=0.49\textwidth]{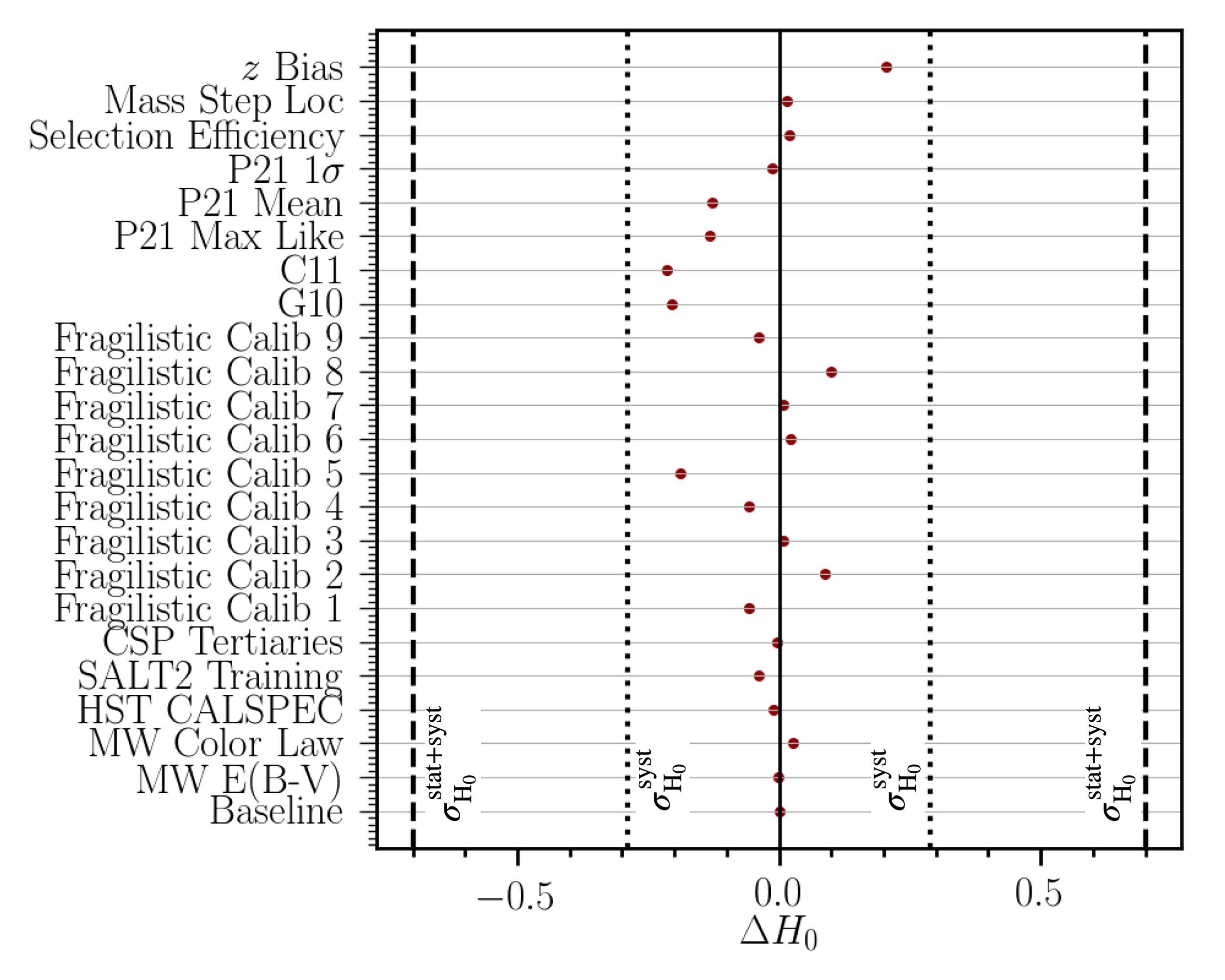}
\caption{The impact on recovery of H$_0$, as explained in Sec.~\ref{sec:Method}, of the systematic uncertainties described in Table~\ref{Tab:inputsys}. The units of these measurements are km\,s$^{-1}$\,Mpc$^{-1}$.  The dashed lines are given at $\Delta {\rm H}_0$ of 0.7, which is the entire contribution of the uncertainty in R22 from SN measurements. }
\label{fig:sysh0}
\vspace{.1in}
\end{figure}

In Figure~\ref{fig:sysh0} we show deviations to the best-fit H$_0$ for each individual source of systematic uncertainty relative to the baseline analysis and assuming $\Lambda$CDM. For reference we also show the full SN contribution to the H$_0$ error bar (dashed). The deviations from the baseline ($\Delta {\rm H}_0$) are small and add in quadrature to 0.32\,km\,s$^{-1}$\,Mpc$^{-1}$. We note that when assessing redshift-specific systematics, because model redshifts are not used for the SN-Cepheid calibration in Eq.~\ref{eq:dmuprime}, they mainly impact the Hubble-flow SNe (third rung of the distance ladder).

Finally, to help visualize the impact of systematic uncertainties, we show in Fig.~\ref{fig:omol} the constraints when including either statistical-only uncertainties or the combined statistical and systematic uncertainties. Error budgets for different cosmological parameterizations can be generated with the delineated files for systematics provided as part of this release.

\begin{table*}
\centering
\caption{Sources of Uncertainty}
\begin{tabular}{llllccc}
\toprule

Description & Baseline  & Systematic (S$_\psi$) & $\sigma_\psi$ &  $\sigma_{w{\rm sys}}$&$\sigma_{w{\rm sys}}/\sigma_{w{\rm stat}}$&$\Delta w_{\rm sys}$ \\
\hline \\

\textbf{\ul{All Systematics}}&&&&0.019&0.79&-0.009\\
\\
\hline \\

\textbf{\ul{Calibration}}\\ %

SALT2 Train \& $^*$LCFIT  & Fragilistic Best Fit & 10 covariance realizations & $1/3$ each&0.009&0.38&0.000\\
SALT2 Method & SALT2-B22 & JLA SALT2 Surface & $1/3$ & 0.008&0.33&0.003\\
CSP Tertiary Stars       & \cite{krisciunis17} &  \cite{Stritzinger2018} & $1$ & 0.003&0.13&$-0.003$\\
{\it HST} & Calspec 2020 Update &  5\,mmag/7000\,\AA & $3$ & 0.003&0.13 &$-0.006$ \\
\\

\textbf{\ul{Redshifts}}\\
$v_{\rm pec}$ Map & 2M++ & 2M++ iLOS \& 2MRS & $0.7$ each &0.002&0.08&0.005\\
Redshift Bias & No $z$-shift & $10^{-4}$ $z$-shift & $1$&0.011&0.46&0.015\\
\\

\textbf{\ul{Astrophysics}}\\

Intrinsic Variations & BS21 dust model &  G10 and C11 & $0.7$ each & 0.002&0.08&$-0.003$ \\
MW $E(B-V)$ & \cite{Schlafly11} & 4\% Scaling & $1.0$ & 0.008&0.33 & $-0.010$\\
MW Color Law & \cite{Fitzpatrick99} & \cite{ccm} & $1/3$ & 0.006&0.25&$-0.006$\\
Mass Step & Split at 10 & Split at 10.2 & $1$ & 0.001&0.04 & 0.000\\
\\

\textbf{\ul{Modeling}}\\

Selection Efficiency & Nominal Exposure Time & 20\% increase & $1$ &0.004&0.17&0.001\\
Populations & BS21 parameters &  3 Sets of Params (P21) & $0.6$  &0.000&0.00&0.003 \\

\\

\toprule\\
\label{Tab:inputsys}
\vspace{-4mm}
\end{tabular}
{\raggedright
$^*$LCFIT denotes zero-points and filter central wavelengths have been varied during light-curve fitting.\\
$^a$Constraints are combined with Planck 2018.\\
\textbf{Notes}: A summary of the systematic uncertainties and the baseline component of the analysis as described in Sec.~\ref{sec:Inputs}, the size of the systematic $S_\Psi$ used to determine the impact of that systematic, the scaling of the systematic $\sigma_\psi$ as constrained in this analysis, and the contribution to the total uncertainty in $w$CDM (can be compared to statistical uncertainty of 0.03), and the shift when allowing the uncertainty on the best-fit cosmological parameter.  The last column shows the simplistic change in best-fit cosmology if a perturbation of size $\sigma_\psi$ is applied with statistical-only uncertainties.  The amount shown is different than seen for the combined shift for best-fit and increase of uncertainty given in the previous columns due to the self-calibration as explained by \cite{binning21}.   \\ 
\vspace{26mm}
}
\end{table*}

  \begin{figure*}[h]
        \centering 
	    \includegraphics[width=.48\textwidth]{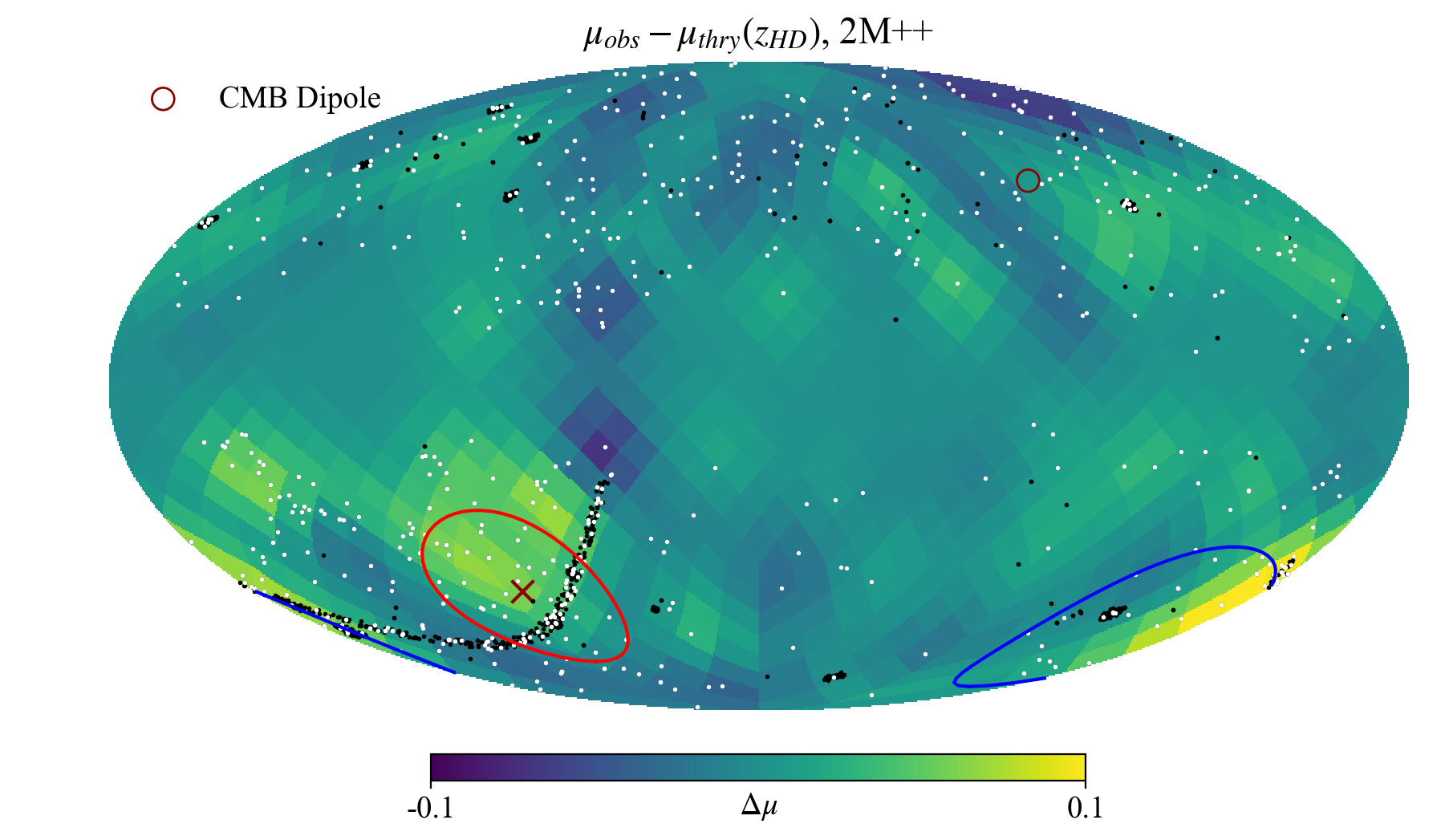}  
	    \includegraphics[width=.48\textwidth]{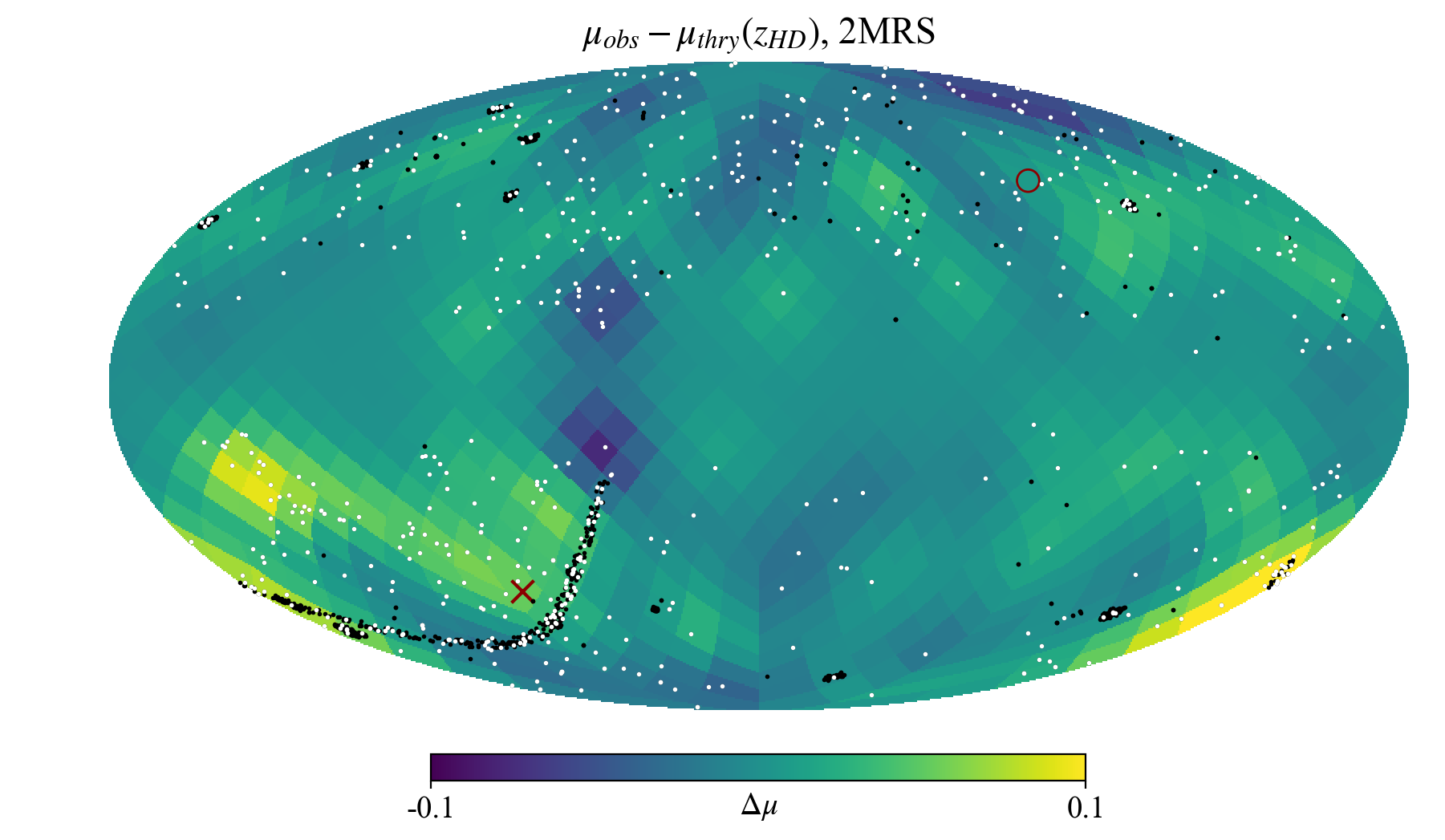}  
	    \includegraphics[width=.48\textwidth]{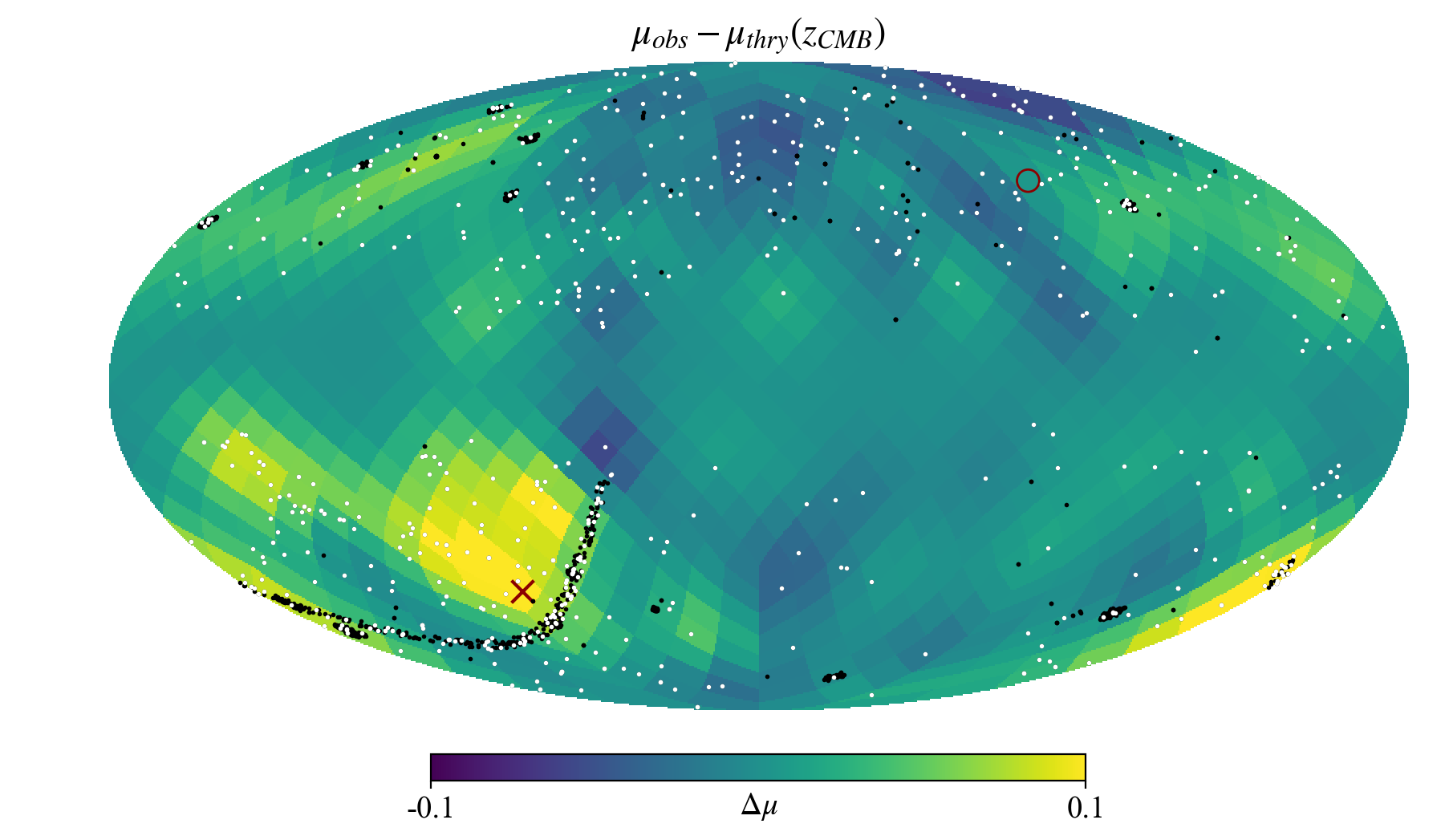}  
	    \includegraphics[width=.48\textwidth]{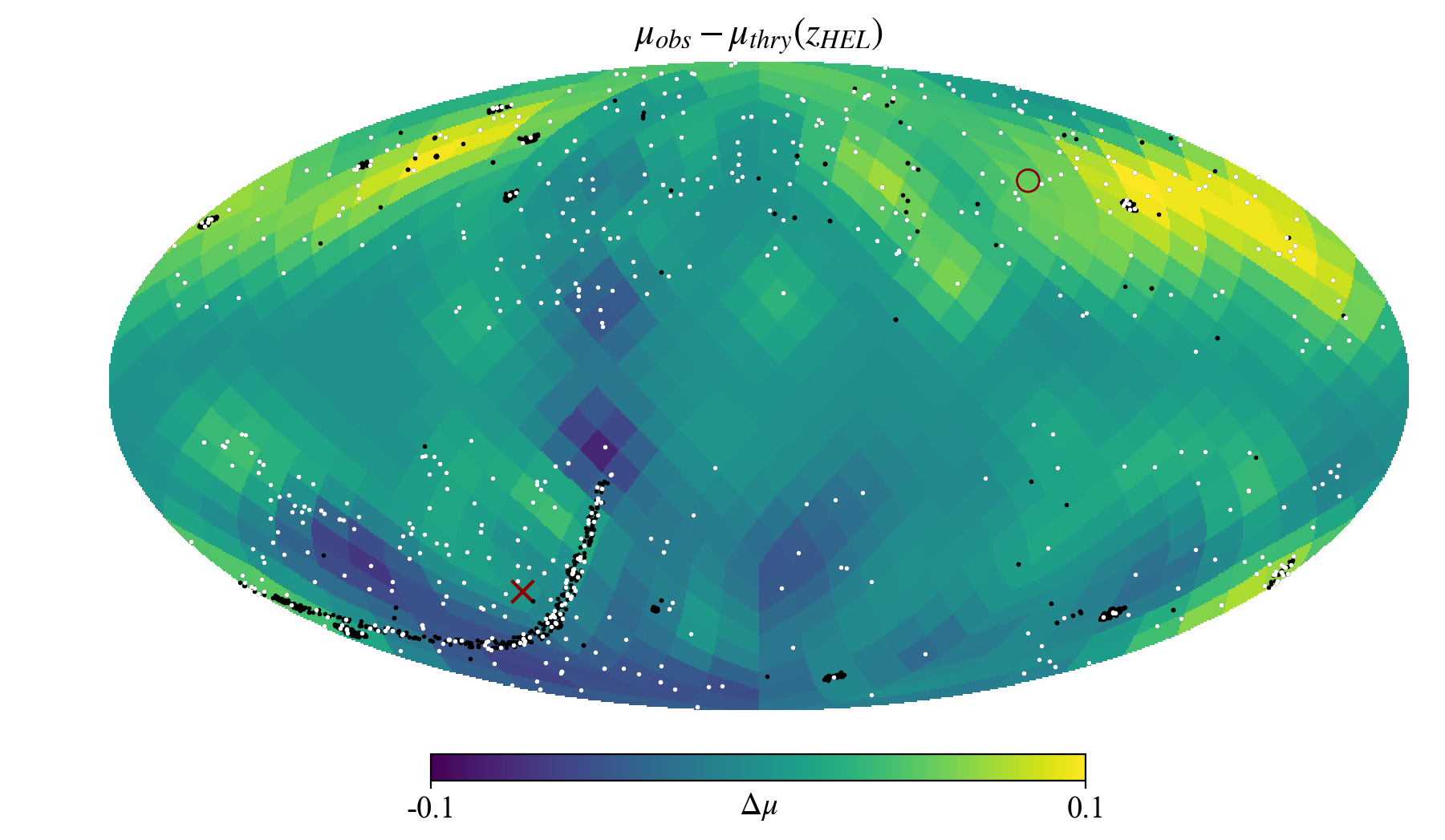}  
	    \includegraphics[width=.48\textwidth]{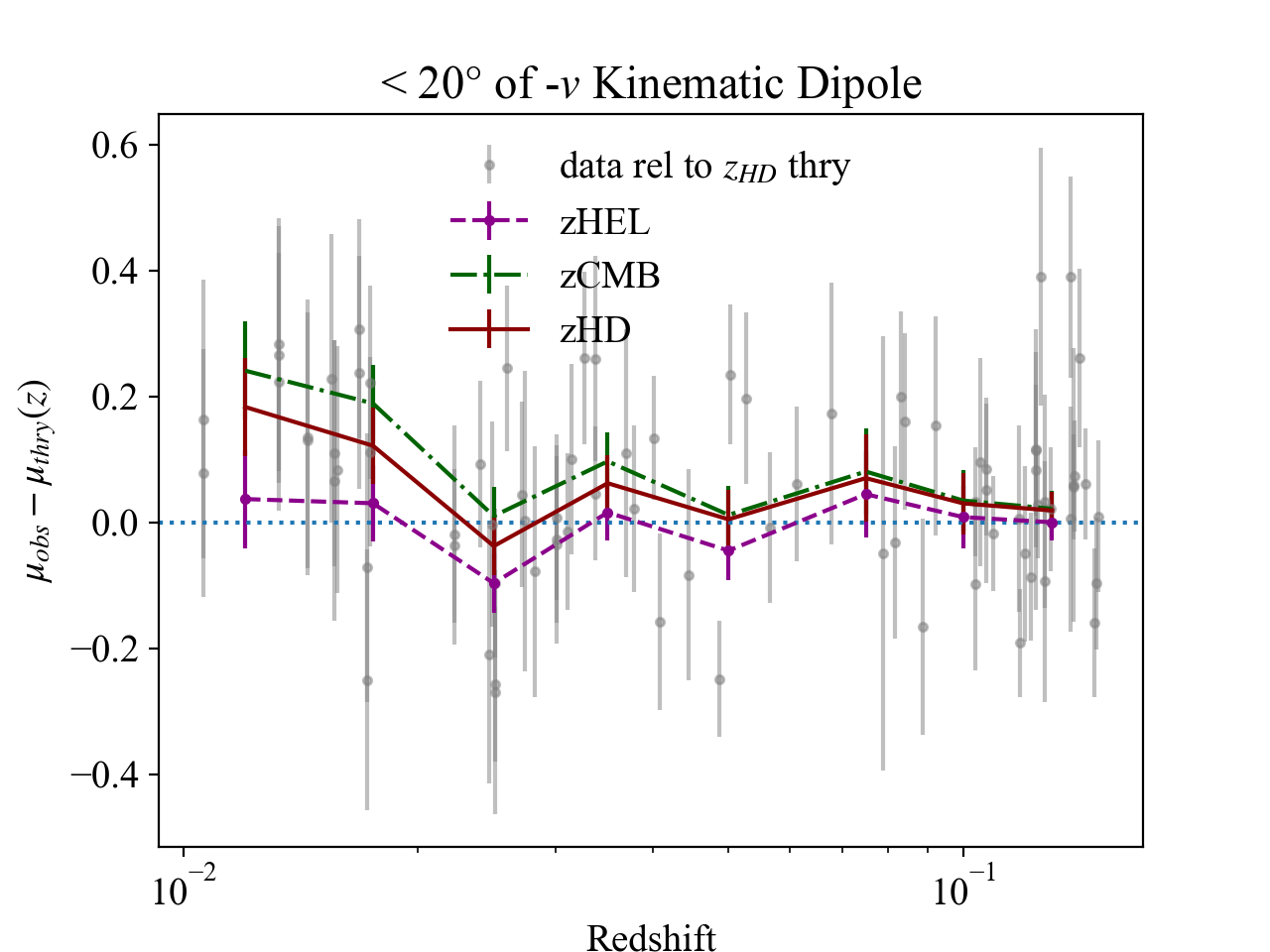}  
	    \includegraphics[width=.48\textwidth]{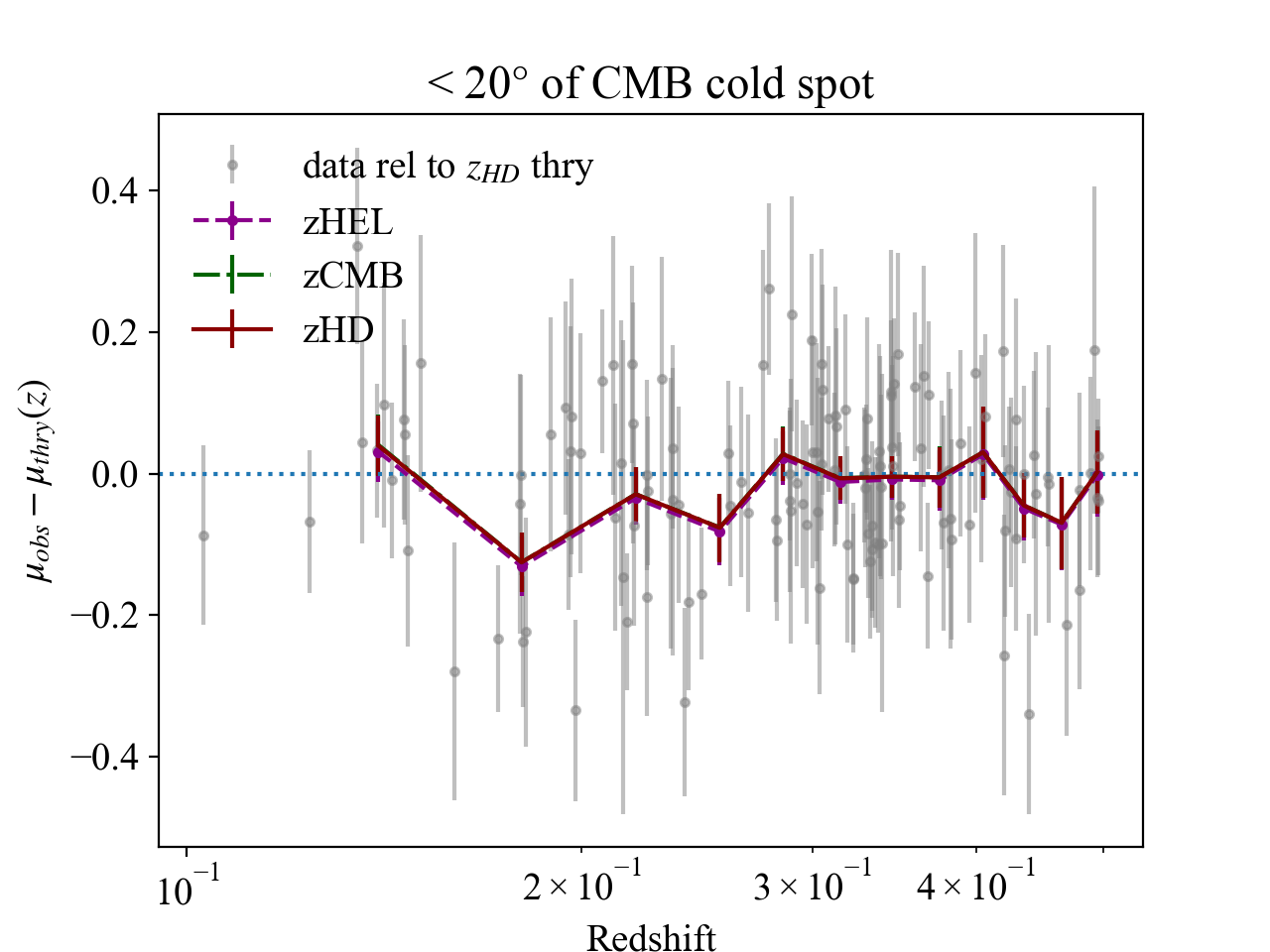}  
        \caption{Healpix (NSIDE=16) Hubble residual sky maps (colorbar is residual magnitudes) with 20 degree 2D-Gaussian kernel smoothing, and Hubble residuals for two selected apertures. $z>0.01$ is applied. { Dots show} the locations of the SNe in the Pantheon+ sample, { with white dots showing the nearby SNe ($z<0.15$) and black dots the distant SNe ($z>0.15$).} \textbf{Top left:}) Hubble diagram corresponding to the baseline analysis utilizing both $z_{\rm CMB}$ dipole corrections and 2M++ peculiar velocity corrections. The circled regions designate the 20 degree regions centred on the negative CMB dipole (red) and CMB cold spot directions (blue). {\red The small circle in the top right (and x in bottom left) of each figure represent the direction (and opposite direction) of the motion causing the CMB dipole} \textbf{Top right:}) same as top left but instead using 2MRS peculiar-velocity corrections \textbf{Middle left:}) same as top left but instead not applying any peculiar-velocity corrections \textbf{Middle right:}) same as top left but instead not applying either peculiar-velocity corrections nor the CMB dipole correction. \textbf{Bottom left:}) 20 degree region aligned with the (opposite) CMB dipole velocity depicting Hubble diagram residuals as a function of redshift. \textbf{Bottom right:}) same as bottom left but with aperture centered at the CMB cold spot ($l=209^\circ$, $b=57^\circ$), { and over a higher redshift range}.}
        \label{fig:skyheatmaps} 
    \end{figure*}

\subsection{Local Structure in the SN~Ia Hubble Diagram}
Large compilations of SN distances have provided impetus for searches of local structure, over/underdensities, and proper motion (e.g., \citealp{Matthews16,Soltis19,Hu20}). As an initial study, we create sky maps of the SN Hubble diagram residuals (see Fig.~\ref{fig:skyheatmaps}) and examine two specific areas on the sky that have been documented in the literature and have sufficient SN statistics in the Pantheon+ sample for study.

\subsubsection{The CMB Kinematic Dipole}

The motion of the Milky Way and Solar System relative to the CMB rest frame ($v= 369.82$\,km\,s$^{-1}$) is corrected for following \cite{carr21} and \cite{peterson21}. The effect of the CMB dipole motion can be seen in the $z_{\rm HEL}$ sky map  (middle right of Fig.~\ref{fig:skyheatmaps}), where $z_{\rm HEL}$ is the heliocentric redshifts. The $z_{\rm CMB}$ skymap (middle left of Fig.~\ref{fig:skyheatmaps}) has the CMB dipole-causing peculiar redshift removed following Eq.~7 of \cite{peterson21}. The direction of the CMB dipole, $l = 264^\circ$ and $b = 48^\circ$ (red o in Fig.~\ref{fig:skyheatmaps}), is shown for reference as well as { its antipole (red x)}.

As discussed in Section~\ref{subsubsection:pvs}, we examine different velocity reconstructions due to local structure that include estimates of the bulk flow; these are the 2M++ \citep{Carrick15} and 2MRS \citep{lilownusser} corrections and are shown in the top row of Fig.~\ref{fig:skyheatmaps}. These corrections also include the CMB dipole correction. \cite{peterson21} show that the peculiar velocity corrections overall reduce the Hubble residual scatter by $\sim10\%$, { and this is qualitatively confirmed in our maps.  The heliocentric map shows a strong dipole as expected; the $z_{\rm CMB}$ map shows the dipole somewhat removed but with an overcorrection (as expected at low-$z$ because local galaxies share some of our motion); and both $z_{\rm HD}$ maps show that the peculiar velocity corrections have removed most of the overcorrection}. 

However, both reconstructions produce a { small} signal that can be seen in the maps in the direction opposite the motion causing the CMB dipole. This signal is found to be local, at $z<0.02$, and grows with decreasing redshift until $z\approx0.01$ (bottom left of Fig.~\ref{fig:skyheatmaps}). A possible reason that there is a residual signal in the negative dipole direction in both the $z_{\rm CMB}$ and peculiar velocity corrected redshifts is that the MW motion is coupled with the motion of nearby galaxies in a way that is not yet sufficiently modelled.  { It is also likely that this is due  to low-number statistics (this is only a $1\sigma$ deviation) and the uneven sky coverage (the SNe in this region are mostly clustered in Stripe-82). Lastly we note that the positive residuals are driven by SNe at $z<0.02$, and thus are not included in the SH0ES \citep{SH0ES21} sample and inference of H$_0$.}

\subsubsection{The CMB Cold Spot}
The ``CMB cold spot," a $5^\circ$ region of $-70\mu K$~centered at ($l\sim209^\circ$ , $b\sim -57^\circ$), was first detected in data from the Wilkinson Microwave Anisotropy Probe \citep{vielva2004,cruz06}, and subsequently in Planck data \citep{planckcold14}.
Evidence for an underdensity aligned with the CMB cold spot was
presented by \cite{Rudnick07}. \cite{szapudi2014cold} and \cite{kovacs22} subsequently found the Eridanus supervoid in the direction of the cold spot at $z \approx 0.15$. However, it is not clear if the alignment of Eridanus and the CMB cold spot is causal or coincidental.

We find a signal in the Pantheon+ Hubble diagram when examining SNe within a $20^\circ$ radius of the location of the CMB cold spot {({ blue circle region in} top-left Fig.~\ref{fig:skyheatmaps})}. The difference in Hubble diagram residuals as a function of redshift is shown in the bottom-right panel of Fig.~\ref{fig:skyheatmaps}. There are 9 SNe in this region of the sky with redshifts on the near side ($0.12<z<0.15$) and there are { 14} SNe on the far side ($0.15<z<0.20$) of the proposed void at $z=0.15$, and there is a Hubble residual difference of { $-0.15 \pm 0.06$}\,mag between these two sets of SNe. For an estimate of the significance, we examine 1000 randomly selected $20^\circ$ apertures across the sky with at least {  8 SNe in each of the near and far redshift ranges split on redshifts between 0.08 and 0.20, and find that deviations with a similar significance occur only 0.2\% of the time.  We note however, that there are not many independent regions that satisfy the selection criteria and the vast majority of the SNe in the cold-spot selection come from the small deep-field patch within that region. Taking 100 random samples of 10 degree radius from the largest densely sampled region in Pantheon+ (Stripe-82 region) we find no other patch has a significance that exceeds 1.6$\sigma$, making the Eridanus patch the most significant step at that redshift in our data.}

\section{Discussion}

~\label{sec:Discussion}

This analysis is the latest in a series of papers that attempt to both grow the compilation of measured SN~Ia light curves and improve on the systematic floor. The two most recent compilations and analyses are those of JLA and Pantheon, which respectively included $\sim40\%$ and $\sim60\%$ of the SN light curves analyzed here. As seen in Fig.~1 of \cite{plussample}, the majority of the statistical increase for Pantheon+ is in the addition of numerous low-redshift samples extending down to $z=0.001$. However, the largest differences in the Hubble diagram are not solely the result of statistical increase, but rather due to improvements in our methodology. 

\begin{figure}[h]
\begin{centering}
\includegraphics[width=0.49\textwidth]{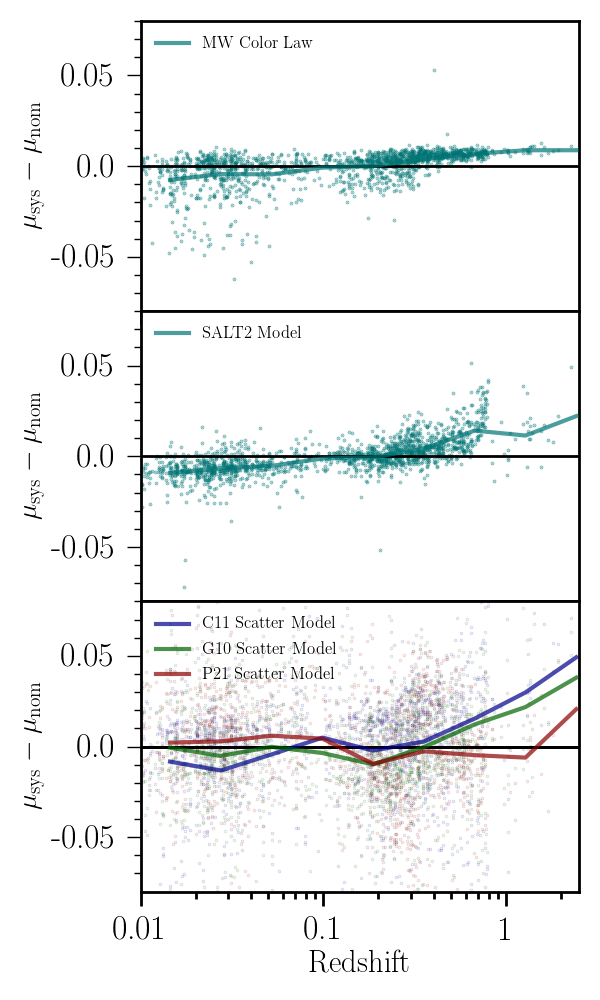}
\caption{Largest differences in analysis compared to \cite{Betoule14} and \cite{Scolnic18}.  (Top panel) Updating the extinction curve used in the light-curve fitting from CCM to F99; (Middle panel) Updating the SALT2 model, as discussed in \cite{fragilistic}; (Bottom panel) Changing the baseline assumption for the intrinsic scatter to the P21, G10, and C11 models.}
\label{fig:muchange}
\end{centering}
\vspace{.1in}
\end{figure}

We show in Fig.~\ref{fig:muchange} the difference in inferred distance-modulus values (marginalized over $M$) for the Pantheon+ sample relative to the assumptions used in the JLA analysis, for the three most significant improvements presented in this work. First is the update in the flux cross calibration to the Fragilistic solution, which impacts both the training of the SALT2 model and the zero-points used in light-curve fitting. Second is the impact from updating the MW extinction curve used in JLA \citep{ccm} to the \cite{Fitzpatrick99} relation that is used here. Third is the change resulting from improved modeling of the SN~Ia intrinsic scatter; while in this work we adopt the BS21 model, we include the models developed for JLA (G10 and C11) as systematics. Each of these changes has been motivated externally by previous works (e.g., \citealt{fragilistic,Schlafly11,bs20}); however, they nonetheless cause shifts in ${\rm d}\mu/{\rm d}z$ of $\sim0.05$, or $\sim0.04$ in $w$. Finally, because all of three of these changes have the same sign of ${\rm d}\mu/{\rm d}z$ slope, rather than canceling each other, when combined in this work they result in a $\sim0.1$ difference in the constraint on $w$ relative to JLA {(after combining with CMB).}

As discussed by \cite{Scolnic_decadal}, the constraining power of large samples of SNe~Ia extends beyond inferences of H$_0$ and $w/\Omega_M$. Large compilations of low-$z$ SNe~Ia enable precision measurements of the local growth-of-structure, typically parameterized by $f\sigma_8$ (e.g., \citealp{Huterer17,  Stahl21}). Work is ongoing for this measurement using the Pantheon+ sample (Boruah et al.,~in prep.), which will include validation with simulations as well as propagation of the covariance matrix, which previously would have limited effect on $\sigma_8$ calculations owing to smoothing/binning over redshift.

\begin{figure}[h]
        \centering 
	    \includegraphics[width=.48\textwidth]{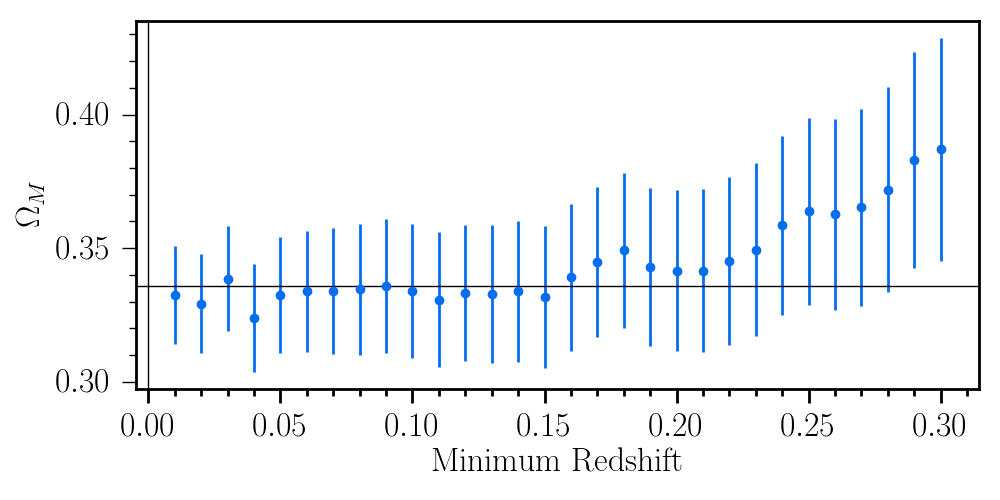}  
	    \includegraphics[width=.48\textwidth]{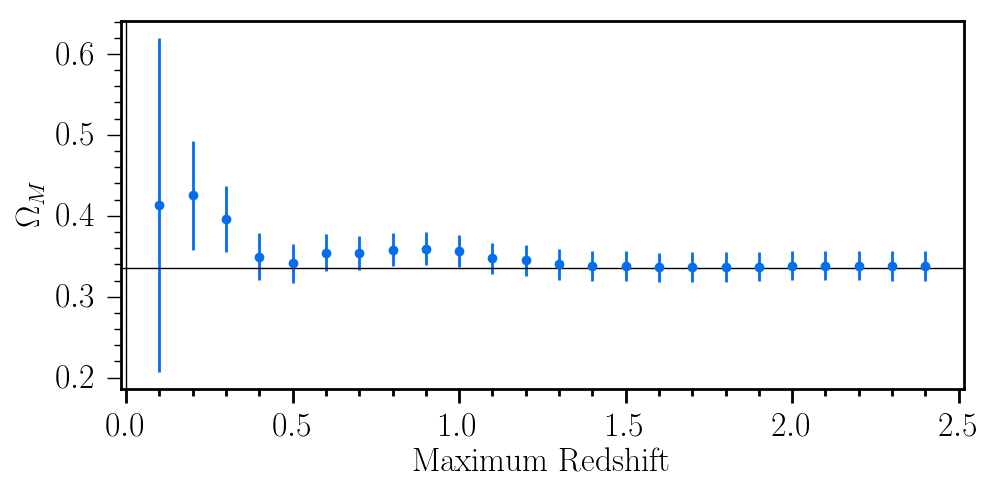}  
        \caption{Constraints on $\Omega_M$ in Flat$\Lambda$CDM when the bounds of the redshift range of the sample are changed. In the top panel, the minimum redshift is varied. The nominal minimum redshift is 0.01 for Pantheon+ cosmology fits without SH0ES.  In the bottom panel, the maximum redshift is varied. The nominal maximum redshift is 2.4 for all fits.}
        \label{fig:omz} 
    \end{figure}

While in Sec.~\ref{sec:Results} we show a Healpix map of Hubble residuals across the sky, there are additional and related tests of anisotropy that can be performed with these data. Previous analyses of the first Pantheon sample (e.g., \citealp{Sarkar,Soltis19,andrade18,Brownsberger19}) typically search for radial or hemispherical residuals across the sky.  The addition of statistics in the low-redshift sample and improved accounting in Pantheon+ would particularly strengthen these types of studies. A search for matter over/underdensities was performed by \cite{Colgain19}, which varied the minimum and maximum redshift in the original Pantheon sample and redetermined cosmological constraints. \cite{Colgain19} found for Pantheon that $\Omega_M$ could be $<0$ for a low maximum $z$ of $\sim0.15$, though with only $\sim2\sigma$ difference compared to the value of $\Omega_M$ from the full sample. We show a similar test in Fig.~\ref{fig:omz} and find relatively stable values of $\Omega_M$ with no signs of the underdensity seen by \cite{Colgain19}.

The main goal of this work, constraints from SNe~Ia alone for a Flat$w$CDM model, results in stat+syst uncertainties of  $^{+0.058}_{-0.063}$ and 0.13 for $\Omega_M$ and $w$, respectively. This represents a factor of \fomimprov~improvement in figure of merit over the original Pantheon {(stat+syst uncertainties 0.072 and 0.22 for $\Omega_M$ and $w$)}. This cannot be explained solely by statistical improvements, but rather is also due to a leap in systematics methodology over the original Pantheon and JLA. As shown by \cite{binning21}, cosmology uncertainty budgets are improved by a factor of $\sim1.5$ when not binning or smoothing data and covariance. In Appendix~B we discuss and show a binned error budget for comparison and find a similar factor of 1.5 improvement from this choice alone. In examining the unbinned error budget in Table~\ref{Tab:inputsys}, it can be seen that several systematics are no longer impacting SN~Ia cosmology analyses as strongly as had previously been thought. One such example is the negligible size of the parent population systematic despite including three additional sources of scatter model uncertainty, as was also seen by \cite{popovic21b}. This, as well as the reduction of a number of other systematics in comparison to their size in binned analyses (also shown in Appendix Table~\ref{Tab:binnedinputsys}), is due to the power of the large datasets themselves to self-constrain the size of systematic uncertainties when the systematic itself is not solely degenerate with the cosmological model parameterization. This is especially important because it brings this work from potentially being dominated by systematics to rather being dominated by statistical uncertainties. Furthermore, as shown by \cite{binning21}, as datasets grow in size, many systematics will continue to shrink without any additional effort. Lastly, it is important to note that approaches such as the Approximate Bayesian Computation method given by \cite{Jennings16} will not be able to make use of this self-constraining benefit unless additional parameters are included to allow the data themselves to scale the input sizes of the systematic uncertainties ($S_{\rm sys}$ in \citealt{binning21}).

While the SN~Ia mass step has received much attention in the last decade, we find here that its contribution to the error budget is exceedingly small. Unlike previous analyses, the mass-step treatment in this work is based on a SN color and dust-dependent model (BS21). We find that this more physical model results in smaller scatter in the Hubble diagram (Table~\ref{tab:nuisance}) and better $\chi^2$ relative to cosmological models which then results in smaller systematic uncertainties. We note that properties of SN~Ia host galaxies other than stellar mass have been seen to correlate with SN~Ia Hubble diagram residuals. Star-formation rate, specific star-formation rate (sSFR), stellar-population age, and metallicity have all been shown to correlate to varying degrees with the distance-modulus residuals after standardization \citep{Sullivan2010,Lampeitl2010,Childress2013,Rose19,Rigault2013}. For this reason, {using sSFR values presented by S22}, we also examined the size of a sSFR step in the subset of the Pantheon+ sample for which we have obtained sSFR measurements ($z<0.2$). Without applying any bias corrections, we find a significant step in sSFR (across the median sSFR) of $0.031\pm0.011$. However, after applying the nominal set of dust and mass-based bias corrections (BS21) used in this analysis, we find {a step in sSFR of} $0.008\pm0.011$, consistent with zero. This is likely due to galaxy properties (i.e., stellar mass) being linked to dust properties, and that applying a dust-mass correction is accounting for most, if not all, of the correlations with sSFR and is also tracing the dust distribution.

Going forward, statistical constraints on $w$ and $\Omega_M$ from SNe will improve significantly owing to upcoming datasets from SN programs of the Dark Energy Survey \citep{D'Andrea18}, Zwicky Transient Facility (ZTF; \citealt{Dhawan22}), Young Supernova Experiment (YSE; \citealp{Jones21}), Legacy Survey of Space and Time (LSST; \citealp{Mandelbaum18,Sanchez21}), Nancy Grace Roman Telescope \citep{Hounsell18}, etc.  It is likely that these future datasets will improve the statistical precision by a factor of 100 \citep{ScolnicKilo}. 

{ The size of systematic errors on cosmological parameter estimates matched the statistical errors for JLA and the original Pantheon.  Systematic uncertainties in this work have been reduced in comparison to Pantheon, and while their impact is still significant, it is no longer the dominant component of the total uncertainty. With the coming surveys,} systematics will also likely improve alongside the increase in statistics, as has been the case for previous analyses over the last two decades {, and as expected from the impact of systematic `self-calibration' described in \cite{binning21}}. 

{ As shown in the systematics error budget Table~\ref{Tab:inputsys}, the dominant sources of systematic uncertainty are now from 1) the combination of SALT2 training and calibration of surveys, 2) potential redshift measurement biases, and 3) Milky Way dust systematics. Fortunately there are paths forward for each of these. For survey flux calibration, dedicated programs are needed and there are currently multiple paths underway to improve the fundamental calibration of SN~Ia samples and how they are tied to various other samples (e.g., \citealp{Regnault2015,Narayan19,stubbs2015precise}). There is also ongoing work (Taylor et al. in prep) to train the SALT2 model with more photometric systems which has already shown promising improvements to systematic uncertainties and the ability to constrain rest frame U-band. The systematic from the redshift measurement floor has the potential to be reduced using improved cosmology fitting methodology, although the extent to which the data itself can constrain the size of this floor remains unproven. Alternatively, future large surveys can use multiple spectroscopic instruments and redshifting codes to mitigate potential sources of redshift measurement bias. The Pantheon+ sample is especially sensitive to Milky Way dust systematics because of the differences in the samples used for low and high redshift. At low redshift, to obtain sufficient statistics in a volume limited sample, we have used SNe across the sky and with up to 0.2 in MWEBV, whereas the high redshift surveys have been carried out in low extinction regions of the sky (MWEBV$<$0.05). Future surveys of larger volumes will be able to mitigate this with a plethora of both low and high redshift in low MW extinction regions on the sky.}  

Throughout this work, there are a number of upstream components of this analysis that impact downstream analysis steps: i.e. new calibration (or MWEBV Maps/Color law) motivates new SALT2 training, which motivates new fitting of the SN parent populations, which motivates new bias corrections. The Pippin framework \citep{Pippin}, used extensively in this work, was intentionally developed to automate and asynchronize this multistep type of analysis; however, it has yet to incorporate aspects such as the SALT2 retraining \citep{taylor21} or population fitting \citep{popovic21b}. Likely, this framework will need to expand for future analyses.

{There is an alternate approach to obtaining cosmology constraints from SNe that has been gaining traction over the last decade. Bayesian Hierarchical Models (BHM) have been developed that utilize
bias-corrected observables \citep{BAHAMAS} and that incorporate selection effects
directly into the model \citep{UNITY} or likelihood \citep{Hinton18}. However, unlike BBC in combination with CosmoSIS, these methods have not been validated with large realistic simulations. As noted in Appendix~\ref{appendix:dr}, we release as part of this analysis 10 realistic simulations of the Pantheon+ dataset for such validations. }

While constraints on $w$ should easily improve with upcoming large SN samples, the road to improving constraints on H$_0$ is more challenging. {There are a limited number of SNe~Ia that will explode in the near future within a $\sim40$\,Mpc radius, a constraint due to {\it HST} discovery limits of Cepheids.} At roughly one SN~Ia per year, it will take several decades to double the current sample of $42$ SNe calibrated by SH0ES Cepheid hosts. Fortunately, we find that the systematics in the measurement of H$_0$ from the SNe are at the scale of 0.3\,km\,s$^{-1}$\,Mpc$^{-1}$ as shown in Fig.~\ref{fig:sysh0}. This is consistent with the general finding of \cite{brownsberger21}, who showed how robust H$_0$ is to systematic uncertainties in comparison to the relatively calibration-sensitive constraints of $w_0$ or $\Omega_M$. Lastly, there is ongoing work that combines the progress used here by \cite{peterson21} and applies it to a ``two-rung" distance-ladder analysis, in which SNe are excluded from the distance ladder \citep{kenworthy22}.

\section{Conclusion}
\label{sec:Conclusions}

This work is the culmination of a number of supporting analyses as part of the Pantheon+ effort. In this work, we summarize the various inputs and analyses required to combine the supporting works and ultimately measure distances and cosmological parameters. For the first time we are able to measure the cosmic expansion history and the local distance ladder H$_0$ simultaneously. We combine our results with additional external probes. Importantly, we release a number of data and analysis products {to facilitate reproducing our work by the community}. This includes a joint covariance of SNe used for measurements of H$_0$ and $w$.  

For our main results, we find $\Omega_M=$~\fLCDMOMPan\ in Flat$\Lambda$CDM from SNe~Ia alone. For a flat$w_0$CDM model, we measure $w_0=$~\wPanSH\ from SNe~Ia alone and $w_0=$~\wPanPlanckBAO\ when combining SNe with constraints on the CMB and allBAO; both are consistent with a cosmological-constant model of dark energy.  We also present the most precise measurements to date on the evolution of dark energy in a Flat$w_0w_a$CDM universe, and measure $w_a=$~\wwawaPan\ from Pantheon+ alone and $w_a=$~\wwawaPanPlanckBAO\ when combining with CMB and BAO data. Finally, while nominal constraints on H$_0$ are presented in a companion paper by the SH0ES team (R22), we perform joint constraints of H$_0$ with expansion history and find H$_0=$~\HPanSH\ in Flat$w$CDM, and we show how systematic uncertainties in measurements of the SN component of the distance ladder cannot account for the current level of the ``Hubble tension."

\bibliography{paper}
\bibliographystyle{aasjournal}
\section{Acknowledgements}
 D.S., D.B., and A.R. thank the John Templeton Foundation for their support of grant \#62314. D.B. acknowledges support for this work provided by NASA through NASA Hubble Fellowship grant HST-HF2-51430.001 awarded by the Space Telescope Science Institute (STScI), which is operated by the Association of Universities for Research in Astronomy, Inc., for NASA, under contract NAS5-26555. D.S. is supported by DOE grant DE-SC0010007, the David and Lucile Packard Foundation, and NASA under Contract No. NNG17PX03C issued through the WFIRST Science Investigation Teams Programme. 
 We acknowledge the generous support of Marc J. Staley, whose fellowship partly funded B.E.S. whilst contributing to the work presented herein as a graduate student. 
A.V.F. is grateful for support from the TABASGO Foundation, the Christopher R. Redlich
Fund, the U.C. Berkeley Miller Institute for Basic Research in Science
(in which he was a Miller Senior Fellow), and many individual donors.
S.N. thanks the STFC Ernest Rutherford Fellowship for support via grant ST/T005009/1
 L.K. thanks the UKRI Future Leaders Fellowship for support through the grant MR/T01881X/1. This work was completed in part with resources provided by the University of Chicago’s Research Computing Center.
 The Katzman Automatic Imaging Telescope (with which the LOSS samples were obtained) and its ongoing operation were made possible by donations from Sun Microsystems, Inc., the Hewlett-Packard Company, AutoScope Corporation, Lick Observatory, the NSF, the University of California, the Sylvia \& Jim Katzman Foundation, and the TABASGO Foundation. Research at Lick Observatory is partially supported by a generous gift from Google.
 
 Simulations, light-curve fitting, BBC, and cosmology pipeline are managed by \texttt{PIPPIN} \citep{Pippin}. Contours and parameter constraints are generated using the \textsc{ChainConsumer} package \citep{Hinton16}. Plots are generated with Matplotlib \citep{matplotlib}. We use astropy \citep{astropy}, SciPy \citep{scipy}, and NumPy \citep{numpy}. Analysis and visualisations provided in part by https://github.com/bap37/Midwayplotter. 
 
 Brout thanks his spouse Isabella and their future daughter for their support as the due date is rapidly approaching!

\appendix
\section{Additional Formalism for Distance and Uncertainty Estimation}
\label{appendix:biascor}

\begin{table*}
   
    \caption{Distance Bias (and Uncertainty) Estimation for Scatter Models}
    \begin{tabular}{lll}
        \toprule
		 & G10/C11 & BS21/P21  \\ 
		\hline
	Dimensionality & 7d ($z,x_1,c,M_{\star},\gamma,\alpha,\beta$)& 4d ($z,x_1,c,M_{\star}$)\\	
	Mass-step correction & $\gamma$ a fitted parameter & $\gamma$ corrected for within $\delta_{\rm bias}$ ($\gamma$ and $\delta_{\rm host}$ consistent with zero) \\
	Intrinsic Scatter Floor & $\sigma_{\rm floor}^2=\sigma_{\rm gray}^2$ & $\sigma_{\rm floor}^2=\sigma^2_{\rm scat}(z_i,c_i,M_{\star})+\sigma^2_{\rm gray}$, applied when $f(z,c,M_\star)>1$ \\
	Selection Effects & $f(z,c)$  &  $f(z,c,M_\star)\leq1$, applied when $\sigma^2_{\rm scat}(z_i,c_i,M_{\star})=0$   \\
    \hline

	\end{tabular}	\vspace{3mm}
	~\\
	\raggedright Notes: Formalism for 4d and 7d bias corrections are described by \cite{popovic21a} that depend on the intrinsic scatter model assumed --- either G10/C11 or BS21/P21.  The statistical and intrinsic scatter uncertainties from Eq.~\ref{eq:TrippErr} are shown here; the other uncertainty components from Eq.~\ref{eq:TrippErr} are independent of the scatter model.
	\label{tab:biassum}
	\end{table*}

As shown in BS21, SN~Ia scatter has both a color and host-mass dependence (increasing scatter) and a redshift dependence that arises from selection effects (decreasing scatter). In this work we introduce a new method of accounting for the uncertainties using the scatter model predictions.
We include $\sigma_{\rm scat}(z,c,M_\star)$ from simulations as an additive uncertainty inside Eq.~\ref{eq:TrippErr} rather than the multiplicative uncertainty $f(z,c,M_\star)$ on the computed $\sigma_{\rm meas}$ that has been used in past analyses. 
{The $\sigma_{\rm scat}(z,c,M_\star)$ term is computed from simulations that use the choice of scatter model.} The BBC process, after correcting distances for selection effects, determines the magnitude of $\sigma_{\rm scat}(z,c,M_\star)$ in each $z,c,M_\star$ bin by requiring that the observed-simulated distance reduced $\chi^2$ in each bin is unity. If the simulations {using a model of intrinsic scatter fully describe the observed scatter in the data, the uncertainty modeling term in Eq.~\ref{eq:TrippErr}, $\sigma_{\rm scat}(z,c,M_\star)$,} will cause $\sigma_{\textrm{gray}}$ to be 0.

{In the case of the decrease in observed scatter at high redshift arising from} only intrinsically bright/blue events being selected at the limits of the telescope \citep{Kessler2015}, we instead apply as a downscaling of $f(z,c,M_\star)$ of the reported measurement uncertainty and set $\sigma_{\rm scat}(z,c,M_\star)=0$. Conversely, for bins of $z,c,M_\star$ with $\chi^2$ greater than unity, the necessary $\sigma_{\rm scat}(z,c,M_\star)$ is applied and $f$ is set to 1. The resulting $f(z,c,M_\star)$ and $\sigma_{\rm scat}(z,c,M_\star)$ found from simulations are applied to the Pantheon+ data.

{The method and dimensionality for the application of bias corrections is dependent on the adopted scatter model.} Table~\ref{tab:biassum} summarizes the differences between the two main methods used in this work, the first of which is applied when assuming the BS21/P21 scatter model, and the other when assuming the G10 or C11 scatter model.  The main difference between these groups of scatter models, as discussed in Sec.~\ref{sec:Inputs}, is whether the intrinsic scatter is driven by diversity in the reddening ratios $R_V$ of the light curves, which affects the application of bias corrections.  For both analysis paths, we follow the methodology introduced by \cite{popovic21a}.

\section{Binned Systematic Error Budget}
\label{appendix:binnedsys}
In Table~\ref{Tab:binnedinputsys} we show a systematic error budget that is nearly identical to what was performed in Table~\ref{Tab:inputsys}, except that the dataset ($\Delta D$) and covariance matrix ($C_{\rm stat+syst}$) are binned in 20 redshift bins. This error budget is similar to the methodology performed in the most recent SN cosmology analyses where binned covariance matrices were used (e.g., Pantheon, DES3YR \citealt{Brout18b}) and where smoothed data vectors and matrices (which were shown to be equivalent to binned) were used (JLA). The total systematic error when binning is a factor of 1.5 larger (0.029) than when not binning the dataset (0.019).

Systematics that improve the most with unbinned matrices are those with smaller $\sigma w^{\rm unbinned}_{\rm sys}/\sigma w^{\rm binned}_{\rm sys}$. Binned analyses collapse valuable information in the Hubble diagram down to a single dimension, redshift. We find that as expected, the redshift bias systematic does not improve much at all. This is because systematics that only exhibit redshift dependence are degenerate with cosmological model parameters and cannot be self-constrained by the data as easily. Systematics that exhibit dependence in other parameters (such as SN color) can be drastically reduced in SN~Ia cosmological parameter error budgets when not performing binned analyses.

\begin{table}
\centering
\caption{Comparison of Binned and Unbinned Systematic Error Budgets}
\begin{tabular}{lccc}
\toprule

Description  & $^a\sigma w^{\rm binned}_{\rm sys}$&$^a\sigma w^{\rm unbinned}_{\rm sys}$ &  $\sigma w^{\rm unbinned}_{\rm sys}/\sigma w^{\rm binned}_{\rm sys}$\\
\hline \\

\textbf{\ul{All Systematics}} & 0.029 & 0.019 & 0.66\\
\\
\hline \\

\textbf{\ul{Calibration}}\\ %

SALT2 Train \& $^b$LCFIT  & 0.019 & 0.009 & 0.47\\
SALT2 Method & 0.009 & 0.008 &  0.88\\
CSP Tertiary Stars  & 0.005 & 0.003 & 0.60\\
$^d${\it HST} & 0.002 & 0.003 & 1.50 \\
\\

\textbf{\ul{Redshifts}}\\
$^cv_{\rm pec}$ Map & N/A & 0.002 & N/A\\
Redshift Bias & 0.012 & 0.011 & 0.92\\
\\

\textbf{\ul{Astrophysics}}\\

Intrinsic Variations & 0.009 & 0.002 & 0.18 \\
MW $E(B-V)$ & 0.012 & 0.008 & 0.67\\
MW Color Law & 0.007 & 0.006 & 0.86\\
Mass Step & 0.001 & 0.001 & 1.00\\
\\

\textbf{\ul{Modeling}}\\

Selection Efficiency & 0.008 & 0.004 & 0.50\\
Populations & 0.011 & 0.000 & 0.00 \\

\\

\toprule\\
\label{Tab:binnedinputsys}
\vspace{-4mm}
\end{tabular}
{\raggedright
\\
$^a$Constraints are combined with Planck prior.\\
$^b$LCFIT denotes zero-points and filter central wavelengths have been varied during light-curve fitting.\\
$^c$Due to implementation methodology of this systematic, it has not been performed in the binned case.\\
$^d$The increase in the ``{\it HST}" systematic is likely due to noise as the values are very small for both binned and unbinned.\\
\vspace{26mm}
}
\end{table}

\section{Products}
\label{appendix:dr}

The following data products {that are provided in part by the full suite of Pantheon+ supporting papers} are released publicly in machine readable format\footnote{{Will be made available after publication}} at \href{pantheonplussh0es.github.io}{pantheonplussh0es.github.io} and as part of \texttt{SNANA} and CosmoSIS (where noted).

\begin{itemize}
 
\item Light-Curve Photometry, Redshifts, and Host-Galaxy Properties; \textit{from S22 and \cite{carr21}}
\item {Trained SALT2-B22 Model; \textit{from \cite{fragilistic}}}
\item SALT2 Fit Parameters; \textit{from S22}
\item 10 Catalog Level Simulations of Pantheon+ Light-Curve Fit Parameters; \textit{this work}
\item SN/Host Redshifts and Peculiar Velocities; \textit{from} \cite{carr21}
\item SN Distance Modulii and Redshifts; \textit{this work, \cite{carr21}}, see Table~\ref{tab:hubblediag}\footnote{$\sigma_{m_B{\rm corr}}^{\rm diag}$ in Table~\ref{tab:hubblediag} is the error on standardized magnitude from the diagonal of the covariance matrix. It is for plotting purposes only and not to be used for cosmological fits.}
\item SN Distance Covariance; \textit{this work}
\item Cepheid Host Distances; \textit{from R22}
\item Cepheid Host Distance Covariance; \textit{from R22}
\item SN~Ia + Cepheid Host Cosmology Likelihood; \textit{this work}
\item SN Cosmology Chains; \textit{this work}

\end{itemize}

\newpage
\newpage
\newpage
\afterpage{
\begin{longtable}{ p{.12\textwidth}  p{.085\textwidth} | p{.05\textwidth}  p{.05\textwidth}  p{.05\textwidth}  p{.05\textwidth}  p{.05\textwidth}  p{.055\textwidth}  p{.05\textwidth}  p{.045\textwidth} p{.05\textwidth}  p{.045\textwidth}  p{.045\textwidth}  p{.045\textwidth}  p{.045\textwidth} } 
\caption{The Pantheon+ Hubble Diagram}\\
\hline\hline
CID & Survey & $z_{\rm HD}$ & $\sigma_{z {\rm HD}}$ & $z_{\rm CMB}$ & $z_{\rm HEL}$ & $m_B^{\rm corr}$ & $\sigma_{m_B{\rm corr}}^{\rm diag}$ & $c$ & $\sigma_c$ & $x_1$ & $\sigma_{x_1}$ & $m_B$ & $\sigma_{m_B}$ \\ \hline\hline
\endhead
        2011fe & LOSS2 & 0.00122 & 0.00084 & 0.00122 & 0.00082 & 9.746 & 1.516 & -0.108 & 0.040 & -0.548 & 0.134 & 9.584 & 0.033 \\
        2011fe & SOUSA & 0.00122 & 0.00084 & 0.00122 & 0.00082 & 9.803 & 1.517 & -0.033 & 0.038 & -0.380 & 0.086 & 9.784 & 0.035 \\
        2012cg & LOSS2 & 0.00256 & 0.00084 & 0.00256 & 0.00144 & 11.470 & 0.782 & 0.101 & 0.018 & 0.492 & 0.024 & 11.816 & 0.024 \\
        2012cg & SOUSA & 0.00256 & 0.00084 & 0.00256 & 0.00144 & 11.492 & 0.799 & 0.122 & 0.039 & 0.713 & 0.084 & 11.880 & 0.036 \\
        1994DRichmond & LOWZ & 0.00299 & 0.00084 & 0.00299 & 0.00187 & 11.523 & 0.881 & -0.112 & 0.026 & -1.618 & 0.050 & 11.533 & 0.032 \\
        1981B & LOWZ & 0.00317 & 0.00084 & 0.0035 & 0.00236 & 11.542 & 0.614 & -0.005 & 0.031 & -0.445 & 0.165 & 11.664 & 0.034 \\
        2013aa & SOUSA & 0.00331 & 0.00085 & 0.00478 & 0.00411 & 11.207 & 0.594 & -0.104 & 0.054 & 0.513 & 0.152 & 10.891 & 0.106 \\
        2013aa & CSP & 0.00331 & 0.00085 & 0.00478 & 0.00411 & 11.300 & 0.580 & -0.158 & 0.036 & 0.633 & 0.139 & 10.844 & 0.100 \\
        2017cbv & CSP & 0.00331 & 0.00085 & 0.00478 & 0.00411 & 11.148 & 0.578 & -0.126 & 0.032 & 0.617 & 0.053 & 10.773 & 0.094 \\
        2017cbv & CNIa0.02 & 0.00331 & 0.00085 & 0.00478 & 0.00411 & 11.258 & 0.578 & -0.096 & 0.035 & 0.819 & 0.066 & 10.914 & 0.099 \\
        ...\\
\hline
\label{tab:hubblediag}
\end{longtable}
\noindent Full table available in machine readable format at\\ \noindent\href{https://iopscience.iop.org/article/10.3847/1538-4357/ac8e04}{https://iopscience.iop.org/article/10.3847/1538-4357/ac8e04}.
}

\end{document}